\documentclass[preprint,12pt,sort&compress]{elsarticle}
\usepackage[utf8]{inputenc}
\usepackage{color}
\usepackage{graphicx}
\usepackage{placeins}
\usepackage[margin=1in]{geometry}
\usepackage{appendix}
\usepackage{amsmath, amssymb, bm}
\usepackage{multirow}
\usepackage{placeins}
\usepackage{xfrac}
\usepackage[colorlinks=true,linkcolor=blue,citecolor=blue]{hyperref}
\usepackage[nameinlink,capitalize]{cleveref}
\usepackage[font=footnotesize,labelfont=bf]{caption}
\usepackage{subcaption}
\usepackage[version=4]{mhchem}
\usepackage{lineno}
\usepackage{bm}
\usepackage{comment}
\usepackage[T1]{fontenc}
\usepackage{anyfontsize}
\usepackage{newtxtext,newtxmath}

\usepackage{tablefootnote}
\usepackage{float}
\usepackage{enumitem}
\setlist{itemsep=0pt,topsep=0pt}
\usepackage{booktabs}

\journal{Journal of Nuclear Materials}

\begin{document}

\begin{frontmatter}

\title{Modeling oxygen-void interactions in uranium nitride}

\author[ncsu]{Mohamed AbdulHameed}
\author[lanl]{Anton J. Schneider}
\author[ncsu,inl]{Benjamin Beeler\corref{a}}
\ead{bwbeeler@ncsu.edu}
\author[lanl]{Michael W.D. Cooper\corref{b}}
\ead{cooper_m@lanl.gov}
\cortext[a]{Corresponding author}
\cortext[b]{Corresponding author}

\address[ncsu]{Department of Nuclear Engineering, North Carolina State University, Raleigh, NC 27695}
\address[lanl]{Los Alamos National Laboratory, Los Alamos, NM 87545}
\address[inl]{Idaho National Laboratory, Idaho Falls, ID 83415}

\begin{abstract}

Oxygen impurities in uranium nitride (UN) are reported to influence its swelling behavior under irradiation, yet the underlying mechanism remains unknown. In this work, we develop a first-principles model that quantifies the interaction of oxygen with voids and fission gas bubbles in UN, leading to a reduction in surface energy that can promote swelling. As an approximation, surface energetics are obtained from planar (001) slab calculations and applied to curved void surfaces. The analysis reveals that segregation of substitutional oxygen at surface nitrogen sites is the primary driver of surface energy reduction, $|\Delta \sigma|$, while oxygen in surface hollow sites plays a minor and sometimes counteracting role. $|\Delta \sigma|$ is most pronounced for small cavities ($R_v \lesssim$ a few nm) at intermediate temperatures that overlap the reported onset range for breakaway swelling in UN. Larger voids require higher temperatures for oxygen adsorption to significantly lower their surface energy. The temperature dependence of $|\Delta \sigma|$ exhibits three regimes: negligible reduction at low temperatures due to sluggish oxygen diffusion, a maximum at intermediate temperatures where oxygen incorporation is optimal, and a decline at high temperatures due to enhanced bulk solubility. A parametric analysis reveals that $|\Delta \sigma|$ depends strongly on oxygen concentration and cavity size, while its porosity dependence remains bounded because the geometric porosity prefactor is partly offset by the surface-site concentration. Our results suggest that oxygen-induced surface energy reduction is essential for reconciling the mechanistic swelling model of UN with experimental observations.

\end{abstract}

\begin{keyword}
Uranium nitride \sep Oxygen segregation \sep Surface energy reduction \sep Void nucleation \sep Fission gas swelling
\end{keyword}

\end{frontmatter}

\newpage

% \linenumbers

\section{Introduction}
\label{intro}

Uranium nitride (UN) is considered a promising fuel for fast nuclear reactors, space reactors, and potentially commercial light water reactors, due to its high uranium density and high thermal conductivity relative to UO$_2$ \cite{Wallenius2020, Uno2020}. However, early use of UN was limited by issues such as rapid oxidation in air, poor tolerance to water and steam, difficulty in sintering, and the need for nitrogen enrichment with N-15 due to the high neutron absorption of N-14 \cite{Wallenius2020, Uno2020}. Furthermore, many aspects of its behavior at high temperatures and under irradiation remain poorly understood, with scarce experimental data and incomplete mechanistic pictures \cite{Wallenius2020, Uno2020}. Several studies have sought to address this knowledge gap by, e.g., elucidating the atomistic mechanisms governing deformation in UN \cite{AbdulHameed2024b, AbdulHameed2025}, which are critical to pellet--clad mechanical interactions, as well as those underlying its nonlinear high-temperature heat-capacity \cite{AbdulHameed2026}. Additionally, a mechanistic multiscale model has been developed to describe its swelling behavior \cite{Rizk2025}.

UN experiences pronounced swelling driven by the growth of fission gas bubbles to sizes larger than those observed in UO$_2$ \cite{Ronchi1975, Ronchi1978, Rizk2025}, making dimensional stability a key concern for this fuel. The mechanistic model by Rizk \textit{et al.} \cite{Rizk2025} gives a multiscale description of fission-gas swelling and release in UN, coupling atomistic defect data to the BISON finite-element framework via the Simple Integrated Fission Gas Release and Swelling (SIFGRS) model. Their calculations, however, employed a constant void/bubble surface energy derived for oxygen-free UN.

A major challenge in utilizing UN as a nuclear fuel is the presence of oxygen impurities, which can markedly affect its swelling characteristics. Rogozkin \textit{et al.} \cite{Rogozkin2003} demonstrated that fuel swelling becomes substantially more pronounced when the oxygen concentration in UN surpasses the range of 1000 to 1500 parts per million (ppm). Consequently, Schuler \textit{et al.} \cite{Schuler2017} recommended limiting the maximum permissible oxygen content to 1000--1500 ppm to maintain dependable performance of fuel elements. Nevertheless, achieving UN powders with oxygen levels below 1500 ppm has thus far only been accomplished under controlled laboratory conditions \cite{Schuler2017}. This highlights the importance of investigating how oxygen interacts with UN's microstructural features, like voids and bubbles. 

Additionally, mechanistic models that aim to predict fission gas swelling in UN require knowledge of surface energies. Surface energy is a core input to the SIFGRS model utilized by Rizk \textit{et al.} \cite{Rizk2025} as it sets both the equilibrium pressure inside a bubble as a function of its radius, and the contact angle of intergranular bubbles. Therefore, understanding whether oxygen significantly alters the surface energy of voids or gas bubbles is essential for accurately modeling swelling behavior.

The effect of light impurities, such as oxygen, on vacancy cluster morphology has been extensively studied in metals like copper, nickel, and stainless steel \cite{Zinkle1987a, Zinkle1987b, Zinkle1990, Igata1998}. These studies have utilized elastic continuum models \cite{Zinkle1987a, Zinkle1987b}, thermodynamic models \cite{Zinkle1990}, and reaction rate theory \cite{Igata1998} to understand the interaction between impurities and void embryos among other vacancy cluster types. Usually, oxygen impurities impact swelling in metals and alloys by segregating around void surfaces \cite{McLean1957}, lowering their surface energy, thereby reducing the critical void radius in the void nucleation stage \cite{Zinkle1987a, Igata1998}. Experimental evidence from metal studies supports the notion that oxygen reduces the surface energy of small voids, typically around 50 vacancies \cite{Zinkle1990}. This reduction in surface energy is a key factor in stabilizing voids and fission-gas bubbles and may similarly affect the swelling behavior in UN. More strikingly, the theoretical studies by Zinkle \textit{et al.} \cite{Zinkle1987a, Zinkle1987b} predict that void formation in high-purity metals is energetically unfavorable, and impurities like oxygen are essential for voids to form. That is, void formation is greatly reduced or even suppressed in low-oxygen metals and alloys. It has been further suggested that oxygen might stabilize void embryos by preventing their collapse into dislocation loops, even if it does not directly reduce surface energy \cite{Zinkle1987b, Zinkle1990}.

Several first-principles studies have explored the properties of oxygen impurities in bulk UN and on its surfaces, calculating the oxygen incorporation, migration, and adsorption energies \cite{Kotomin2008, Kotomin2009, Zhukovskii2009JNM, Zhukovskii2009SS, Bocharov2011JNM, Bocharov2011SS, Bocharov2013, Lopes2016, Schuler2017, Zergoug2018, Sikorski2021}. These studies have demonstrated that oxygen atoms prefer to incorporate into pre-existing nitrogen vacancies, both in bulk UN and on the surface, rather than occupying tetrahedral interstitial sites \cite{Kotomin2008, Bocharov2013}. Oxygen impurities also reduce the migration energy of nearby nitrogen vacancies \cite{Kotomin2009}, indicating that oxygen can influence behaviors that depend on defect dynamics, such as creep. Additionally, it has been observed that O$_2$ molecules parallel to the (001) surface can spontaneously dissociate when centered over a hollow site or a nitrogen atom \cite{Zhukovskii2009JNM}.

A particularly relevant study on the effect of oxygen on swelling in UN is that by Schuler \textit{et al.} \cite{Schuler2017}, who investigated the thermodynamic and transport properties of oxygen in UN using a self-consistent mean-field (SCMF) model. The input energies for their model were derived from density functional theory with the addition of a Hubbard $U$ term (DFT+$U$). Schuler \textit{et al.} found that oxygen is thermodynamically more favorable in the bulk of UN rather than on the (001) surface, leading them to conclude that oxygen segregation on surfaces would be driven by kinetic factors rather than by thermodynamics. However, Kocevski \textit{et al.} \cite{Kocevski2022I} later demonstrated that UN modeled using DFT+$U$ exhibits imaginary phonon frequencies, indicating a dynamically unstable crystal structure. Consequently, the defect energetics used in the SCMF model by Schuler \textit{et al.} might need reconsideration. % as they were based on a dynamically unstable structure.

Oxygen occurs in fresh UN fuel in solution or as a separate UO$_2$ phase, dependent on O concentration and temperature \cite{Lyubimov2014}. UO$_2$ precipitates have been observed experimentally in regions of UN samples where the local oxygen concentration apparently exceeded the solubility limit \cite{Mishchenko2021}. Additionally, DFT studies indicate that oxygen interaction with UN surfaces can lead to the formation of oxynitrides (UO$_x$N$_y$) or pseudo-UO$_2$ structures \cite{Bocharov2013, Lopes2016}.

In this work, we use DFT data to parametrize a thermodynamic model based on defect reactions to study how oxygen impurities interact with voids and fission-gas bubbles in UN. Our model focuses on oxygen adsorption onto void surfaces, where it may reduce surface energy and potentially influence swelling. Throughout this work, we assume that inert gas atoms have no impact on surface energy; therefore, the surface energies of voids and fission-gas bubbles are considered identical. To the best of our knowledge, this is the first study that attempts to quantify the extent to which oxygen interaction with void surfaces affects the void nucleation behavior in UN. We assume that oxygen is the only impurity present, and do not account for the formation of UO$_2$. These simplifications allow us to concentrate on the primary thermodynamic factors that govern the interaction of oxygen with voids, providing an initial step toward understanding the role of oxygen in UN swelling.

\section{Methodology}

\subsection{DFT calculations}

DFT calculations performed in this work utilize the Vienna Ab-initio Simulation Package (VASP) \cite{Kresse1993, Kresse1996a, Kresse1996b} using the Perdew-Burke-Ernzerhof (PBE) \cite{Perdew1996} generalized gradient approximation (GGA) to the exchange-correlation (XC) functional. The pseudopotential models of uranium, nitrogen, and oxygen are based on the projector augmented wave (PAW) method \cite{Kresse1999} where the valence electron configuration of U is $6s^2$ $6p^6$ $6d^2$ $5f^2$ $7s^2$ (14 electrons), that of N is $2s^2$ $2p^3$ (5 electrons), and that of O is $2s^2$ $2p^4$ (6 electrons). Methfessel and Paxton’s smearing method \cite{Methfessel1989} of the first order is used with a width of 0.1 eV to determine the partial occupancies for each wavefunction.

UN adopts the NaCl structure (space group $Fm\bar{3}m$). $3 \times 3 \times 3$ supercells of ferromagnetic (FM) UN, containing 108 formula units (216 atoms), are used to model bulk UN and to study the properties of defects and impurities. The relaxed lattice parameter is $a_0$ = 4.861~\AA\ with a site-projected uranium magnetic moment of $\mu_\text{U} = 1.21 \mu_\text{B}$, reproducing the PBE FM benchmarks of Kocevski \textit{et al.} \cite{Kocevski2022I} and deviating by $-0.6$\% from the experimental value of 4.89 \AA\ \cite{Hayes1990I}. As an additional validation, a seven-point Birch-Murnaghan fit gives $V_0 = 3104.07$~\AA$^3$ for the supercell with $B_0 = 219.9$~GPa and $B_0' = 4.77$. The calculated formation enthalpy is $\Delta H_f(\mathrm{UN}) = -2.697$~eV/f.u. ($-260.18$~kJ/mol-f.u.), in direct agreement with the PBE FM value of $-2.696$~eV/f.u. reported by Kocevski \textit{et al.} \cite{Kocevski2022I}.

% This value is less exothermic than the experimental calorimetric value of $-288.8 \pm 11.8$~kJ/mol-f.u. reported by Goncharov \textit{et al.} \cite{Goncharov2022}, and also less exothermic than the thermodynamically optimized value of $-294.35$~kJ/mol-f.u. reported by Chevalier \textit{et al.} \cite{Chevalier2000}. Thus, PBE reproduces the Kocevski benchmark exactly, but underbinds UN relative to experimental and thermodynamic-assessment values by approximately $0.30$--$0.35$~eV/f.u.

To study surface properties, 6-layer and 8-layer symmetric slabs based on the same $3 \times 3$ surface unit cell are used, containing 216 and 288 atoms, respectively, where the vacuum gap in each is twice the length of the slab along the $z$-direction. It was shown by Kocevski \textit{et al.}
\cite{Kocevski2022I} that the FM ordering provides a suitable model of UN at 0 K. Brillouin zone sampling is performed using Monkhorst-Pack \cite{Monkhorst1976} $3 \times 3 \times 3$ and $3 \times 3 \times 1$ \textit{k}-point meshes for the $3 \times 3 \times 3$ supercells and the symmetric slabs, respectively. In all energy minimization calculations, we set the cutoff energy of the plane-wave basis as 520 eV, the electronic relaxation convergence criterion as $10^{-4}$ eV, and the ionic relaxation convergence criterion as $10^{-2}$ eV/\AA. The nudged elastic band (NEB) method \cite{Henkelman2000a} with the climbing image algorithm \cite{Henkelman2000b} is used to calculate the migration energies of O$_i$ and $\{\text{O}_\text{N} \! : \! v_\text{N}\}$. The structures and ionic minimization trajectories are visualized using OVITO \cite{Stukowski2010} and VESTA \cite{Momma2008}.

% Equations

Point defect formation energies are calculated from:
\begin{equation}
E_f = E_d - E_p - \sum_i n_i \mu_i,
\end{equation}
where $E_d$ and $E_p$ are the energies of the supercell with and without the defect, respectively, $n_i$ is the number of atoms of type $i$ removed ($n_i < 0$) or added ($n_i > 0$) to form the defect, and $\mu_i$ is the chemical potential of the \textit{i}th species.

The U and N chemical potentials are calculated from \cite{AbdulHameed2024}:
\begin{equation}
E_c(\text{U}_{x}\text{N}_{y}) = x \mu_{\ce{U}} + y \mu_{\ce{N}},
\label{Eq:ChemPot}
\end{equation}
where $E_c(\ce{U$_{x}$N$_{y}$})$ is the cohesive energy (energy per formula unit) of compound U$_x$N$_y$. \cref{Eq:ChemPot} is solved for UN and $\alpha$-U to get the U-rich chemical potentials, and for UN and $\alpha$-\ce{U2N3} to get the N-rich chemical potentials \cite{Huang2020, Yang2021}. The chemical potentials at the intermediate near-stoichiometric conditions are the averages of these two bounds \cite{Woodward1998}.

The binding energy of a defect cluster $\{\text{A}_\text{X} \! : \! \text{B}_\text{Y}\}$ is calculated from:
\begin{equation}
E_b = E(\{\text{A}_\text{X} \! : \! \text{B}_\text{Y}\}) + E_p - E(\text{A}_\text{X}) - E(\text{B}_\text{Y}),
\end{equation}
where $E(\cdot)$ is the energy of the supercell containing the relevant defect. Based on this formulation, a negative $E_b$ means binding is favorable.

The incorporation energy of an impurity B in a defect $v_\text{X}$ to form B$_\text{X}$ is calculated from:
\begin{equation}
E_\text{inc} = E( \text{B}_\text{X} ) - E( v_\text{X} ) - \mu_\text{B},
\end{equation}
where $\mu_\text{B}$ is the chemical potential of the impurity in its reference state. The oxygen chemical potential is discussed in \cref{Sec:uO}. Note that $E( v_i ) = E_p$, where $v_i$ denotes a vacant interstitial site. % The solution energy of the same impurity is $E_s = E_i + E_f$ where $E_f$ is the formation energy of $v_\text{X}$. Note also that for an impurity in $v_i$, $E_\text{sol} = E_\text{inc}$ because $v_i$ has no formation energy.

The surface energy is calculated from \cite{Finnis2005}:
\begin{equation}
\sigma = \frac{1}{A} \left( E^* - \sum_{i} N_i \mu_i \right),
\label{Eq:Sigma}
\end{equation}
where $E^*$ is the total DFT energy of the supercell containing the surface, $A$ is the total surface area, $N_i$ is the number of atoms of species $i$ in the supercell containing the surface, and $\mu_i$ is the corresponding chemical potential. Note that $i$ $\in$ \{U, N, O\}.

The adsorption energy is calculated from:
\begin{equation}
E_\text{ad} = \frac{1}{N_\text{ad}} \left( E_\text{surf+ad} - E_\text{surf} \right) - \mu_\text{ad},
\label{Eq:Ead}
\end{equation}
where $E_\text{surf+ad}$ and $E_\text{surf}$ are the energies of the symmetric slabs with and without the adsorbed atom(s), $N_\text{ad}$ is the number of adsorbed atoms, and $\mu_\text{ad}$ is their chemical potential. 

Finally, it is of interest to estimate the change in the surface energy upon oxygen adsorption or surface defect formation. This is accomplished by using the following expression, which is derived in \ref{App2} (using geometric identities derived in \ref{App1}):
\begin{equation}
\Delta \sigma = \frac{8}{3} \frac{1-p}{p} \frac{R_v}{a^3} c_\text{O} \left[ \alpha_1 \frac{ [ \text{O}_\text{N}^{\text{(s)}} ] }{ [ \text{O}_\text{N}^{\text{(b)}} ] }  E_\text{ad}( \text{O}_\text{N}^{\text{(s)}} ) + \alpha_2 \frac{ [ \text{O}_i^{\text{(s)}} ] }{ [ \text{O}_\text{N}^{\text{(b)}} ] }  E_\text{ad}( \text{O}_i^{\text{(s)}} ) \right] + \frac{2}{a^2} [ v_\text{N}^{\text{(s)}} ] E_f ( v_\text{N}^{\text{(s)}} ),
\label{Eq:DeltaSigma1}
\end{equation}
where $p$ is the porosity, $R_v$ is the average void radius, $a$ is the lattice constant, $c_\text{O}$ is the atomic oxygen concentration, and $\alpha_1$ and $\alpha_2$ are the kinetic corrections for the defects $\text{O}_\text{N}^\text{(s)}$ and $\text{O}_i^{\text{(s)}}$, respectively. For a defect $i$, $\alpha_i$ is defined as \cite{Zinkle1990}:
\begin{equation}
\alpha_i =
\begin{cases}
    \sqrt{D_i t}/\lambda_v & \text{if } \sqrt{D_i t}/\lambda_v < 1, \\
    1 & \text{if } \sqrt{D_i t}/\lambda_v \geq 1,
\end{cases}
\label{Eq:alpha}
\end{equation}
where $D_i$ is the diffusivity of the relevant defect, $t$ is the void nucleation time, and $\lambda_v$ is the effective capture length for diffusion to a void surface. If the void number density is
\begin{equation}
n_v = \frac{p}{\frac{4}{3}\pi R_v^3},
\end{equation}
then the center-to-center void spacing is $n_v^{-1/3}$. We take the relevant diffusion distance to be the matrix gap from the void surface to the midplane between neighboring voids,
\begin{equation}
\lambda_v
=
0.5\,n_v^{-1/3} - R_v
=
\left[0.5-\left(\frac{3p}{4\pi}\right)^{1/3}\right]n_v^{-1/3}.
\label{Eq:lambda_eff}
\end{equation}
This definition is positive for $p<\pi/6\approx0.52$, which is satisfied for all porosities considered here. We assume a void nucleation time $t$ = 42 hours. This estimate is justified by taking a typical dose rate, $\dot{G}$, in a fast reactor core as on the order of $10^{-6}$~dpa/s~\cite{Mansur1983, Saha2018}. In stainless steels, stable voids are fully formed by a dose, $G$, in the range of 0.1--0.2~dpa~\cite{Surh2004}. Using the relation $t = G / \dot{G}$, this corresponds to a void nucleation time between 28 and 56 hours, with an average of 42 hours. Due to the lack of experimental data on early-stage void nucleation in UN fuels, this estimated time serves as a reasonable placeholder. A discussion of the model sensitivity to the kinetic parameter is given in \cref{Sec:Discuss}.

Note that \cref{Eq:DeltaSigma1} predicts that the surface energy reduction scales approximately linearly with oxygen concentration through the explicit $c_\text{O}$ prefactor; a weak additional $c_\text{O}\ln c_\text{O}$ dependence enters via the impurity-referenced $E_\text{ad}$ (see \ref{App:Tpeak}). Note also that because $\Delta \sigma$ depends on the adsorption and defect-formation energies linearly rather than on $\sigma$ itself, curvature corrections to the absolute surface energy do not propagate into $\Delta \sigma$. Any residual curvature or coordination dependence enters only through $E_\text{ad}(\text{O}_\text{N}^{(\text{s})})$ and $E_f(v_\text{N}^{(\text{s})})$.

\subsection{Oxygen chemical potential}
\label{Sec:uO}

DFT calculations with GGA exchange-correlation functionals are known for being inaccurate for gases like \ce{O2} \cite{Bocharov2013}. This is usually remedied by using hybrid functionals or applying semi-empirical corrections \cite{Sargeant2021}. Alternatively, Finnis \textit{et al.} \cite{Finnis2005} suggested using common oxides as the oxygen reference state. For an oxide M$_a$O$_b$, the formation reaction is:
\begin{equation}
a \text{M(s)} + \frac{b}{2} \text{O$_2$(g)} \rightleftharpoons \text{M$_a$O$_b$(s)}.
\end{equation}
The Gibbs free energy balance for this reaction is:
\begin{equation}
\Delta_f G^0 (\text{M$_a$O$_b$}) = g^0(\text{M$_a$O$_b$}) - a \mu_\text{M}^0 - b \mu_\text{O}^0,
\label{Eq:OChemPot}
\end{equation}
where $\Delta_f G^0 (\text{M$_a$O$_b$})$ is the standard Gibbs free energy of formation for the oxide, which is available in thermochemical tables \cite{Linstrom2024}, $g^0$(M$_a$O$_b$) is approximated by the DFT energy per formula unit of the oxide, $\mu_\text{M}^0$ is approximated by the DFT energy per atom of the metal, and $\mu_\text{O}^0$ is the oxygen chemical potential. The superscript 0 on any quantity denotes its value at the standard pressure, $P^0$ = 1 bar, and standard temperature, $T^0$ = 298.15 K. Approximating $T^0$ with 0 K in the context of DFT calculations introduces little error \cite{Finnis2005}. Rearranging \cref{Eq:OChemPot}, the oxygen chemical potential is:
\begin{equation}
\mu_\text{O}^0 = \frac{1}{b} \left[ g^0(\text{M$_a$O$_b$}) - a \mu_\text{M}^0 - \Delta_f G^0 (\text{M$_a$O$_b$}) \right].
\label{Eq:OxidePot}
\end{equation}

Finnis \textit{et al.} state that the chemical potential of oxygen should be independent of the specific oxide used. Using \ce{MgO} and \ce{Al2O3} as the initial oxide, we found that the chemical potential for both is $\mu_\text{O} = - 4.18$ eV. This value is used as the reference chemical potential to report all the oxygen incorporation and adsorption energies in \cref{Sec:Results}. However, we are concerned with studying how oxygen interacts with voids in UN, where oxygen atoms are introduced as impurities during manufacturing. For this setting, the oxygen reference state is not the oxide, but rather the most dominant oxygen impurity type, which is $\text{O}_\text{N}^\text{(b)}$ as will be shown in \cref{Sec:Results}. Note that we use the Kröger-Vink notation \cite{Kroger1956} to describe point defects, but with charges omitted because UN exhibits metallic properties and defects have no associated charge \cite{Cooper2023}. Superscripts (b) and (s) are added to the Kröger-Vink notation to differentiate bulk and surface defects, respectively.

The reference state of oxygen impurities in UN is based on the following defect reaction:
\begin{equation}
v_\text{N}^\text{(b)} + \frac{1}{2} \text{O}_2 \text{(g)} \rightleftharpoons \text{O}_\text{N}^\text{(b)},
\end{equation}
which has the following mass-action law:
\begin{equation}
K_1 = \frac{ [ \text{O}_\text{N}^\text{(b)} ] }{ [ v_\text{N}^\text{(b)} ] P_{\text{O}_2}^{1/2} } = \text{exp}\! \left[ - \frac{E( \text{O}_\text{N} ) - E( v_\text{N} )}{kT} \right],
\label{Eq:Kmu}
\end{equation}
where $E( \text{O}_\text{N} )$ and $E( v_\text{N} )$ are the DFT energies of the supercells containing $\text{O}_\text{N}$ and $v_\text{N}$, respectively. This difference is termed the ``raw incorporation energy'' \cite{AbdulHameed2024}, i.e., the incorporation energy without considering the impurity's reference state. The dependence of the oxygen chemical potential on oxygen partial pressure, $P_{\text{O}_2}$, and temperature, $T$, is given by \cite{Finnis2005}:
\begin{equation}
\mu_\text{O}(P_{\text{O}_2}, T) = g(T) + \frac{1}{2} kT \, \text{ln} \!\left( P_{\text{O}_2} / P^0 \right),
\label{Eq:muTPO2}
\end{equation}
where the temperature dependence, $g(T)$, is extracted from thermochemical tables \cite{Linstrom2024}. Focusing on the dependence on the oxygen partial pressure:
\begin{equation}
\mu_\text{O}(P_{\text{O}_2}) = \frac{1}{2} kT \, \text{ln} \!\left( P_{\text{O}_2} / P^0 \right),
\label{Eq:muPO2}
\end{equation}
and rearranging \cref{Eq:muPO2,Eq:Kmu}, it can be easily shown that the oxygen chemical potential is:
\begin{equation}
\mu_\text{O} = E( \text{O}_\text{N} ) - E( v_\text{N} ) + kT \, \text{ln} \! \left( [ \text{O}_\text{N}^\text{(b)} ] / [ v_\text{N}^\text{(b)} ] \right).
\label{Eq:OxygenTruePot}
\end{equation}

\subsection{Oxygen diffusivity}
\label{Sec:Diff}

The diffusivity of oxygen impurities is also explored in this work. The diffusivity of a defect $d$ is calculated from:
\begin{equation}
D_d = \frac{1}{6} z \lambda^2 \nu \, \text{exp}\! \left( - \frac{E_m}{ k T } \right),
\label{Eq:Diffusion}
\end{equation}
\noindent where $d$ $\in$ \{O$_i$, $\{\text{O}_\text{N} \! : \! v_\text{N}\}$\}, $z$ is the number of equivalent sites the atom can jump to, and $\lambda$ is the jump distance. Both $z$ and $\lambda$ are determined from the crystal structure. For O$_i$, $z$ = 6 and $\lambda$ = $a/2$, with $a$ being the lattice constant. For $\{\text{O}_\text{N} \! : \! v_\text{N}\}$, $z$ = 12 and $\lambda$ = $a/\sqrt{2}$. $E_m$ is the effective migration energy. For O$_i$, $E_m$ = 2.41 eV. For $\{\text{O}_\text{N} \! : \! v_\text{N}\}$, net bulk diffusion requires two steps \cite{Kocevski2022II}: (\textit{a}) O$_\text{N}$ and $v_\text{N}$ exchange, and (\textit{b}) $v_\text{N}$ rotation around O$_\text{N}$. Note the $\text{O}_\text{N}$ is immobile by itself and can only move via this two-step process with the assistance of a nearby $v_\text{N}$. The saddle point energy of the exchange step is 2.87 eV, whereas that of the rotation step is 3.06 eV. Based on the highest barrier approximation \cite{Claisse2016}, the effective migration energy is 3.06 eV. % All migration energies calculated in this work are shown in \cref{Tab:Mig}. 

In \cref{Eq:Diffusion}, $\nu$ is the attempt frequency, which can be estimated from phonon calculations. Because these calculations are computationally expensive, we use a semi-classical approach to estimate $\nu$ as \cite{Olander2017}:
\begin{equation}
\nu = \left( \frac{E_m}{2m\lambda^2} \right)^{1/2},
\label{Eq:nu}
\end{equation}
where $m$ is the mass of the diffusing atom. For $\{\text{O}_\text{N} \! : \! v_\text{N}\}$, the limiting step is the rotation of $v_\text{N}$  around O$_\text{N}$ and the mass of the nitrogen atom is used. For O$_i$, the mass of the oxygen atom is used. For the diffusion of $\{\text{O}_\text{N} \! : \! v_\text{N}\}$, $\nu = 9.45~\times~10^{12}$ Hz, whereas for $\text{O}_i$, $\nu = 1.11~\times~10^{13}$ Hz.

% \Cref{Eq:nu} is based on two simplifications: First, at the equilibrium position, the energy of the diffusing impurity is closely represented by the harmonic approximation \cite{Olander2017}:
% \begin{equation}
% U(r) = U(r_0) + \frac{1}{2}  \left( \frac{d^2U}{dr^2} \right)_{r_0} (r - r_0)^2.
% \end{equation}

% \noindent Then, $\nu$ has the form:
% \begin{equation}
% \nu = \frac{1}{2 \pi} \left[ \frac{1}{m} \left( \frac{d^2U}{dr^2} \right)_{r_0} \right]^{1/2}.
% \end{equation}

% \noindent Second, between two equilibrium positions, the energy barrier that the impurity encounters has the following form:
% \begin{equation}
% U(r) = U(r_0) + E_m \, \text{sin}^2 \! \left[ \frac{ \pi (r - r_0) }{ \lambda } \right].
% \end{equation}

The diffusivity of an impurity X by a mechanism that depends on defect $d$ is calculated from \cite{Cooper2023}:
\begin{equation}
D_{X, d} = f \frac{c_d}{c_\text{X}} D_d,
\end{equation}
where $f$ is the correlation factor. For $\{\text{O}_\text{N} \! : \! v_\text{N}\}$, $f$ = 0.7815 \cite{Laughlin2014}, which is the value for vacancy-mediated migration in the FCC nitrogen sublattice. For O$_i$, $f$ = 1 \cite{Laughlin2014}. $c_d$ is the atomic concentration of the defect $d$ and $c_\text{X}$ is the total atomic concentration of impurity X. As an approximation, we assume that $c_\text{O} = [\text{O}_\text{N}]$ which will be justified in \cref{Sec:Oxygen} of the results.

\subsection{Defect reactions}

In this section, we outline the defect reactions used to estimate the relative concentrations of various oxygen defect types. In general, there are three categories of defect reactions: (\textit{a}) reactions that relate defects in the bulk, (\textit{b}) reactions that relate defects on the void surfaces, and (\textit{c}) reactions that thermodynamically dictate defect transitions from the bulk to the void surfaces, and vice versa. A limitation of the latter category of reactions is that it neglects the kinetic aspects of the defect transitions. That is, the kinetics of reactions in (\textit{c}) category are assumed to be faster than those in the (\textit{a}) and (\textit{b}) categories. % and only gives concentrations in the limit of infinite time.

The first bulk defect reaction is:
\begin{equation}
\text{O}_\text{N}^\text{(b)} \rightleftharpoons v_\text{N}^\text{(b)} + \text{O}_i^\text{(b)},
\end{equation}
and its mass-action law is:
\begin{equation}
K_2 = \frac{ [ \text{O}_i^\text{(b)} ]  [ v_\text{N}^\text{(b)} ] } { [ \text{O}_\text{N}^\text{(b)} ] } = \text{exp}\! \left[ - \frac{ E_\text{inc}( \text{O}_i^\text{(b)} ) - E_\text{inc}( \text{O}_\text{N}^\text{(b)} ) } { k T } \right].
\end{equation}

The second bulk defect reaction is:
\begin{equation}
v_\text{N}^\text{(b)} + \text{O}_\text{N}^\text{(b)} \rightleftharpoons \{\text{O}_\text{N} \! : \! v_\text{N}\}^\text{(b)},
\end{equation}
and its mass-action law is:
\begin{equation}
K_3 = \frac{ [ \{\text{O}_\text{N} \! : \! v_\text{N}\}^\text{(b)} ] } { [ v_\text{N}^\text{(b)} ] [ \text{O}_\text{N}^\text{(b)} ] } = \text{exp}\! \left[ - \frac{ E_b( \{\text{O}_\text{N} \! : \! v_\text{N}\}^\text{(b)} ) } { k T } \right].
\end{equation}

The defect reaction at void surfaces is:
\begin{equation}
v_i^\text{(s)} + \text{O}_\text{N}^\text{(s)}  \rightleftharpoons  v_\text{N}^\text{(s)} + \text{O}_i^\text{(s)},
\end{equation}
and its mass-action law is:
\begin{equation}
K_4 = \frac{ [ v_\text{N}^\text{(s)} ]  [ \text{O}_i^\text{(s)} ] } { [ v_i^\text{(s)} ] [ \text{O}_\text{N}^\text{(s)} ] } = \text{exp}\! \left[ - \frac{ E_\text{ad}( \text{O}_i^\text{(s)} ) - E_\text{ad}( \text{O}_\text{N}^\text{(s)} ) } { k T } \right].
\end{equation}

The first reaction that relates bulk UN and the void surface is:
\begin{equation}
v_\text{N}^\text{(b)} + \text{N}_\text{N}^\text{(s)}  \rightleftharpoons v_\text{N}^\text{(s)} + \text{N}_\text{N}^\text{(b)},
\end{equation}
and its mass-action law is:
\begin{equation}
K_5 = \frac{ [ v_\text{N}^\text{(s)} ]  [ \text{N}_\text{N}^\text{(b)} ] } { [ v_\text{N}^\text{(b)} ] [ 
\text{N}_\text{N}^\text{(s)} ] } = \text{exp}\! \left[ - \frac{ E_f ( v_\text{N}^\text{(s)} ) - E_f ( v_\text{N}^\text{(b)} ) } { k T } \right].
\label{Eq:VN-VNs}
\end{equation}
$[ \text{N}_\text{N}^\text{(s)} ]$, the atomic concentration of the nitrogen sites on the surface, can be estimated from knowledge of the porosity, $p$, and the average void radius, $R_v$ by the following expression which is derived in \ref{App1}:
\begin{equation}
[ \text{N}_\text{N}^\text{(s)} ] = \frac{3}{2} \frac{p}{1-p} \frac{a}{R_v},
\label{Eq:NNs1}
\end{equation}
where $a$ is the lattice constant. Note that $[ \text{N}_\text{N}^\text{(b)} ] = 1 - [ \text{N}_\text{N}^\text{(s)} ]$. 

The second reaction is:
\begin{equation}
v_\text{N}^\text{(b)} + \text{O}_i^\text{(s)}  \rightleftharpoons v_i^\text{(s)} + \text{O}_\text{N}^\text{(b)},
\end{equation}
and its mass-action law is:
\begin{equation}
K_6 = \frac{ [ v_i^\text{(s)} ]  [ \text{O}_\text{N}^\text{(b)} ] } { [ v_\text{N}^\text{(b)} ] [ \text{O}_i^\text{(s)} ] } = \text{exp}\! \left[ - \frac{ E_\text{inc}( \text{O}_\text{N}^\text{(b)} ) - E_\text{ad}( \text{O}_i^\text{(s)} ) } { k T } \right].
\end{equation}
As derived in \ref{App1}, $[ v_i^\text{(s)} ]$ is given by:
\begin{equation}
[ v_i^\text{(s)} ] = \frac{3p}{1-p} \frac{a}{R_v}.
\label{Eq:v_i^s}
\end{equation}

To find the relative concentration $[ \text{O}_\text{N}^\text{(s)} ] / [ \text{O}_\text{N}^\text{(b)} ]$, it is not necessary to resort to a defect reaction. Instead, the ratio is calculated from: 
\begin{equation}
\frac{ [ \text{O}_\text{N}^\text{(s)} ] }{ [ \text{O}_\text{N}^\text{(b)} ] } = \frac{ [ \text{O}_i^\text{(s)} ] }{ [ \text{O}_\text{N}^\text{(b)} ] } \div \frac{ [ \text{O}_i^\text{(s)} ] }{ [ \text{O}_\text{N}^\text{(s)} ] } = \frac{ [v_\text{N}^\text{(s)}] }{ [v_\text{N}^\text{(b)}] } \, \text{exp}\! \left[ - \frac{ E_\text{ad}( \text{O}_\text{N}^\text{(s)} ) - E_\text{inc}( \text{O}_\text{N}^\text{(b)} ) } { k T } \right].
\label{Eq:ONs_ONb}
\end{equation}

\section{Results}
\label{Sec:Results}

\subsection{Defect energetics}
\label{Sec:DefE}

The formation energies of U and N vacancies in UN are shown in \cref{Tab:EfVN}. The bulk formation energies compare very well with those calculated by Yang and Kaltsoyannis \cite{Yang2021} and Kocevski \textit{et al.} \cite{Kocevski2022I}, which acts as a validation of our DFT model. It is noticed that there is a consistent difference of about 0.76 eV between N vacancy formation on the surface and in the bulk, independent of stoichiometry. This difference is termed the segregation energy of the N vacancy. It is easier to form an N vacancy on the surface than in the bulk, and the number of $v_\text{N}^\text{(s)}$ sites is limited by the available surface area.

\begin{table}[h!]
\scriptsize
\centering
\caption{Formation energies (eV) of U and N vacancies in UN. Reference energies are DFT values calculated by Yang and Kaltsoyannis \cite{Yang2021} and Kocevski \textit{et al.} \cite{Kocevski2022I}.}
\begin{tabular}{l|cc|cc|cc}
\hline
 & \multicolumn{2}{c|}{U-rich} & \multicolumn{2}{c|}{Intermediate} & \multicolumn{2}{c}{N-rich} \\
& Calc. & Ref. & Calc. & Ref. & Calc. & Ref. \\
\hline
$v_\text{U}^\text{(b)}$ & 3.28 & 3.17--3.43 \cite{Yang2021}, 3.27--3.86 \cite{Kocevski2022I} & 2.79 & 2.75--3.01 \cite{Yang2021} & 2.30 & 2.34--2.60 \cite{Yang2021}, 2.09--2.58 \cite{Kocevski2022I} \\
$v_\text{N}^\text{(b)}$ & 1.64 & 1.76--1.90 \cite{Yang2021}, 0.62--1.86 \cite{Kocevski2022I} & 2.13 & 2.18--2.31 \cite{Yang2021} & 2.62 & 2.59--2.72 \cite{Yang2021}, 1.42-2.82 \cite{Kocevski2022I} \\
$v_\text{N}^\text{(s)}$ & 0.88 & --   & 1.37 & -- & 1.86 & --   \\
\hline
\end{tabular}
\label{Tab:EfVN}
\end{table}

The incorporation energies of oxygen in UN are shown in \cref{Tab:Einc}, all given relative to $\mu_\text{O} = - 4.18$ eV. The most favorable oxygen defect in the bulk is $\text{O}_\text{N}$ followed by $\text{O}_i$. We also attempted to calculate the incorporation energy of the $\langle 111 \rangle$ and $\langle 110 \rangle$ O-N dumbbells. The $\langle 111 \rangle$ O-N dumbbell has a larger incorporation energy ($-1.11$ eV) than the tetrahedral site, whereas the $\langle 110 \rangle$ O-N dumbbell is unstable and relaxes to the $\langle 111 \rangle$ configuration. The largest incorporation energy is that of $\text{O}_\text{U}$, which indicates that O impurities exist almost exclusively in the N-sublattice.
% as confirmed by visual inspection of the ionic minimization trajectory using OVITO.

\begin{table}[h!]
\centering
\caption{Incorporation energies of oxygen in UN for $\mu_\text{O} = - 4.18$ eV. This value of $\mu_\text{O}$ is calculated according to the Finnis \textit{et al.} \cite{Finnis2005} oxide method.}
\footnotesize
\begin{tabular}{lc}
\hline
Defect & Incorporation energy [eV] \\
\hline
O$_\text{N}$ & $-7.23$ \\
O$_i$          & $-2.94$ \\
$\langle 111 \rangle$ O-N dumbbell & $-1.11$ \\
$\langle 110 \rangle$ O-N dumbbell & Unstable \\
O$_\text{U}$ & $-0.40$ \\
\hline
\end{tabular}
\label{Tab:Einc}
\end{table}

The binding energies of the explored defect clusters are shown in \cref{Tab:Eb}, where negative values indicate favorable binding. The most relevant defect cluster to our study, i.e., $\{\text{O}_\text{N} \! : \! v_\text{N}\}$, has a binding energy of 0.03 eV, which is practically zero. That is, binding is neither favorable nor unfavorable, and the N vacancy can roam around independently of $\text{O}_\text{N}$. On the other hand, we found a strong tendency for binding between $\text{O}_\text{N}$ and $v_\text{U}$, $\text{Kr}_\text{U}$, and $\text{Xe}_\text{U}$, all with an average binding energy of about $-0.4$ eV. We propose that this binding may promote the initial aggregation of defects into void or bubble embryos. This finding is consistent with experimental observations by Turos \textit{et al.} \cite{Turos1990}, who reported that Kr and Xe atoms form impurity-defect complexes in UN, serving as precursors for gas bubble formation. A comparable phenomenon is observed with helium in copper \cite{Zinkle1987b}, where helium does not reduce the void surface energy but stabilizes void embryos by binding with vacancies and vacancy clusters. A similar role may be played by oxygen in UN, potentially contributing to the stabilization of voids and bubbles. Due to the low concentrations of $\text{Kr}_\text{U}$ and $\text{Xe}_\text{U}$, we anticipate that this effect is less significant than the direct reduction of surface energy through oxygen adsorption.

\begin{table}[h!]
\centering
\caption{Binding energies of defect clusters in UN. Negative values mean that binding is favorable and vice versa.}
\footnotesize
\begin{tabular}{lc}
\hline
Defect cluster & Binding energy [eV] \\
\hline
$\{\text{O}_\text{N} \! : \! v_\text{N}\}$  & $0.03$  \\
$\{\text{O}_\text{N} \! : \! v_\text{U}\}$  & $-0.40$ \\
$\{\text{O}_\text{N} \! : \! \text{Kr}_\text{U}\}$ & $-0.41$ \\
$\{\text{O}_\text{N} \! : \! \text{Xe}_\text{U}\}$ & $-0.39$ \\
\hline
\end{tabular}
\label{Tab:Eb}
\end{table}

The migration energies of defects relevant to this study are shown in \cref{Tab:Mig}. As can be expected, the migration energy of $\text{O}_i$ is smaller than that of $\{ \text{O}_\text{N} \! : \! v_\text{N} \}$, which is 3.06 eV based on the highest barrier approximation. It is predicted that the migration energy of $v_\text{N}$ is slightly reduced from 3.10 to 3.06 eV when an $\text{O}_\text{N}$ is in a first nearest neighbor position. A similar observation has been made by Kotomin \textit{et al.} \cite{Kotomin2009}.

\begin{table}[h!]
\centering
\caption{Migration energies of some defects and/or migration paths in UN.}
\footnotesize
\begin{tabular}{lc}
\hline 
Defect and/or migration path & Migration energy [eV] \\
\hline
$v_\text{N}$ & 3.10 \\
O$_i$          & 2.41 \\
O$_\text{N}$-$v_\text{N}$ exchange & 2.87 \\
$v_\text{N}$ rotation around O$_\text{N}$ & 3.06 \\
\hline
\end{tabular}
\label{Tab:Mig}
\end{table}

\subsection{UN surface properties}

Upon relaxation, the 8-layer UN (001) surface undergoes a slight adjustment of the interlayer spacing, with the maximum interlayer spacing change remaining below 1\%. The energy of the UN (001) surface under various stoichiometric conditions with and without defects is presented in \cref{Tab:SurfE}. Our calculated pristine surface energy of approximately 1.59 J/m$^2$ falls within the range of 1.22--1.70 J/m$^2$ reported by Bocharov \textit{et al.} \cite{Bocharov2011SS}. To our knowledge, no experimental measurements of UN surface energy currently exist, making this computational comparison the only benchmark available. It is evident that the surface energies of the 6-layer and 8-layer slabs are very similar, indicating that the 6-layer slab is sufficient to achieve a converged description of the UN (001) surface. Therefore, the 6-layer slab is used in all subsequent adsorption calculations. Several observations can be made regarding the impact of adsorption on surface energy. Introducing $\text{O}_i^{\text{(s)}}$ onto the pure 6-layer surface decreases its energy by 0.182 J/m$^2$, irrespective of the stoichiometry. Furthermore, the surface energy of the 6-layer surface with $\text{O}_\text{N}^{\text{(s)}}$ is reduced relative to that of the 6-layer surface with $v_\text{N}^{\text{(s)}}$ by 0.256 J/m$^2$, also independent of stoichiometry.

\begin{table}[h!]
\centering
\caption{Surface energies of the UN (001) surface in units of J/m$^2$ as calculated by \cref{Eq:Sigma}.}
\footnotesize
\begin{tabular}{lccc}
\hline
 & U-rich & Intermediate & N-rich \\
\hline
6-layer pure surface & 1.586 & 1.586 & 1.586 \\
8-layer pure surface  & 1.592 & 1.592 & 1.592 \\
6-layer surface with O$_i^\text{(s)}$ & 1.404 & 1.404 & 1.404 \\
6-layer surface with $v_\text{N}^\text{(s)}$ & 1.620 & 1.638 & 1.656 \\
6-layer surface with O$_\text{N}^\text{(s)}$ & 1.364 & 1.382 & 1.400 \\
\hline
\end{tabular}
\label{Tab:SurfE}
\end{table}

\begin{figure}[h!]
    \centering
    \includegraphics[width=0.45\textwidth]{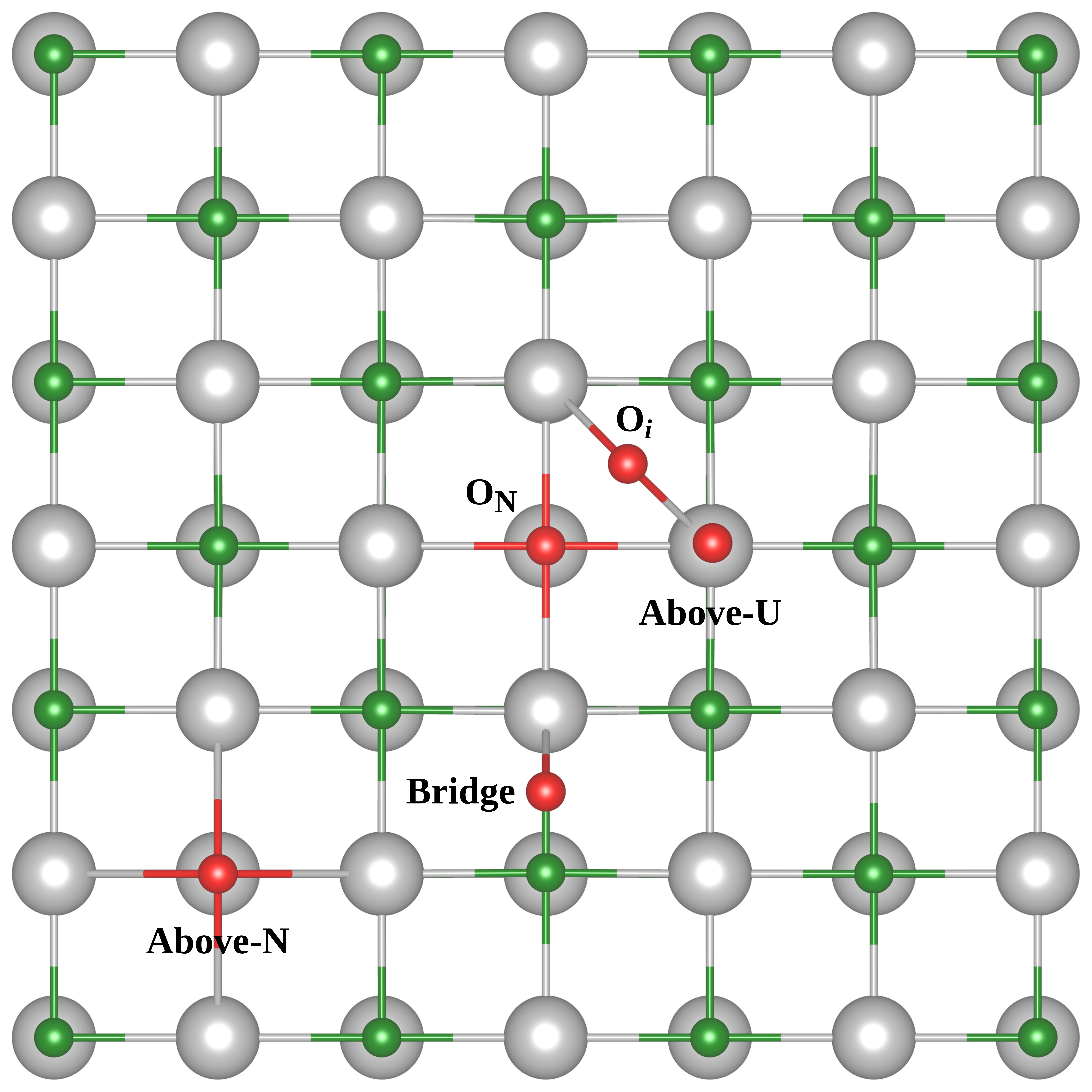}
    \caption{(Color online) A view of the UN surface along the $z$-axis (i.e., surface normal) showing the different oxygen adsorption sites. U atoms are gray, N atoms are green, and O atoms are red.}
    \label{Fig:Oad-2}
\end{figure}

The oxygen adsorption sites explored in this work are shown in \cref{Fig:Oad-2} generated using VESTA. Oxygen adsorption energies in different sites on the 6-layer slab are given in \cref{Tab:Ead}. The most stable site is a N surface vacancy, followed by the hollow site, i.e., $\text{O}_i^\text{(s)}$, as was also found in previous studies \cite{Kotomin2008, Bocharov2013}. The adsorption energy of the oxygen in the hollow site does not change if the oxygen atoms are on one or both sides of the slab, which indicates the absence of any polarity effects. Thus, considering only one side of the slab is sufficient for obtaining accurate adsorption energies. Oxygen adsorption energy in a hollow site slightly decreases from $-4.84$ eV to $-4.85$ eV when an oxygen atom is introduced in a nearest neighbor hollow site and increases to $-4.72$ eV for 4 neighboring O atoms forming a square, which is due to the lateral repulsive interaction between closely packed oxygen adsorbate atoms. This suggests that the saturation coverage of oxygen is probably smaller than 1 ($\theta_s < 1$). However, in this model, oxygen-oxygen interactions are neglected, and a saturation coverage of 1 is assumed for all impurity sites. If the actual saturation coverage was, say, $\theta_s = 0.5$, the effect of this would be to divide $[ v_i^\text{(s)} ]$ by 2, which is negligible on a log scale.

Note that the adsorption energy of O atoms in above-U sites is the same as that of O in a hollow site (\cref{Tab:Ead}). To confirm that the above-U site is stable, we shifted the oxygen atom by 0.2 {\AA} toward the hollow site, and it relaxed back to the above-U site. One way to account for the above-U sites in our model is to multiply $[ v_i^\text{(s)} ]$ by 2, as outlined in \ref{App1}. The bridge site is marked as unstable since oxygen atoms in this configuration relax to the above-U position. The largest adsorption energy is that of the above-N site (equivalent to a site on the U sublattice). % which is expected as oxygen does not prefer to exist on the U sublattice, either in bulk (\cref{Tab:Einc}) or on the surface. % This effect is nearly canceled out by the fact that, based only on hollow-site considerations, $\theta_s \sim 0.5$. Thus, \cref{Eq:Vis} is used to account for $[ v_i^\text{(s)} ]$ with no modifications.

\begin{table}[h!]
\centering
\caption{Adsorption energy of oxygen into sites on the symmetric 6-layer slab model of the UN (001) surface at a coverage of $\theta = 1/9$~ML (one O atom on one side of the $3 \times 3$ surface unit cell). These values correspond to $\mu_\text{O} = -4.18$~eV, which is calculated according to the Finnis \textit{et al.}~\cite{Finnis2005} oxide method.}
\footnotesize
\begin{tabular}{lc}
\hline
Adsorption site & Adsorption energy [eV] \\
\hline
$v_\text{N}^\text{(s)}$                          & $-6.80$ \\
Hollow site (one side), i.e., O$_i^\text{(s)}$   & $-4.84$ \\
Hollow site (both sides)                         & $-4.84$ \\
2 nearest-neighbor hollow sites (dumbbell)       & $-4.85$ \\
4 nearest-neighbor hollow sites (square)         & $-4.72$ \\
Bridge site                                      & Unstable \\ % Relaxes to the above-U site 
Above-U site                                     & $-4.83$ \\
Above-N site                                     & $-2.77$ \\
\hline
\end{tabular}
\label{Tab:Ead}
\end{table}

\begin{figure}[h!]
    \centering
    \includegraphics[width=0.45\textwidth]{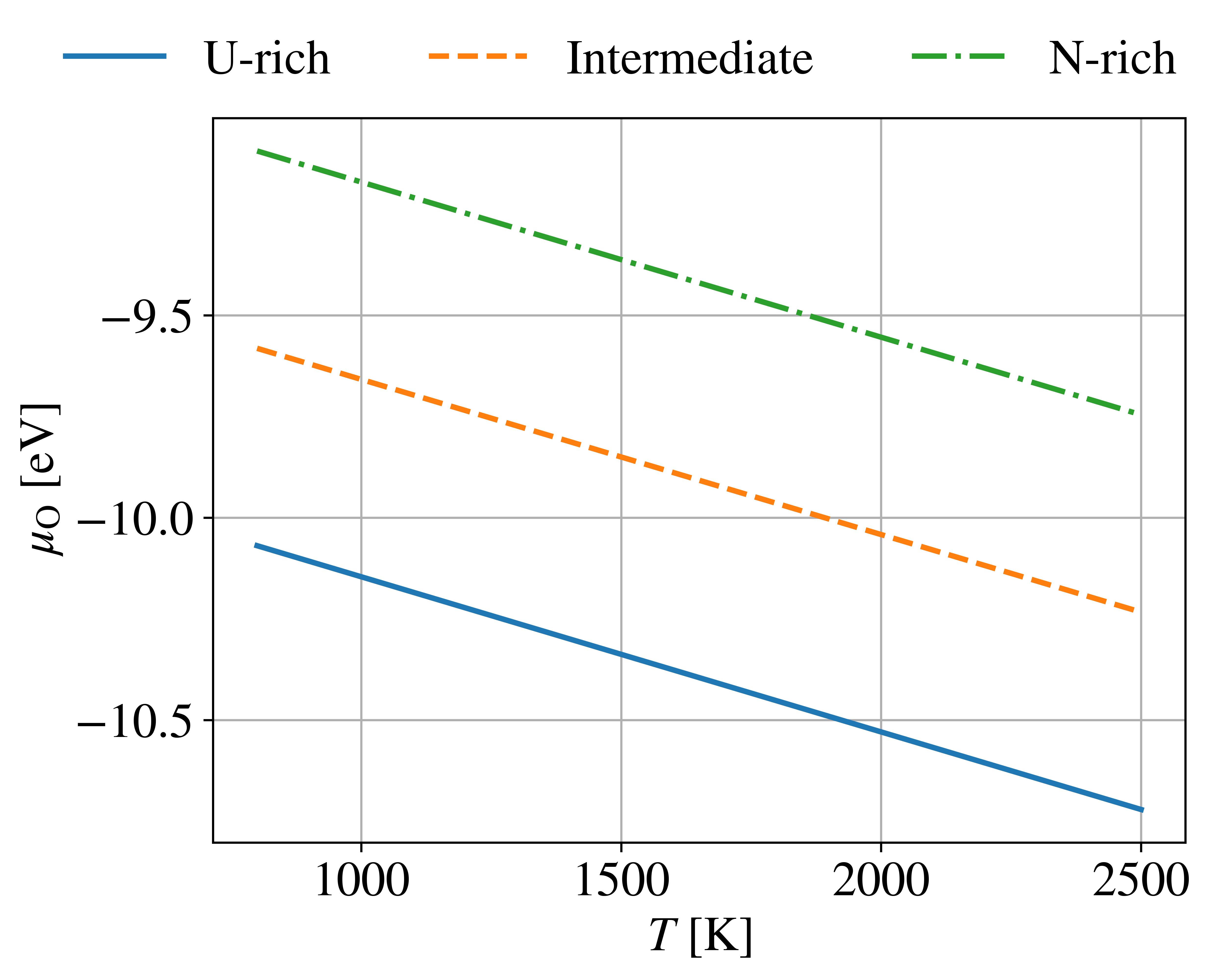}
    \caption{Oxygen chemical potential as calculated based on \cref{Eq:OxygenTruePot}.}
    \label{Fig:uO}
\end{figure}

\begin{figure}[h!]
\centering
\begin{subfigure}{0.48\textwidth}
    \includegraphics[width=\textwidth]{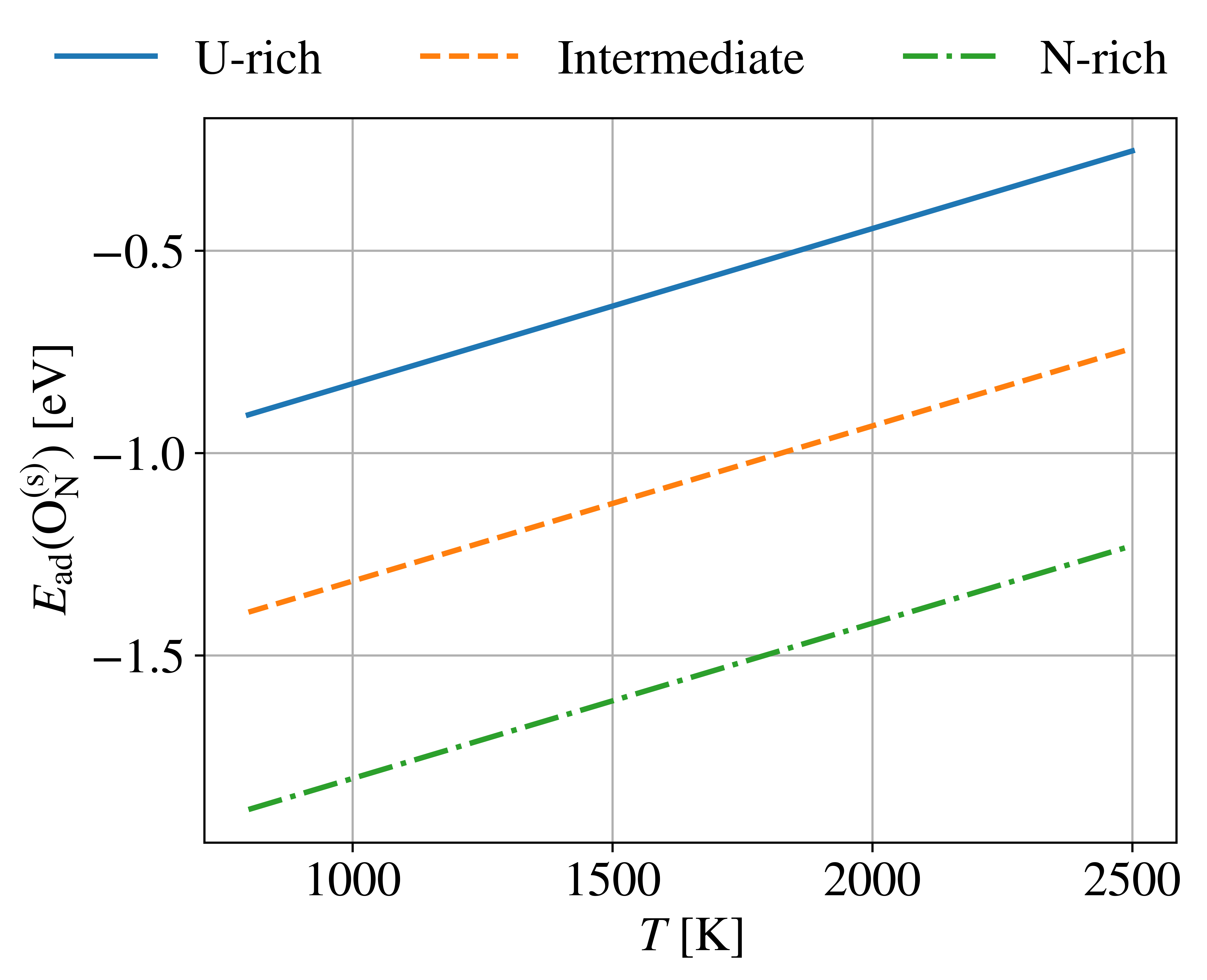}
    \caption{}
    \label{Fig:EadONs}
\end{subfigure}
\hfill
\begin{subfigure}{0.48\textwidth}
    \includegraphics[width=\textwidth]{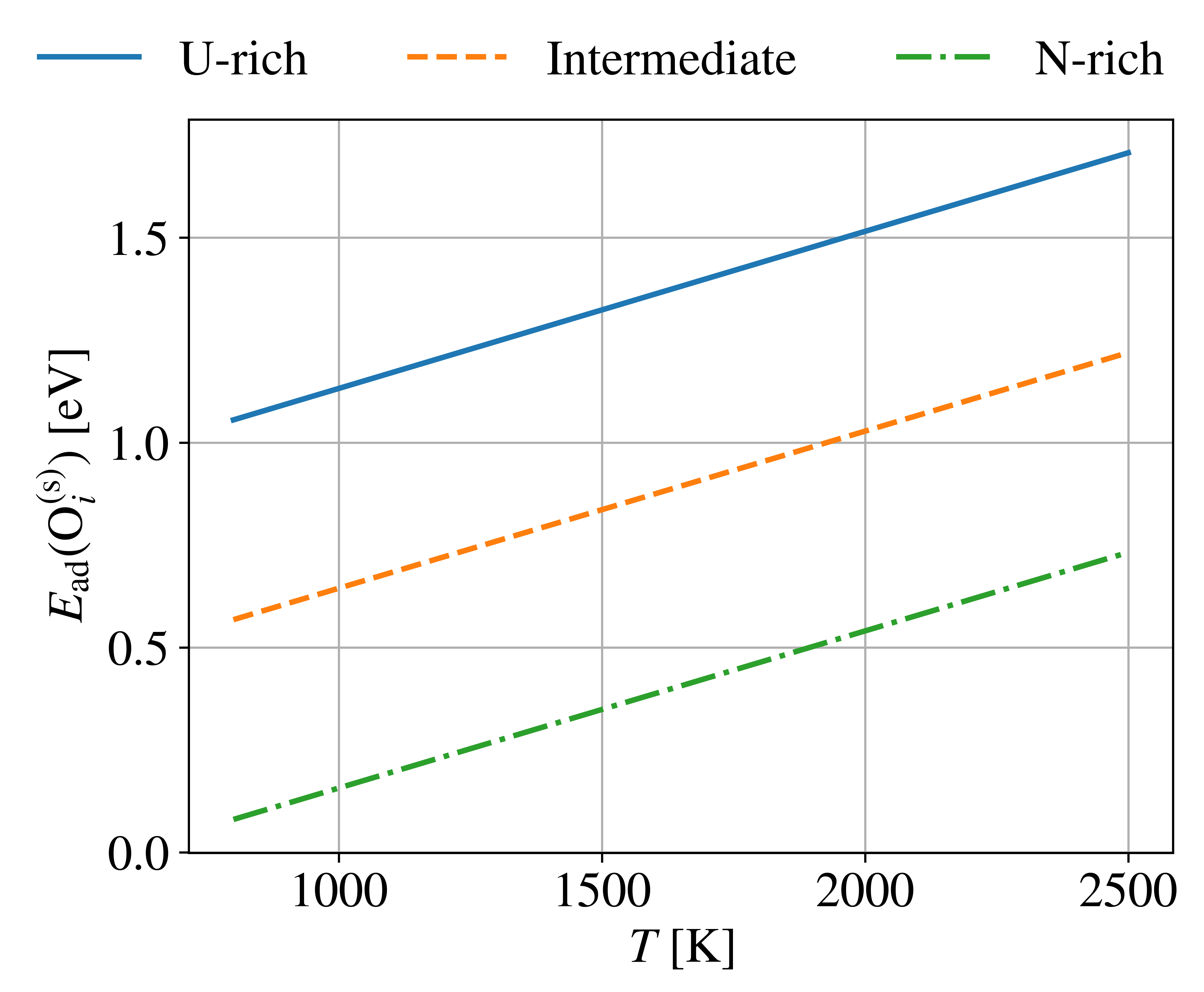}
    \caption{}
    \label{Fig:EadOis}
\end{subfigure}
\caption{Temperature evolution of oxygen adsorption energy relative to the chemical potential shown in \cref{Fig:uO} for \textbf{(a)} $\text{O}_\text{N}^\text{(s)}$ and \textbf{(b)} $\text{O}_i^\text{(s)}$.}
\label{Fig:EduO}
\end{figure}

The oxygen adsorption energies in \cref{Tab:Ead} are reported with the chemical potential of oxygen taken from the oxide as the reference state (i.e., \cref{Eq:OxidePot}). As explained earlier, however, the more appropriate oxygen reference state is oxygen pre-existing in UN as an impurity. In this case, the relevant chemical potential is that shown in \cref{Fig:uO}. The adsorption energies of oxygen relative to that chemical potential are shown in \cref{Fig:EadONs,Fig:EadOis} for $\text{O}_\text{N}^\text{(s)}$ and $\text{O}_i^\text{(s)}$, respectively. Comparing these adsorption energies to those in \cref{Tab:Ead}, we can differentiate between two cases: While oxygen adsorption on external surfaces reduces surface energy both as $\text{O}_\text{N}^\text{(s)}$ and $\text{O}_i^\text{(s)}$, pre-existing oxygen in UN adsorbing on internal void surfaces reduces the surface energy only when adsorbing as $\text{O}_\text{N}^\text{(s)}$. In contrast, $\text{O}_i^\text{(s)}$ increases the surface energy of internal void surfaces.

% This behavior is due to the different reference states of oxygen in both cases.

\subsection{Electronic structure of surface oxygen configurations}

\begin{figure}[h!]
\centering
\begin{subfigure}{0.48\textwidth}
    \includegraphics[width=\textwidth]{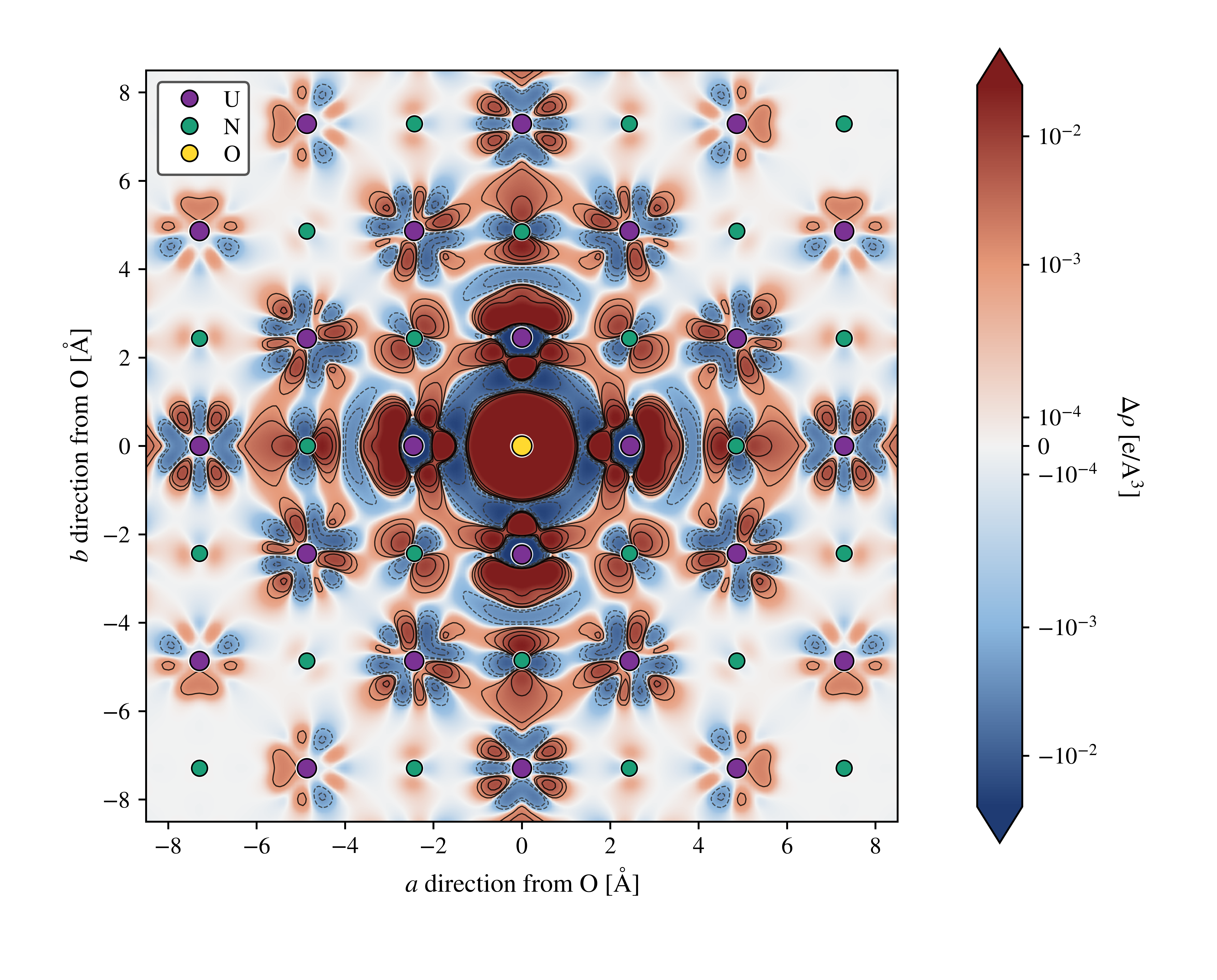}
    \caption{}
    \label{Fig:drho_ONs_ab}
\end{subfigure}
\hfill
\begin{subfigure}{0.48\textwidth}
    \includegraphics[width=\textwidth]{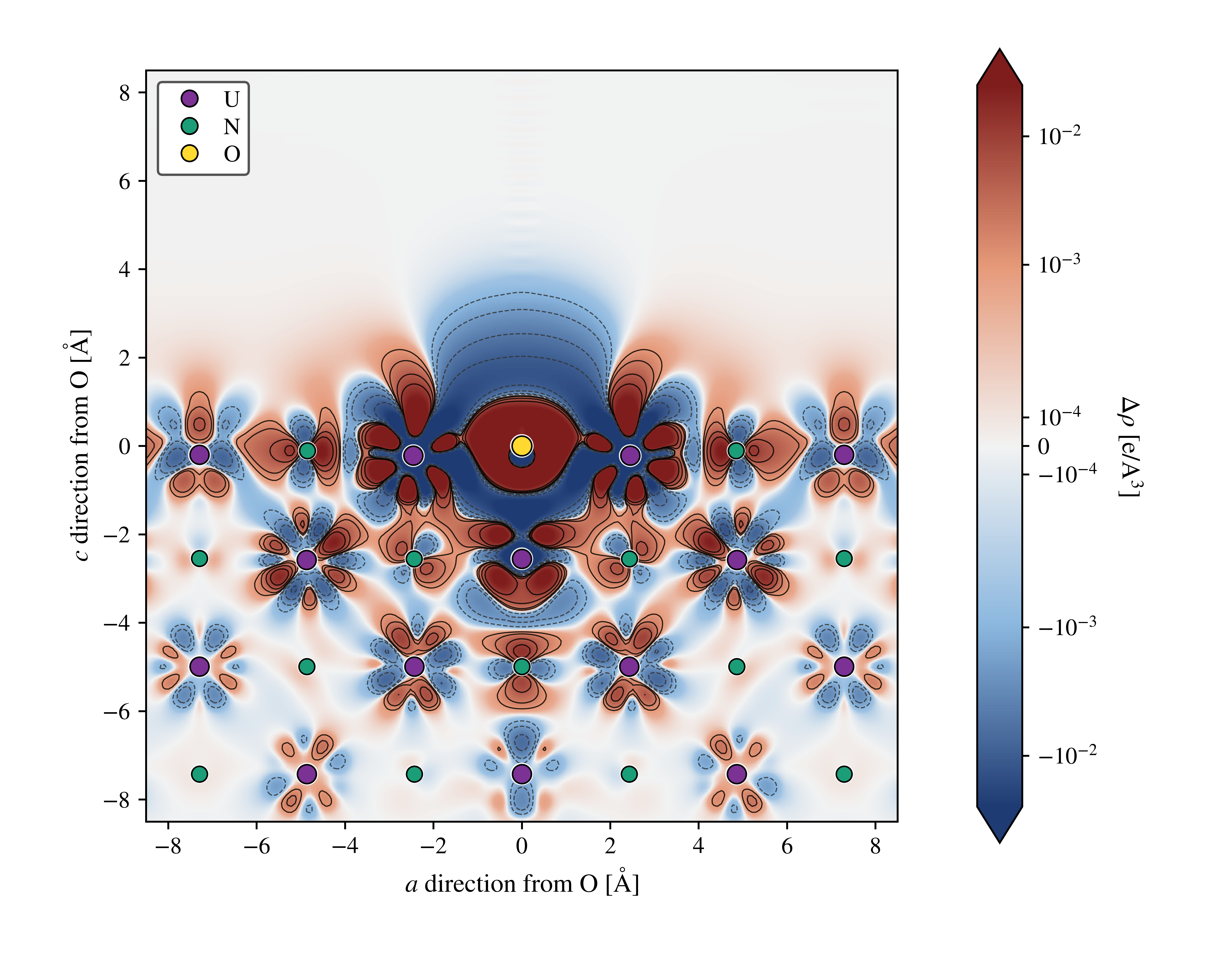}
    \caption{}
    \label{Fig:drho_ONs_ac}
\end{subfigure}
\hfill
\begin{subfigure}{0.48\textwidth}
    \includegraphics[width=\textwidth]{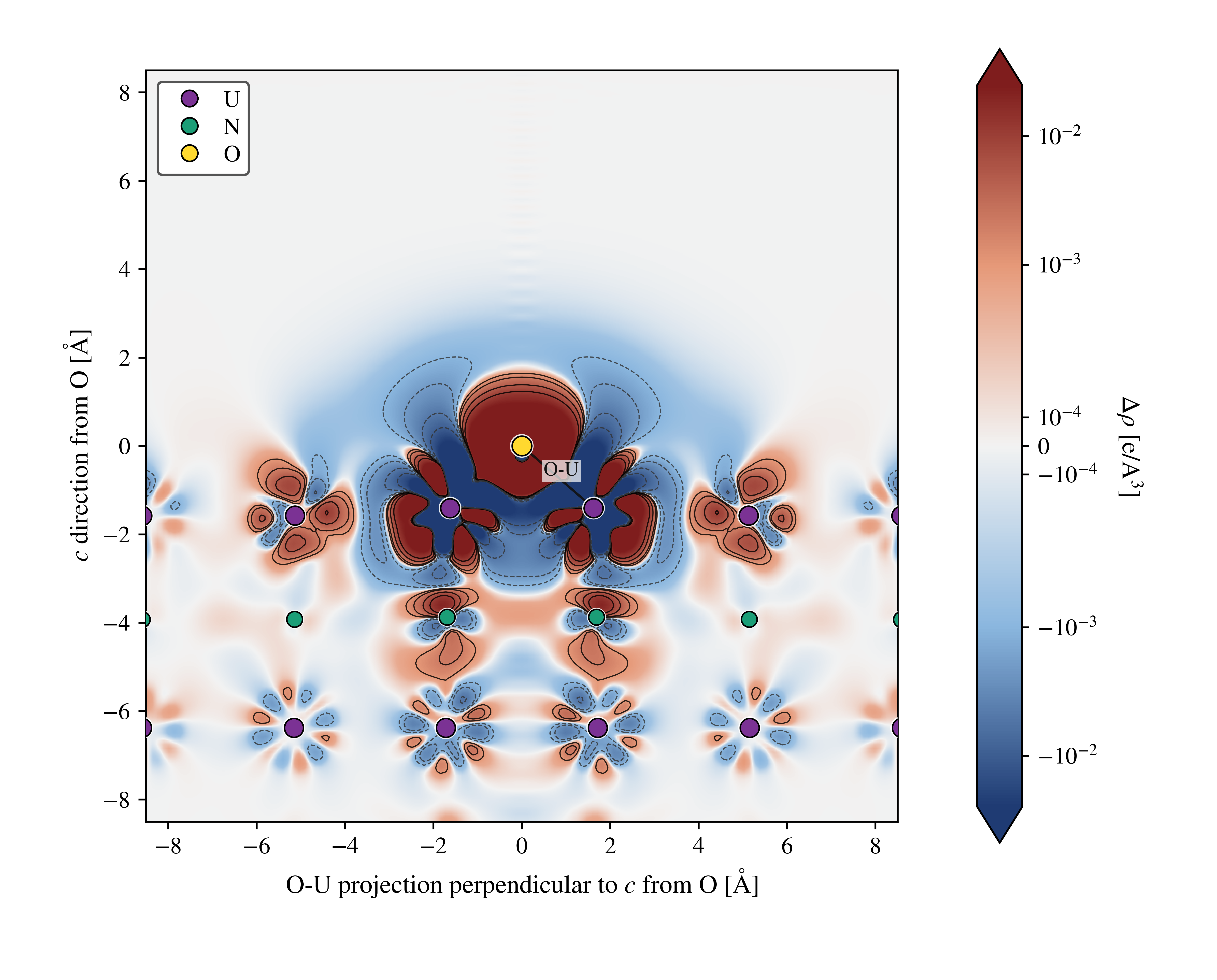}
    \caption{}
    \label{Fig:drho_Ois_OUc}
\end{subfigure}
\caption{
Charge-density difference, $\Delta\rho$, around surface oxygen in UN.
(\textbf{a}) $\mathrm{O}_{\mathrm{N}}^{\text{(s)}}$ in the $a$-$b$ plane.
(\textbf{b}) $\mathrm{O}_{\mathrm{N}}^{\text{(s)}}$ in the $a$-$c$ plane.
(\textbf{c}) $\mathrm{O}_i^{\text{(s)}}$ in the plane spanned by the nearest O-U direction and the crystallographic $c$ axis, with $c$ vertical.
Red denotes electron accumulation and blue denotes electron depletion in $\Delta\rho$.
Solid and dashed contours mark positive and negative $\Delta\rho$, respectively.
U, N, and O atoms are shown as purple, green, and yellow markers.
}
\label{Fig:CDD}
\end{figure}

\begin{table}[h!]
\centering
\scriptsize
\caption{
O-U charge-density metrics for surface oxygen configurations.
$\rho_\mathrm{min}$ is the minimum total charge density along the O-U bond path.
$r_\mathrm{O}$ and $r_\mathrm{U}$ are the O- and U-centered sphere radii defined by the position of this bond-density minimum.
$Q_\mathrm{O}$ and $Q_\mathrm{U}$ are the corresponding integrals of the total charge density inside those spheres.
$Q^\mathrm{abs}_{\Delta\rho,\mathrm{O}}$ and $Q^\mathrm{abs}_{\Delta\rho,\mathrm{U}}$ are the integrals of $|\Delta\rho|$ inside the same spheres.
Note that these quantities are local charge-density metrics and are not formal oxidation states or Bader charges.
The net $\Delta\rho$ integrals inside the O-centered spheres are discussed in the text to indicate local electron accumulation around oxygen.
}
\begin{tabular}{lcccccccc}
\hline
Configuration/path &
$d_{\text{O-U}}$ [\AA{}] &
$\rho_\mathrm{min}$ [$e$ \AA{}$^{-3}$] &
$r_\mathrm{O}$ [\AA{}] &
$r_\mathrm{U}$ [\AA{}] &
$Q_\mathrm{O}$ [$e$] &
$Q_\mathrm{U}$ [$e$] &
$Q^\mathrm{abs}_{\Delta\rho,\mathrm{O}}$ [$e$] &
$Q^\mathrm{abs}_{\Delta\rho,\mathrm{U}}$ [$e$]
\\
\hline
$\mathrm{O}_{\mathrm{N}}^{\text{(s)}}$, in-plane O-U & 2.4635 & 0.3910 & 1.1332 & 1.3303 & 6.4670 & 10.6259 & 0.3199 & 0.1524 \\
$\mathrm{O}_{\mathrm{N}}^{\text{(s)}}$, $c$-axis O-U & 2.5548 & 0.3238 & 1.1701 & 1.3847 & 6.6247 & 11.0186 & 0.3271 & 0.0798 \\
$\mathrm{O}_i^{\text{(s)}}$, tilted O-U & 2.1475 & 0.8403 & 0.9664 & 1.1811 & 5.7744 & 9.4373 & 0.4108 & 0.3035 \\
\hline
\end{tabular}
\label{Tab:OU_charge_density_metrics}
\end{table}

The energetic preference of $\mathrm{O}_{\mathrm{N}}^{\text{(s)}}$ over $\mathrm{O}_i^{\text{(s)}}$ (\cref{Tab:Ead}) is consistent with the way the UN surface electronic density accommodates oxygen in the two geometries. To examine this directly, we evaluated the charge-density difference
\begin{equation}
\Delta\rho(\mathbf{r}) = \rho_{\mathrm{surf+O}}(\mathbf{r}) - \rho_{\mathrm{surf}}(\mathbf{r}) - \rho_{\mathrm{O}}(\mathbf{r}),
\label{Eq:drho}
\end{equation}
where $\rho_{\mathrm{surf+O}}$ is the charge density of the oxygen-containing surface, $\rho_{\mathrm{surf}}$ is the charge density of the corresponding surface reference after removing oxygen, and $\rho_{\mathrm{O}}$ is the charge density of the isolated oxygen atom in the same cell. Positive values of $\Delta\rho$ indicate electron accumulation relative to the separated reference densities, and negative values indicate electron depletion.

For $\mathrm{O}_{\mathrm{N}}^{\text{(s)}}$, the $a$-$b$ and $a$-$c$ sections in \cref{Fig:drho_ONs_ab,Fig:drho_ONs_ac}, respectively, show that the charge redistribution is centered on the substituted oxygen and the adjacent uranium coordination shell. In the $a$-$b$ section, the redistribution appears around four nearly equivalent in-plane O-U paths, reflecting the local symmetry of the nitrogen sublattice site. In the $a$-$c$ section, the strongest features remain concentrated near the oxygen atom and the first U layer. This response follows directly from the local geometry: oxygen occupies a surface N vacancy and is therefore placed on an anion sublattice site coordinated by U atoms. The O-U bond-density metrics in \cref{Tab:OU_charge_density_metrics} show that the four in-plane O-U paths are equivalent within numerical precision, with $d_{\text{O-U}}=2.4635$~\AA{} and $\rho_\mathrm{min}=0.3910~e$/\AA{}$^{3}$ along the O-U path. The corresponding sphere-integrated total charges are $Q_\mathrm{O}=6.4670~e$ and $Q_\mathrm{U}=10.6259~e$, with $Q^\mathrm{abs}_{\Delta\rho,\mathrm{O}}=0.3199~e$ and $Q^\mathrm{abs}_{\Delta\rho,\mathrm{U}}=0.1524~e$. The additional O-U path along the $c$ direction has a comparable partition geometry, with $d_{\mathrm{O-U}}=2.5548$~\AA{}, $\rho_\mathrm{min}=0.3238~e$/\AA{}$^{3}$, $Q_\mathrm{O}=6.6247~e$, and $Q_\mathrm{U}=11.0186~e$. Thus, $\mathrm{O}_{\mathrm{N}}^{\text{(s)}}$ is accommodated through a regular set of O-U interactions associated with the surface anion site.

The hollow-site $\mathrm{O}_i^{\text{(s)}}$ configuration has a different electronic response. Its nearest O-U contacts are shorter, with $d_{\text{O-U}}=2.1475$~\AA{}, and the corresponding O-U/$c$ section in \cref{Fig:drho_Ois_OUc} isolates the tilted adsorption geometry. In this section, the redistribution extends from the interstitial oxygen toward the neighbouring U layer. The total charge density along this O-U path has a larger bond-path minimum, $\rho_\mathrm{min}=0.8403~e$/\AA{}$^{3}$, and the local $|\Delta\rho|$ integrals inside the O- and U-centered spheres are also larger than for an individual $\mathrm{O}_{\mathrm{N}}^{\text{(s)}}$ O-U path: $Q^\mathrm{abs}_{\Delta\rho,\mathrm{O}}=0.4108~e$ and $Q^\mathrm{abs}_{\Delta\rho,\mathrm{U}}=0.3035~e$. This shows that $\mathrm{O}_i^{\text{(s)}}$ forms a shorter and more strongly perturbed O-U adsorption motif.

The net $\Delta\rho$ integrals inside the O-centered partition spheres further indicate electron accumulation around oxygen in both configurations, consistent with charge transfer toward O. For the in-plane $\mathrm{O}_{\mathrm{N}}^{\text{(s)}}$ O-U paths, the O-centered partition sphere gains $0.3122~e$ of charge-density difference, while the corresponding value for the tilted $\mathrm{O}_i^{\text{(s)}}$ O-U path is $0.4095~e$. These values are local partition-dependent charge-density metrics, rather than formal oxidation states or Bader charges.

The distinction between the two configurations is therefore the way the electronic response is incorporated into the surface lattice. In $\mathrm{O}_{\mathrm{N}}^{\text{(s)}}$, oxygen occupies the surface anion vacancy and is accommodated by several comparable O-U interactions within the existing U coordination network. In $\mathrm{O}_i^{\text{(s)}}$, oxygen occupies a hollow/interstitial site and forms shorter O-U contacts that produce a stronger local charge-density response along selected directions. Although the tilted $\mathrm{O}_i^{\text{(s)}}$ O-U contacts show larger local $|\Delta\rho|$ integrals than an individual $\mathrm{O}_{\mathrm{N}}^{\text{(s)}}$ O-U path, this stronger local response is concentrated into only two nearest O-U contacts. By contrast, $\mathrm{O}_{\mathrm{N}}^{\text{(s)}}$ occupies the anion-sublattice site and is coordinated by four nearly equivalent in-plane O-U paths together with an additional O-U path along the $c$ direction. In this sense, the preference for $\mathrm{O}_{\mathrm{N}}^{\text{(s)}}$ can be viewed as a dangling-bond-like coordination-saturation effect: oxygen at the surface N site satisfies several under-coordinated U neighbours at once, whereas $\mathrm{O}_i^{\text{(s)}}$ forms shorter and locally stronger O-U contacts but only with a smaller subset of neighbouring U atoms. Thus, $\mathrm{O}_i^{\text{(s)}}$ is better described as a coordination-frustrated hollow-site configuration, while $\mathrm{O}_{\mathrm{N}}^{\text{(s)}}$ realizes the fuller U coordination shell available at the anion-sublattice site.

This real-space picture is consistent with previous DOS, Bader-charge, and charge-density-difference analyses of oxygen on UN surfaces, which attributed the stability of oxygen at surface nitrogen sites to O~$2p$-U~$5f$ interaction and charge transfer toward oxygen within the uranium coordination shell~\cite{Bocharov2011JNM,Zhukovskii2009SS,Zhukovskii2009JNM}. The present CDD maps and O-U charge-density metrics therefore support the adsorption-energy ordering in \cref{Tab:Ead}: $\mathrm{O}_{\mathrm{N}}^{\text{(s)}}$ is preferred because it satisfies several under-coordinated U neighbours at the anion-sublattice site, whereas $\mathrm{O}_i^{\text{(s)}}$ corresponds to a higher-energy coordination-frustrated hollow-site adsorption geometry with stronger but fewer tilted O-U contacts. Consequently, $\mathrm{O}_{\mathrm{N}}^{\text{(s)}}$ supplies the dominant surface-energy-lowering channel for oxygen on internal void surfaces, while $\mathrm{O}_i^{\text{(s)}}$ remains a secondary surface oxygen configuration.

\subsection{Oxygen behavior}
\label{Sec:Oxygen}

\subsubsection{Oxygen defect populations and transport}

Several parameters must be defined to study the behavior of oxygen using our model. We assume an initial oxygen concentration of $w_\text{O}$ = 1500 ppm, a porosity of $p$ = 5\%, and an average void radius of $R_v$ = 1 nm. The chosen $w_\text{O}$ = 1500 ppm is the upper limit of the tolerated oxygen impurity concentration in UN \cite{Rogozkin2003, Schuler2017}. The optimal initial (closed) porosity for UN fuels to both suppress oxidation and accommodate gaseous fission products is estimated to be 4\% \cite{Johnson2018}. Note that the porosity in irradiated fuels is typically as high as 15\% \cite{Nichenko2014}, which exceeds the 5\% porosity assumed here. Lastly, the lower limit of reported fission gas bubble sizes in nitride and carbide fuels is close to $R_v$ = 1 nm \cite{Matzke1986}, which is also the critical void radius in metals \cite{Zinkle1987b}.

\begin{figure}[h!]
\centering
\begin{subfigure}{0.48\textwidth}
    \includegraphics[width=\textwidth]{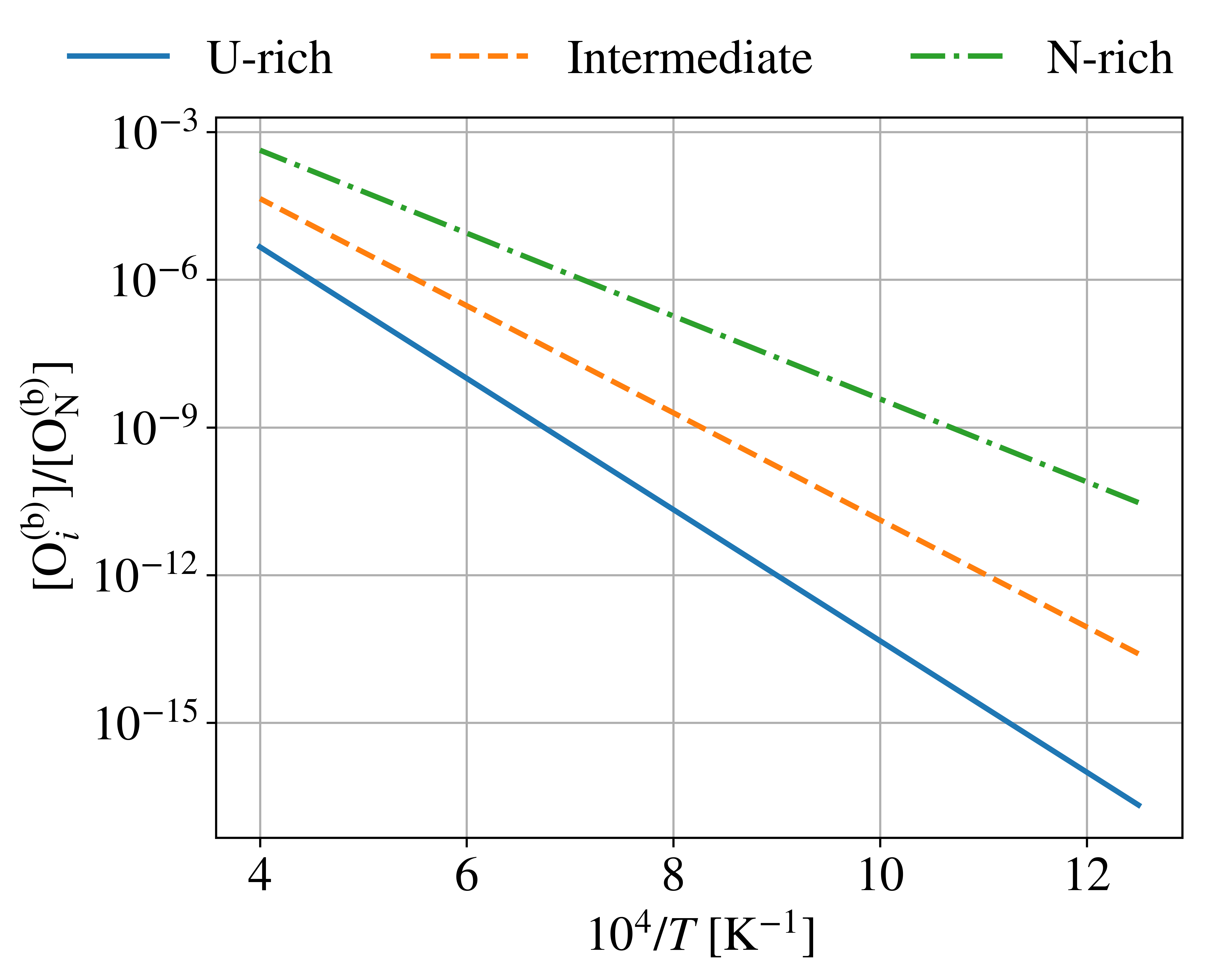}
    \caption{}
    \label{Fig:Oi_ON}
\end{subfigure}
\hfill
\begin{subfigure}{0.48\textwidth}
    \includegraphics[width=\textwidth]{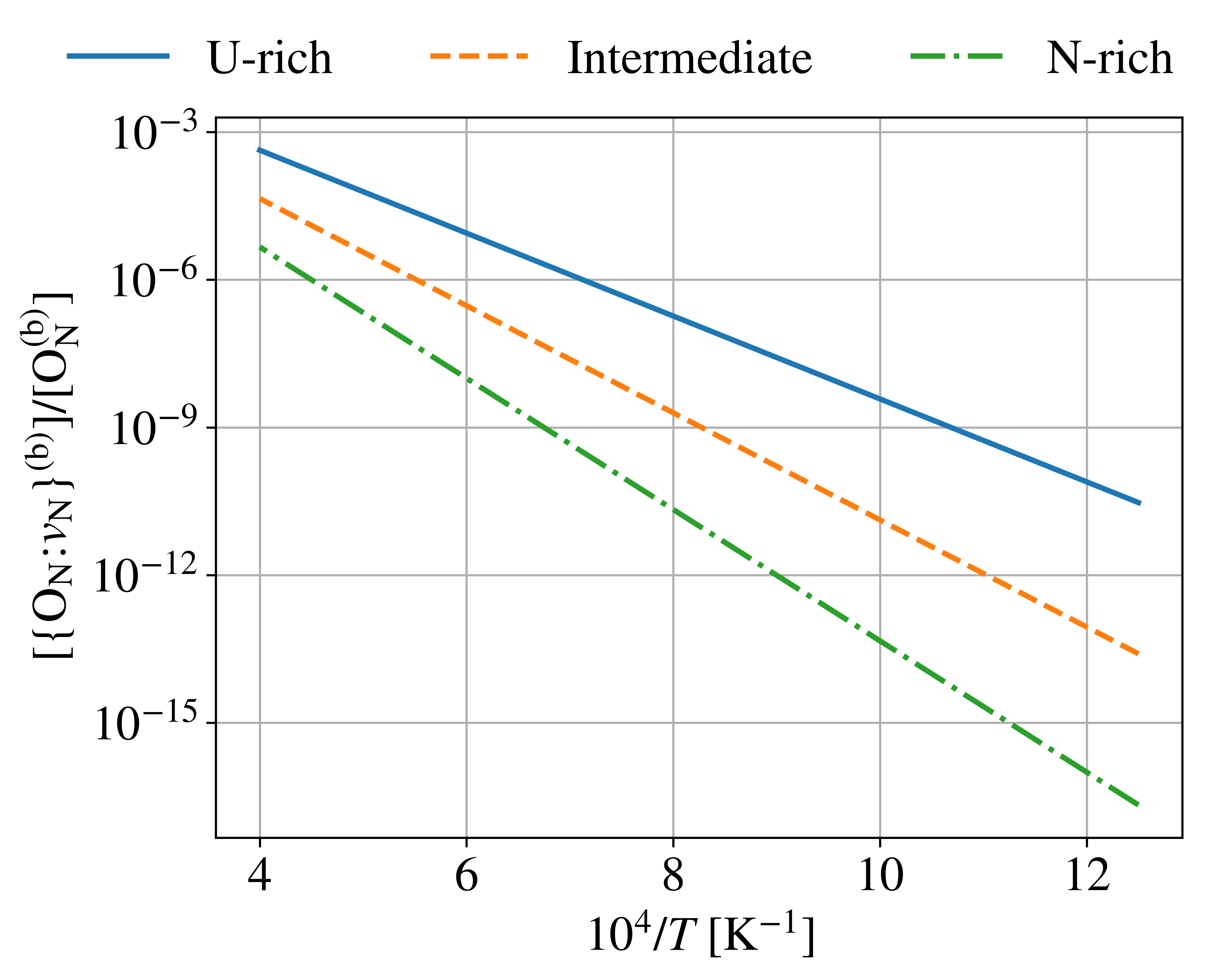}
    \caption{}
    \label{Fig:ONVN_ON}
\end{subfigure}
\caption{Relative concentrations of \textbf{(a)} $\text{O}_i^\text{(b)}$ and \textbf{(b)} $\{\text{O}_\text{N} \! : \! v_\text{N}\}^\text{(b)}$.}
\label{2}
\end{figure}

The atomic concentrations of the N vacancy in the bulk and on the void surfaces are calculated with the aid of formation energies in \cref{Tab:EfVN} and \cref{Eq:VN-VNs} and are shown in \cref{Fig:VN,Fig:VNs}, respectively. The concentrations of $\text{O}_i^\text{(b)}$ and $\{\text{O}_\text{N} \! : \! v_\text{N}\}^\text{(b)}$, relative to $\text{O}_\text{N}^\text{(b)}$ are shown in \cref{Fig:Oi_ON,Fig:ONVN_ON}, respectively. It is clear that $\text{O}_\text{N}^\text{(b)}$ is the dominant defect type for oxygen impurities. Thus, it is justified to set $[ \text{O}_\text{N}^\text{(b)} ] \approx c_\text{O}$ where $c_\text{O}$ is the atomic fraction of oxygen impurities, given by \cref{Eq:cO}. $\text{O}_\text{N}^\text{(b)}$ is an immobile defect. Surface energy reduction requires O migration to void surfaces, which can only occur via the mobile defects $\text{O}_i^\text{(b)}$ and $\{\text{O}_\text{N} \! : \! v_\text{N}\}^\text{(b)}$. However, the immobile $\text{O}_\text{N}^\text{(b)}$ defects can stabilize void embryos via another mechanism by binding to $v_\text{U}$, $\text{Kr}_\text{U}$, and $\text{Xe}_\text{U}$, as explained earlier.
 
\begin{figure}[h!]
\centering
\begin{subfigure}{0.48\textwidth}
    \includegraphics[width=\textwidth]{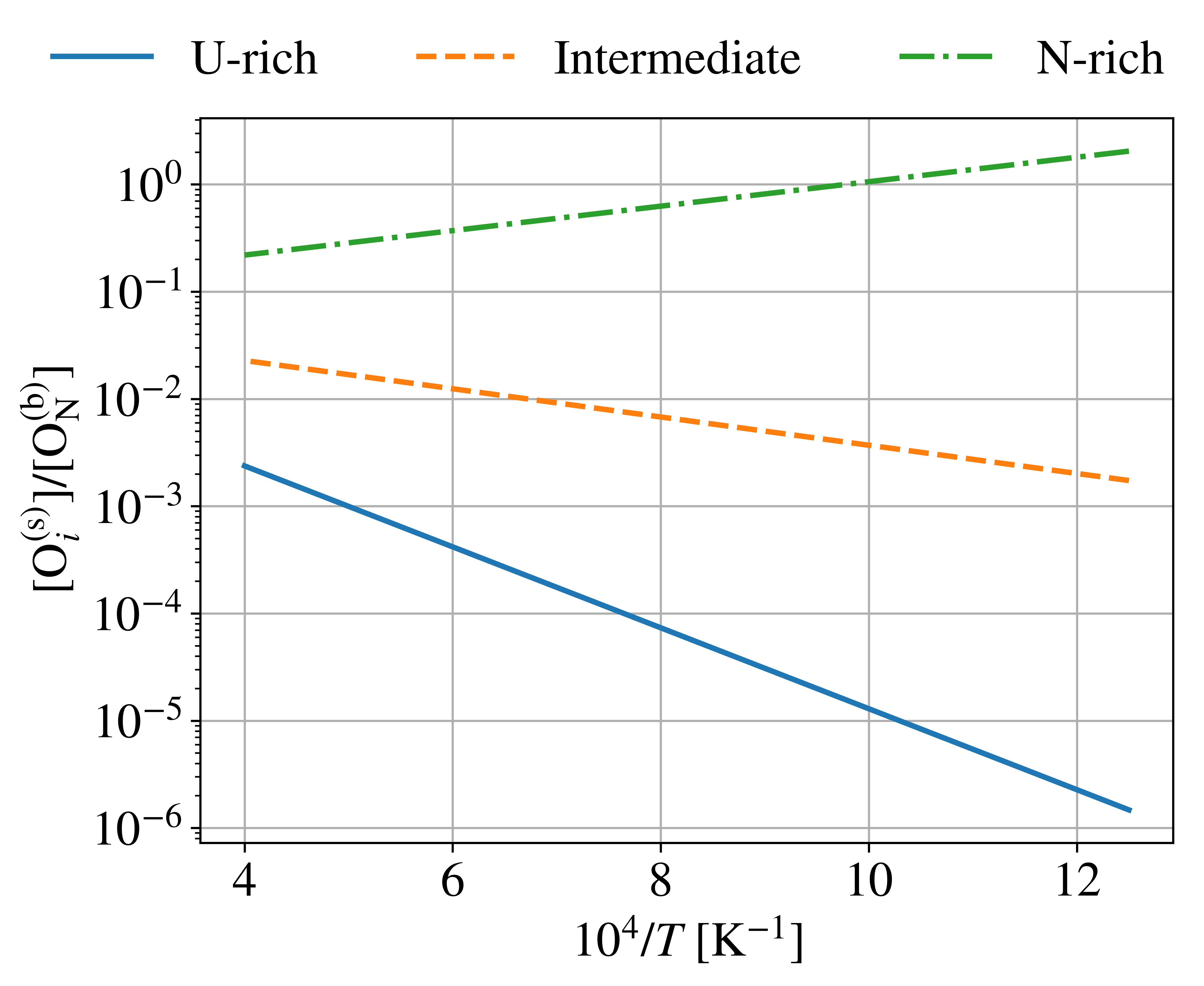}
    \caption{}
    \label{Fig:Ois_ONb}
\end{subfigure}
\hfill
\begin{subfigure}{0.48\textwidth}
    \includegraphics[width=\textwidth]{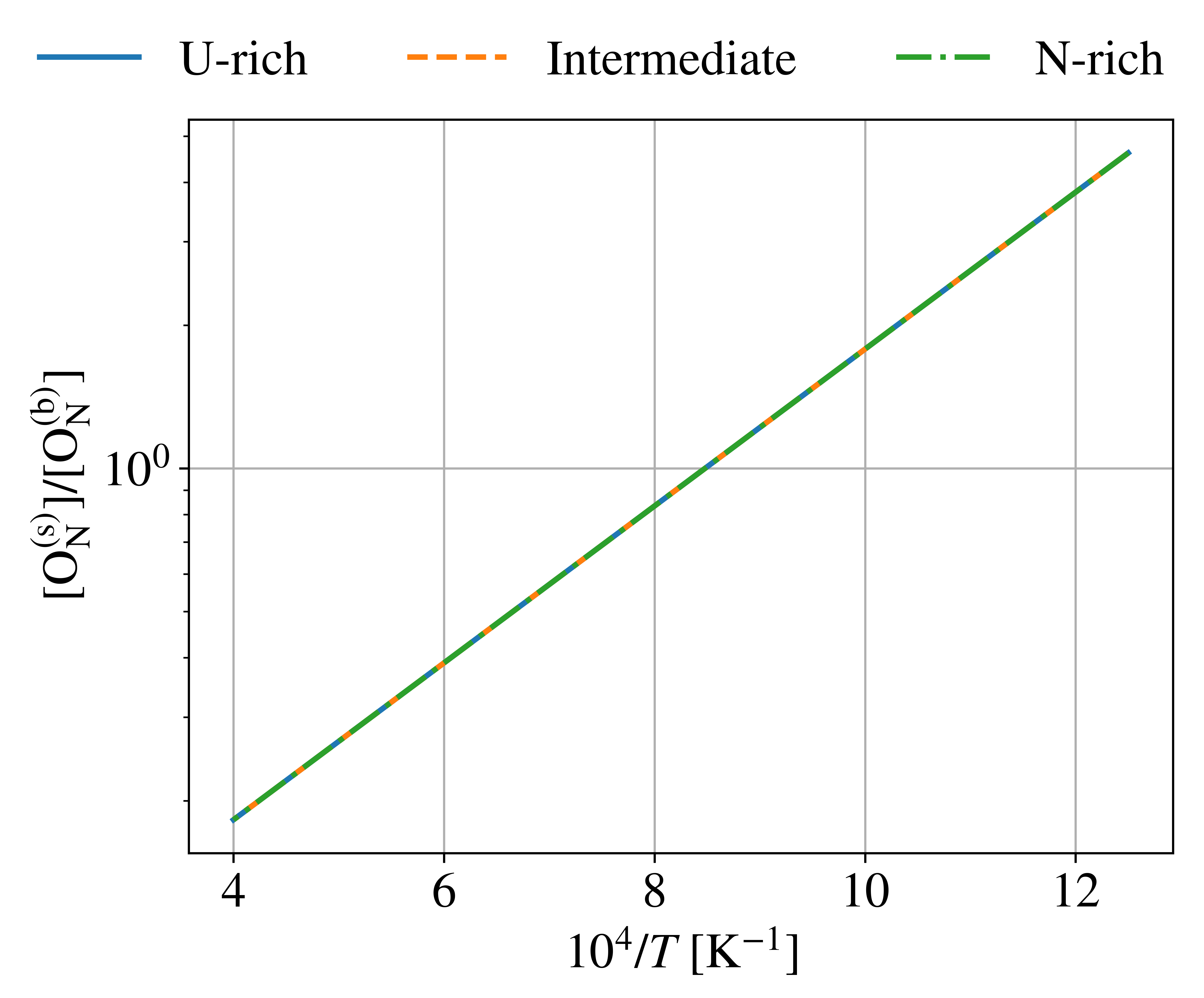}
    \caption{}
    \label{Fig:ONs_ONb}
\end{subfigure}
\hfill
\begin{subfigure}{0.48\textwidth}
    \includegraphics[width=\textwidth]{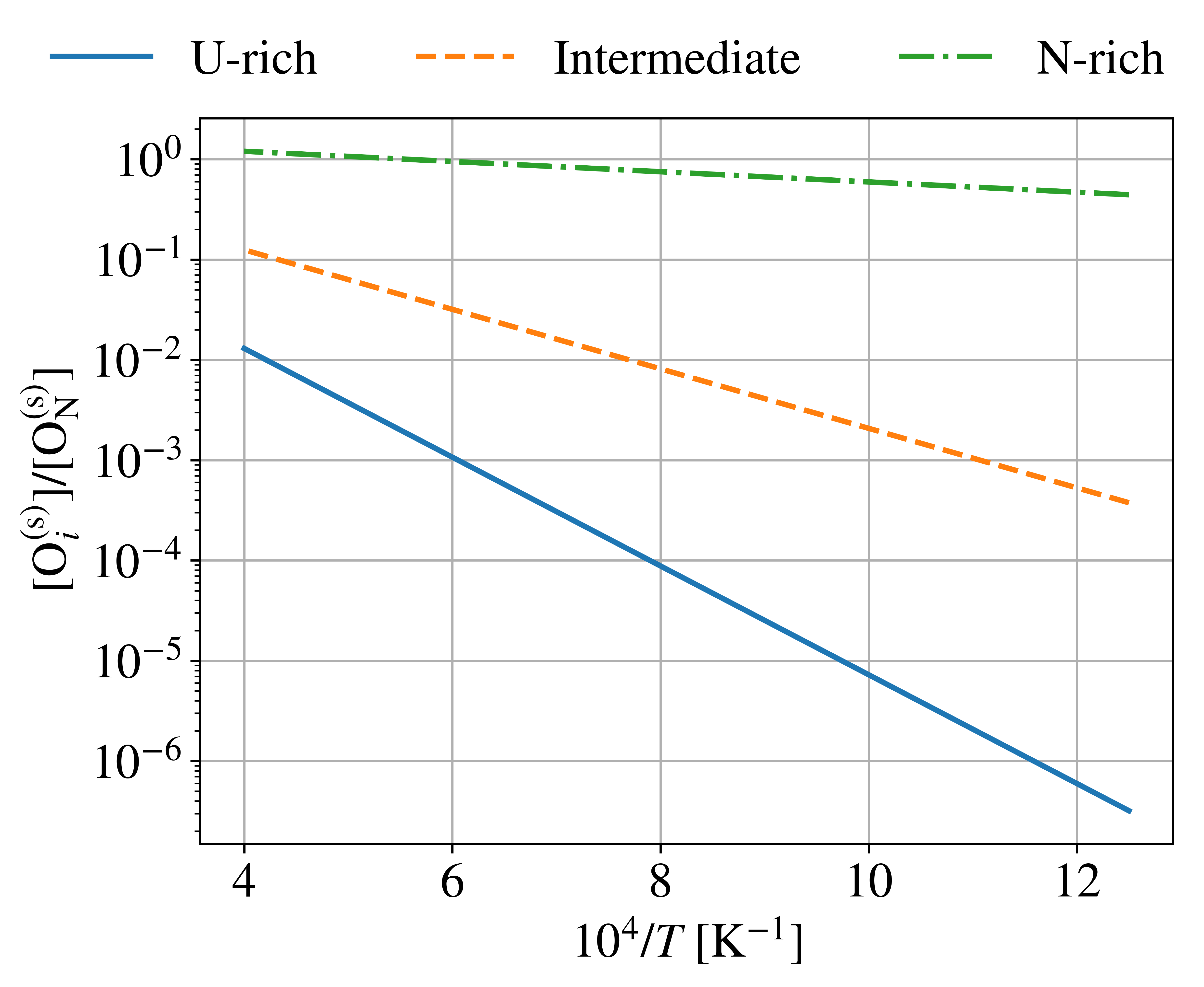}
    \caption{}
    \label{Fig:Ois_ONs}
\end{subfigure}
\caption{Relative concentrations of \textbf{(a)} $\text{O}_i^\text{(s)}$ and \textbf{(b)} $\text{O}_\text{N}^\text{(s)}$, as well as \textbf{(c)} the ratio of their concentrations.}
\label{3}
\end{figure}

The relative concentration $ [ \text{O}_i^\text{(s)} ] / [ \text{O}_\text{N}^\text{(b)} ] $ is illustrated in \cref{Fig:Ois_ONb}. Under U-rich and intermediate conditions, this relative concentration exhibits Arrhenius behavior and remains relatively low. However, under N-rich conditions, which are more relevant in irradiation environments, the relative concentration is significantly higher and follows an anti-Arrhenius trend. Around 1600 K, for instance, for every ten oxygen atoms in the bulk, approximately four segregate to the surface as $\text{O}_i^\text{(s)}$. This behavior arises because, in N-rich environments, there are fewer nitrogen vacancies available to form $ \text{O}_\text{N}^\text{(s)} $, leading to a preference for oxygen atoms to segregate to the surface. Similarly, in \cref{Fig:ONs_ONb}, an anti-Arrhenius trend is also observed for the relative concentration $ [ \text{O}_\text{N}^\text{(s)} ] / [ \text{O}_\text{N}^\text{(b)} ] $. This behavior is independent of stoichiometry, as it is solely governed by the segregation energy. As expected, this anti-Arrhenius behavior indicates that the solubility of oxygen in the bulk increases with increasing temperature. Lastly, \cref{Fig:Ois_ONs} depicts the relative concentration $ [ \text{O}_i^\text{(s)} ] / [ \text{O}_\text{N}^\text{(s)} ] $. As expected, $ \text{O}_\text{N}^\text{(s)} $ is the dominant oxygen impurity site on the void surfaces. Interestingly, despite the large difference in adsorption energies, this relative concentration exceeds 0.1 in N-rich conditions across all considered temperatures.

% At very low temperatures, this ratio exceeds 1, meaning that, if kinetic limitations were to be neglected, for every oxygen atom in the bulk, approximately two atoms would segregate to the surface as $ \text{O}_\text{N}^\text{(s)} $.

The diffusivity of oxygen impurities in bulk UN through mechanisms dependent on the $\{\text{O}_\text{N} \! : \! v_\text{N}\}$ and $\text{O}_i$ defects is illustrated in \cref{Fig:DO}. Under U-rich conditions, the defect cluster $ \{\text{O}_\text{N} \! : \! v_\text{N}\} $ dominates oxygen diffusion. This can be explained by the fact that U-rich environments have a higher concentration of nitrogen vacancies, allowing oxygen to exist as $ \text{O}_\text{N} $ in the bulk. These nitrogen vacancies also facilitate the diffusion process. In intermediate conditions, $\text{O}_i$ exhibits higher diffusivity. In contrast, under N-rich conditions, the availability of nitrogen vacancies decreases, and oxygen predominantly exists as an interstitial $\text{O}_i$. Consequently, the diffusion mechanism via $ \text{O}_i $ becomes dominant in this regime.

It is important to note that we assume $\text{O}_\text{N}^\text{(s)}$ defects reach the surface predominantly through the diffusion of $\{\text{O}_\text{N} \! : \! v_\text{N}\}$. This assumption is later utilized when calculating the reduction in surface energy due to surface adsorption. Specifically, the diffusivity used to calculate $\alpha_1$ in \cref{Eq:DeltaSigma1} is that of $\{\text{O}_\text{N} \! : \! v_\text{N}\}$.

\subsubsection{Surface-energy reduction and peak-temperature mechanism}

\begin{figure}[h!]
\centering
\begin{subfigure}{0.48\textwidth}
    \includegraphics[width=\textwidth]{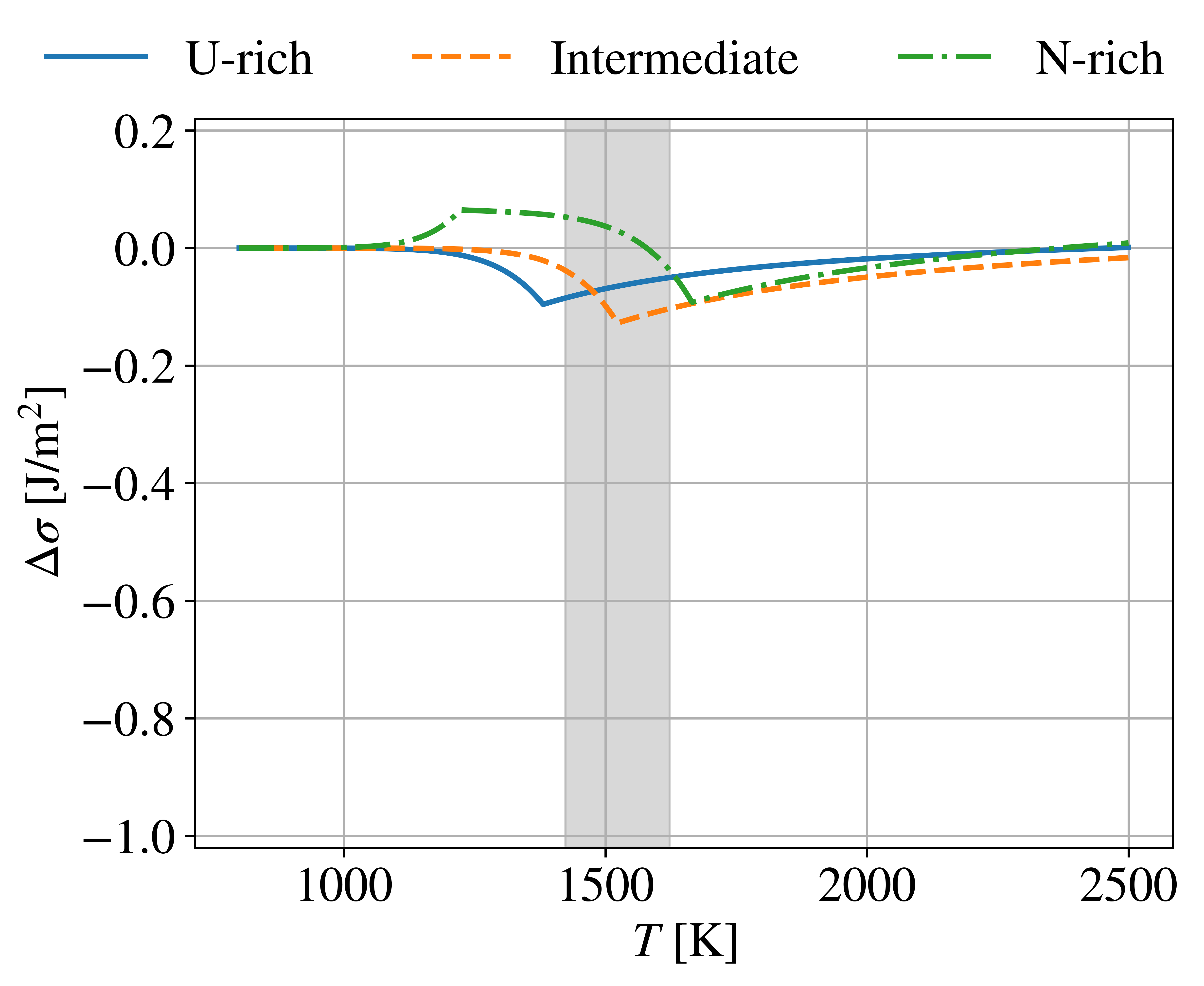}
    \caption{$w_\text{O}$ = 500 ppm, $p$ = 5\%, $R_v$ = 1 nm}
    \label{Fig:500_0.05_1}
\end{subfigure}
\hfill
\begin{subfigure}{0.48\textwidth}
    \includegraphics[width=\textwidth]{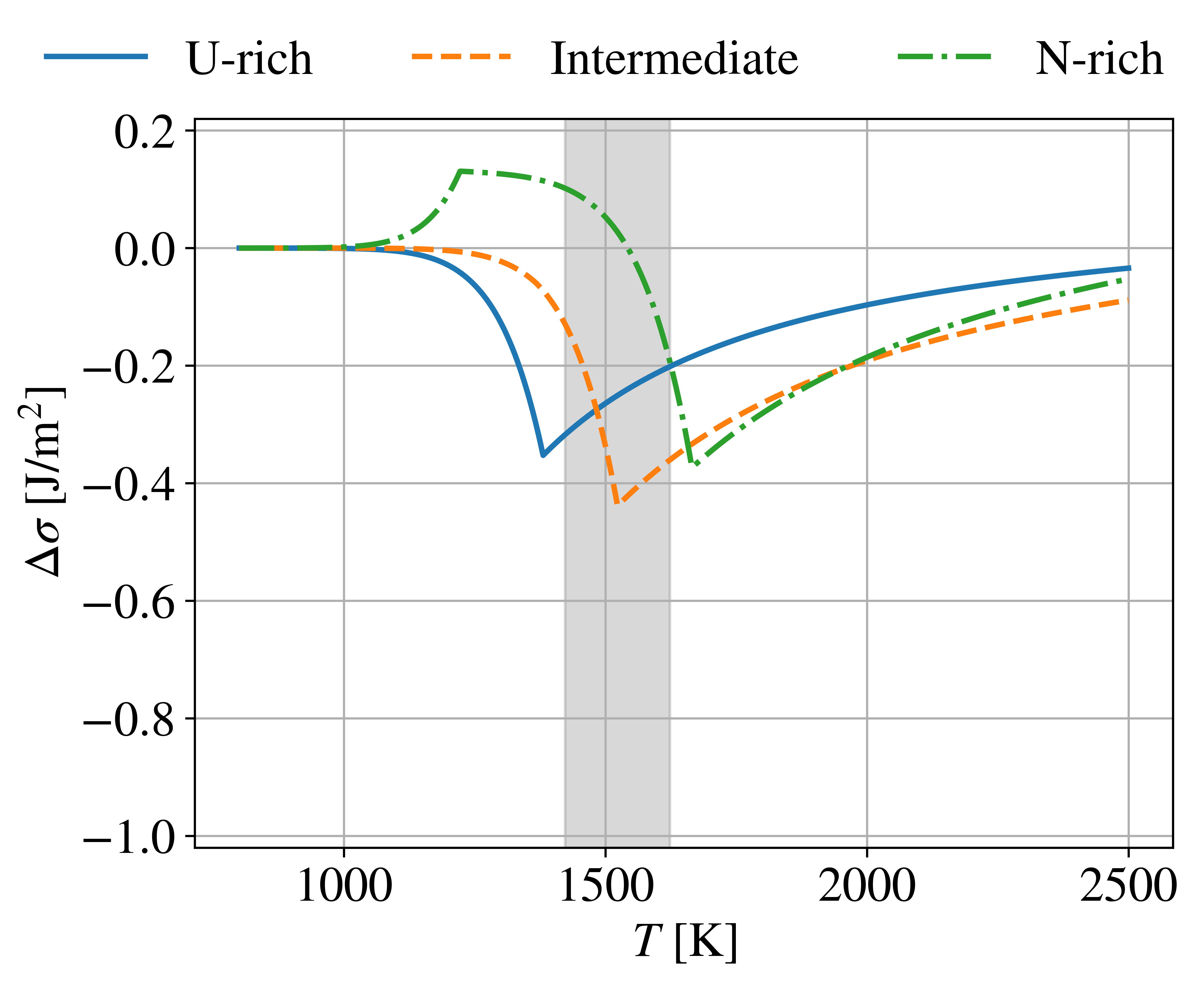}
    \caption{$w_\text{O}$ = 1500 ppm, $p$ = 5\%, $R_v$ = 1 nm}
    \label{Fig:1500_0.05_1}
\end{subfigure}
\hfill
\begin{subfigure}{0.48\textwidth}
    \includegraphics[width=\textwidth]{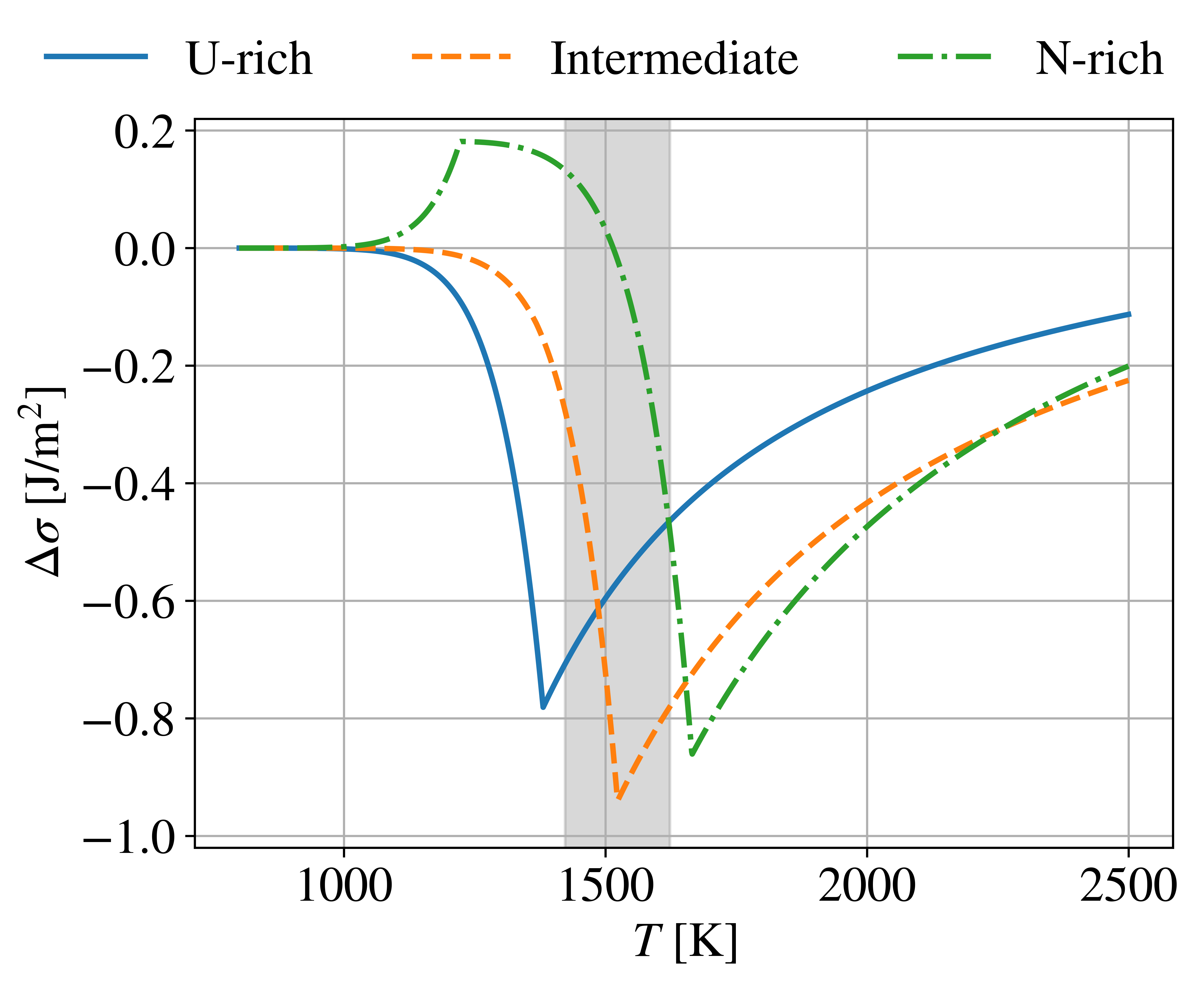}
    \caption{$w_\text{O}$ = 3000 ppm, $p$ = 5\%, $R_v$ = 1 nm}
    \label{Fig:3000_5_1}
\end{subfigure}
\begin{subfigure}{0.48\textwidth}
    \includegraphics[width=\textwidth]{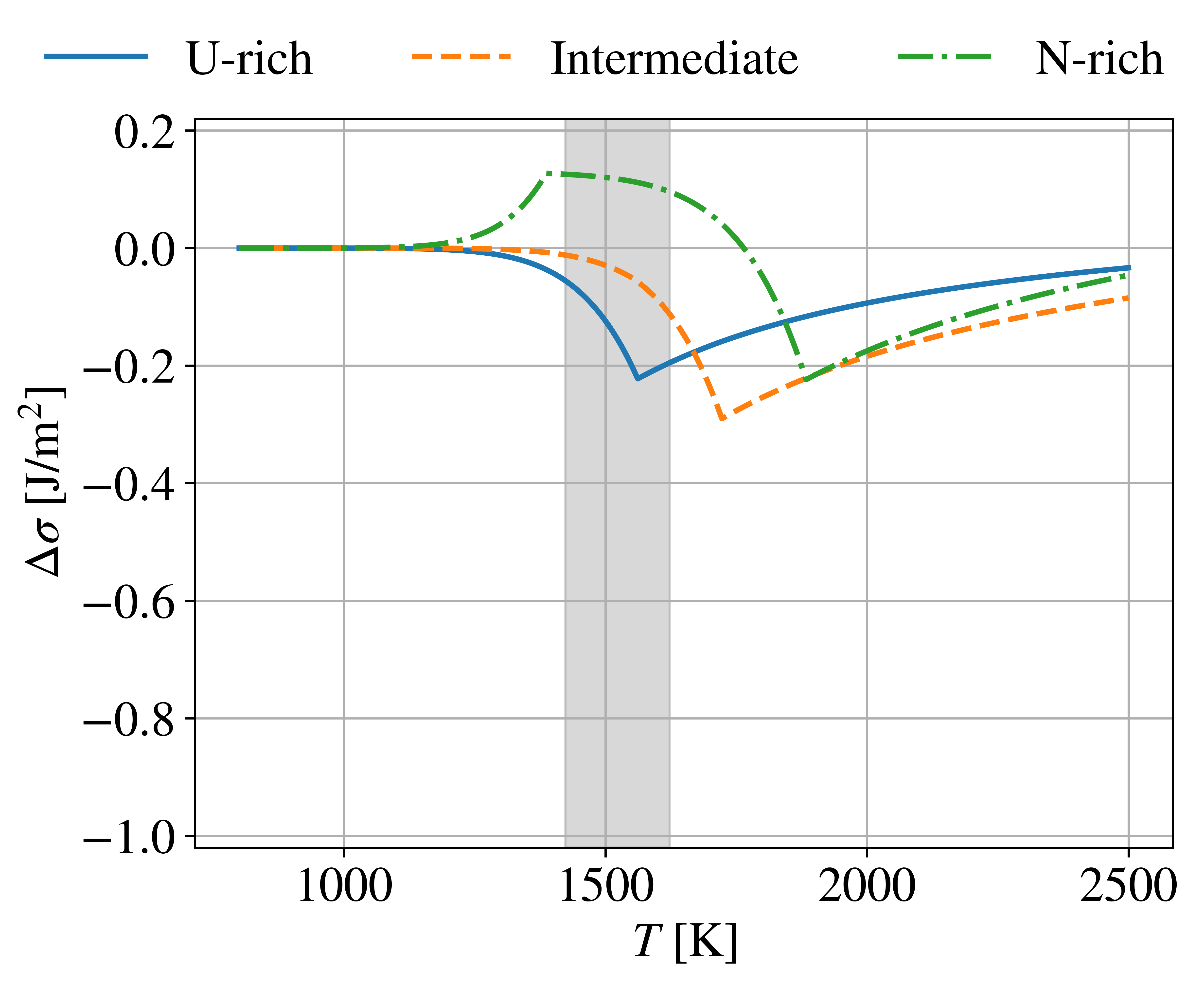}
    \caption{$w_\text{O}$ = 1500 ppm, $p$ = 5\%, $R_v$ = 10 nm}
    \label{Fig:1500_0.05_10}
\end{subfigure}
% \hfill
% \begin{subfigure}{0.3\textwidth}
%     \includegraphics[width=\textwidth]{Delta_sigma_0.1_1500.0_1e-09.png}
%     \caption{$w_\text{O}$ = 1500 ppm, $p$ = 10\%, $R_v$ = 1 nm}
%     \label{Fig:1500_0.10_1}
% \end{subfigure}
% \hfill
% \begin{subfigure}{0.3\textwidth}
%     \includegraphics[width=\textwidth]{Delta_sigma_0.1_1500.0_1e-08.png}
%     \caption{$w_\text{O}$ = 1500 ppm, $p$ = 10\%, $R_v$ = 10 nm}
%     \label{Fig:1500_0.10_10}
% \end{subfigure}
\caption{Temperature evolution of surface energy change due to oxygen adsorption and vacancy formation for various porosities, oxygen concentrations, and average void radii. The gray shaded band indicates the breakaway swelling critical temperature range for mixed nitrides (1423--1623~K), based on $T_c = 1473$--$1573$~K with $\pm$50~K uncertainty~\cite{Ronchi1978}.}
\label{Fig:Delta_sigma}
\end{figure}

The variation in surface energy due to oxygen adsorption and vacancy formation on void surfaces is depicted in \cref{Fig:Delta_sigma} for various porosities, oxygen concentrations, and average void radii. At low temperatures, the change in surface energy is negligible. This is because oxygen diffusivity, incorporated in the model via the kinetic correction, $\alpha_i$, is very slow at low temperatures. Within the void nucleation time, the mobility of oxygen impurities is insufficient for them to reach the void surfaces. For the reference case of $w_\text{O}$ = 1500 ppm, $p$ = 5\%, $R_v$ = 1 nm (\cref{Fig:1500_0.05_1}), the surface energy change is most significant near its peak at 1524~K, remaining above half-maximum over 1465--1914~K and overlapping the breakaway swelling regime in nitride fuels. Ronchi \textit{et al.}~\cite{Ronchi1978} reported the breakaway swelling critical temperature $T_c$ for mixed nitrides in the interval 1473--1573~K, with an uncertainty of $\pm$50~K. This range, shown as a shaded band in \cref{Fig:Delta_sigma}, overlaps with the temperature of maximum $|\Delta \sigma|$ predicted by our model. The mechanistic modeling by Rizk \textit{et al.}~\cite{Rizk2025} corroborates this picture, attributing the breakaway transition in UN to a change in the gas atom diffusion mechanism occurring around 1500~K. Noting that the range of operating temperatures in liquid metal-cooled fast reactors is 800--2000~K~\cite{Turos1990}, these findings highlight oxygen adsorption as a critical factor in facilitating void nucleation under conditions typical of reactor environments. Beyond the representative cases shown in \cref{Fig:Delta_sigma}, the peak reduction $\Delta\sigma_\text{min}$ and the corresponding $T_\text{peak}$ for a denser set of void radii ($R_v$ = 1--50~nm) and porosities are compiled in \cref{Tab:App:Tpeak}.

% \begin{figure}[h!]
%     \centering
%     \includegraphics[width=0.5\textwidth]{Delta_sigma.png}
%     \caption{Temperature evolution of surface energy change due to oxygen adsorption and vacancy formation for oxygen weight concentration of $w_\text{O}$ = 1500 ppm, porosity of $p$ = 5\%, and average void radius of $R_v$ = 1 nm.}
%     \label{Fig:Delta_sigma}
% \end{figure}

\begin{figure}[h!]
\centering
\begin{subfigure}{0.48\textwidth}
    \includegraphics[width=\textwidth]{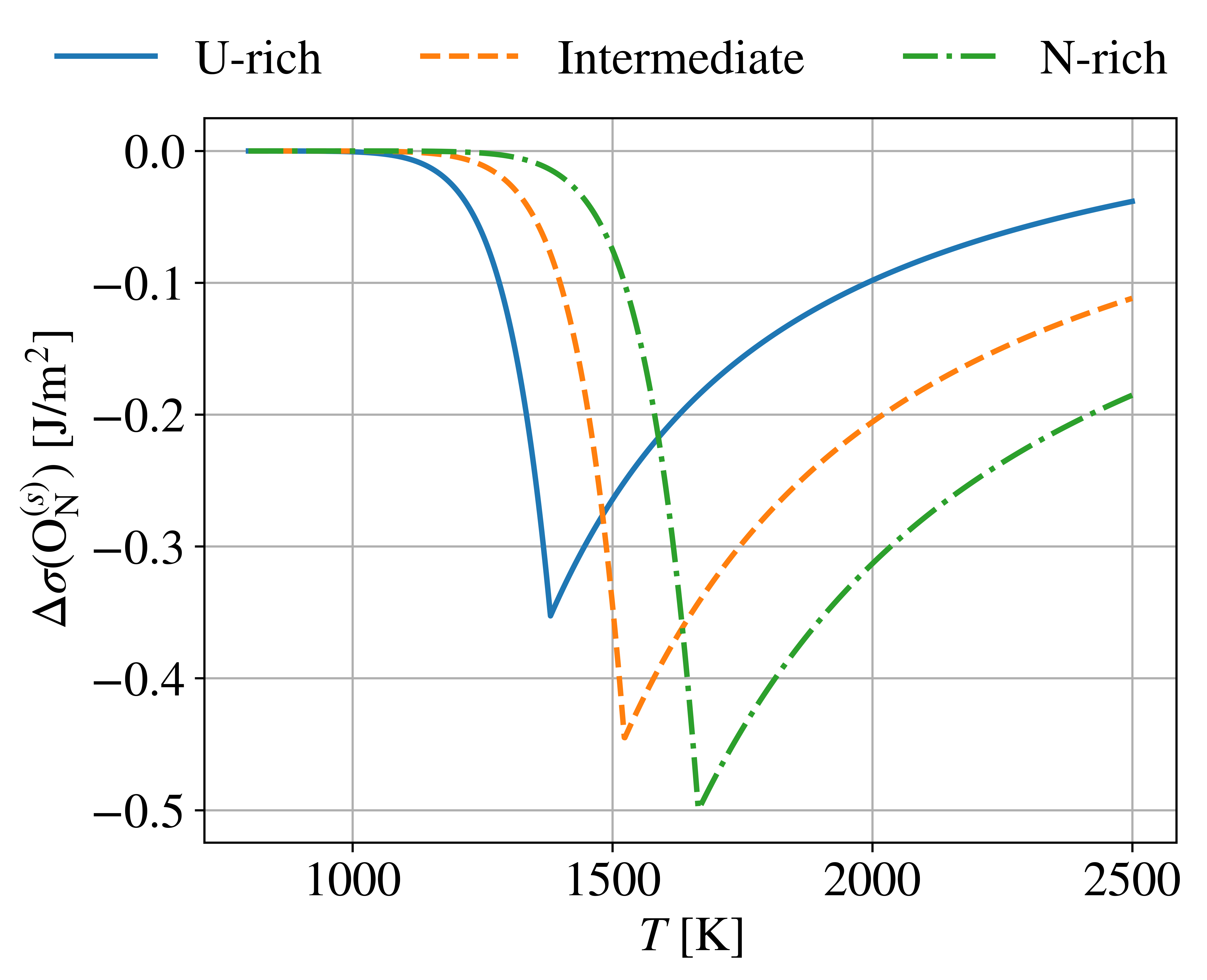}
    \caption{}
    \label{Fig:D1}
\end{subfigure}
\hfill
\begin{subfigure}{0.48\textwidth}
    \includegraphics[width=\textwidth]{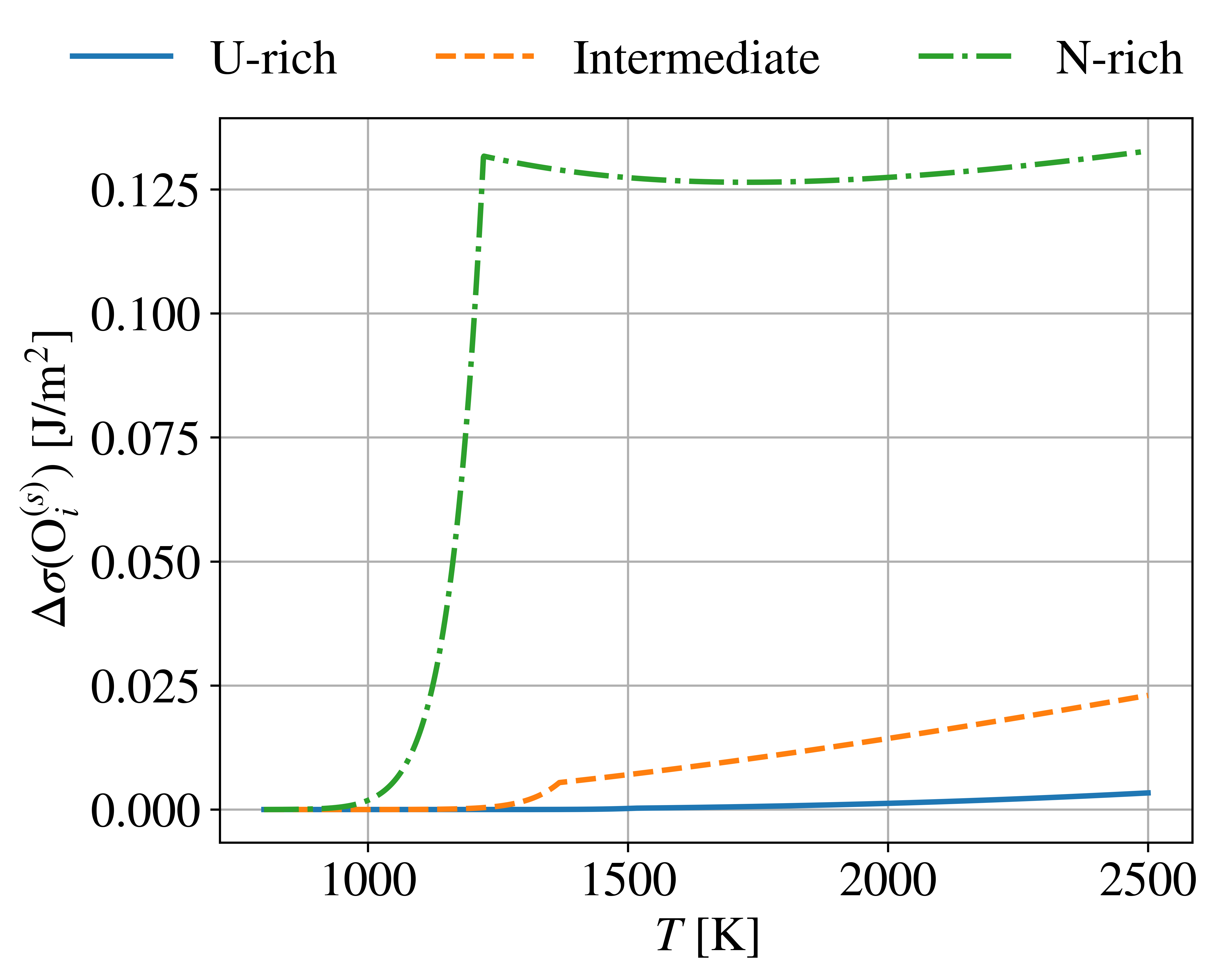}
    \caption{}
    \label{Fig:D2}
\end{subfigure}
\hfill
\begin{subfigure}{0.48\textwidth}
    \includegraphics[width=\textwidth]{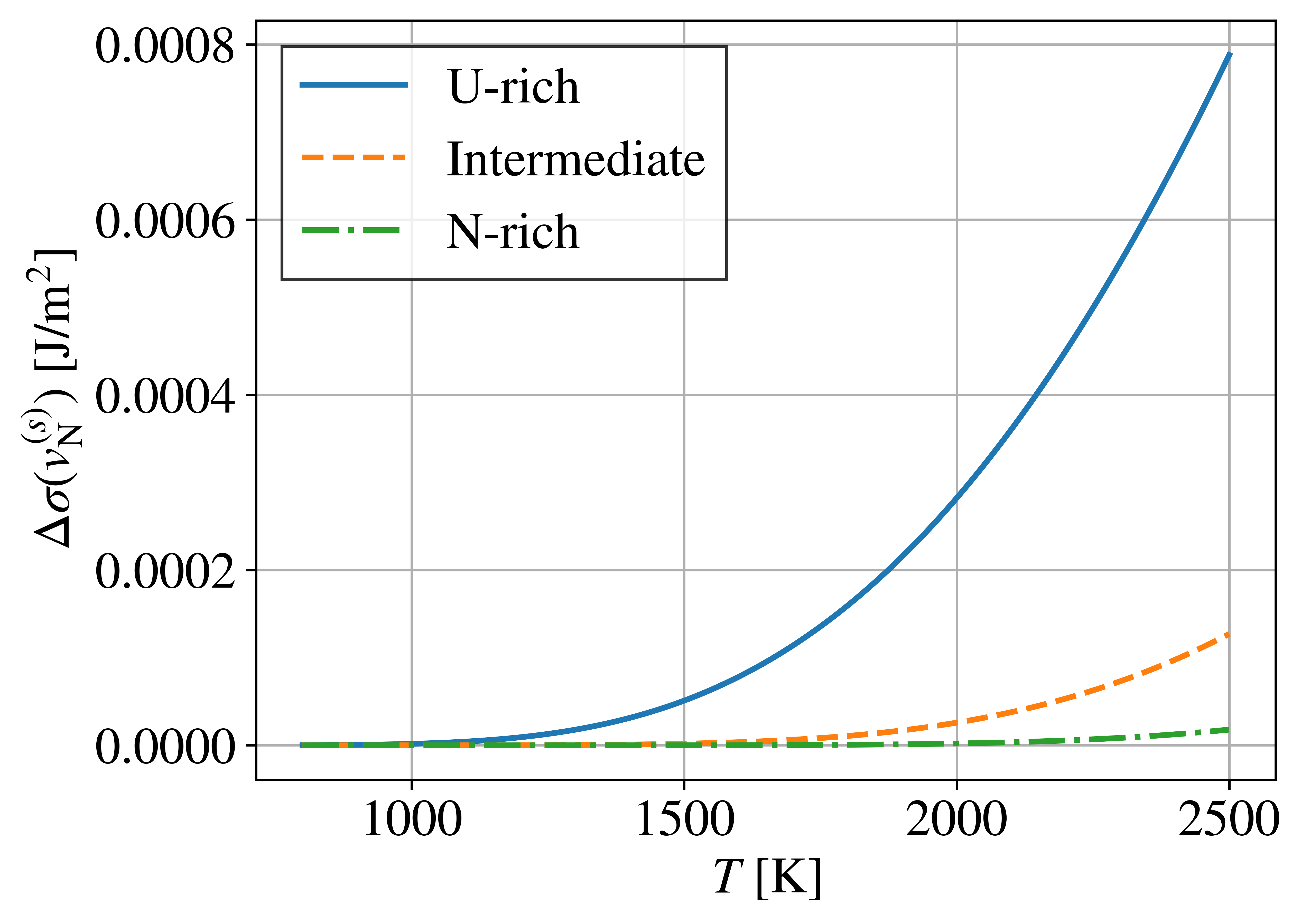}
    \caption{}
    \label{Fig:D3}
\end{subfigure}
\caption{Different terms of the surface energy change due to oxygen adsorption and vacancy formation on void surfaces for $w_\text{O}$ = 1500 ppm, $p$ = 5\%, and $R_v$ = 1 nm. \textbf{(a)} $\Delta \sigma ( \text{O}_\text{N}^{\text{(s)}} )$, \textbf{(b)} $\Delta \sigma ( \text{O}_i^{\text{(s)}} )$, and \textbf{(c)} $\Delta \sigma (v_\text{N}^{\text{(s)}})$ are given in \cref{Eq:D1,Eq:D2,Eq:D3}, respectively.}
\label{Fig:D}
\end{figure}

It is of interest to examine the contribution of the different terms to the change in surface energy, where $\Delta \sigma ( \text{O}_\text{N}^\text{(s)} )$ (\cref{Eq:D1}) represents the contribution of $\text{O}_\text{N}^\text{(s)}$ to $\Delta \sigma$, with analogous definitions for $\Delta \sigma ( \text{O}_i^\text{(s)} )$ (\cref{Eq:D2}) and $\Delta \sigma ( v_\text{N}^\text{(s)} )$ (\cref{Eq:D3}). 

\begin{equation}
\Delta \sigma ( \text{O}_\text{N}^{\text{(s)}} ) = \frac{8}{3} \frac{1-p}{p} \frac{R_v}{a^3} c_\text{O} \alpha_1 \frac{ [ \text{O}_\text{N}^{\text{(s)}} ] }{ [ \text{O}_\text{N}^{\text{(b)}} ] }  E_\text{ad}( \text{O}_\text{N}^{\text{(s)}} ).
\label{Eq:D1}
\end{equation}

\begin{equation}
\Delta \sigma ( \text{O}_i^{\text{(s)}} ) = \frac{8}{3} \frac{1-p}{p} \frac{R_v}{a^3} c_\text{O} \alpha_2 \frac{ [ \text{O}_i^{\text{(s)}} ] }{ [ \text{O}_\text{N}^{\text{(b)}} ] }  E_\text{ad}( \text{O}_i^{\text{(s)}} ).
\label{Eq:D2}
\end{equation}

\begin{equation}
\Delta \sigma (v_\text{N}^{\text{(s)}}) = \frac{2}{a^2} [ v_\text{N}^{\text{(s)}} ] E_f ( v_\text{N}^{\text{(s)}} ).
\label{Eq:D3}
\end{equation}
\noindent These terms are plotted in \cref{Fig:D}. Comparing \cref{Fig:Delta_sigma,Fig:D1}, it can be seen that the adsorption of oxygen as $\text{O}_\text{N}^\text{(s)}$ dominates the $\Delta \sigma$ profile. This is expected since $\text{O}_\text{N}^\text{(s)}$ is the most dominant oxygen adsorption site (refer to \cref{Fig:Ois_ONs}) and it leads to a reduction of surface energy at all temperatures (also see \cref{Fig:EadONs}). From \cref{Fig:D2}, it can be observed that $\text{O}_i^\text{(s)}$ is responsible for the slight surface energy increase in the range of 1020--1540~K in the N-rich profile of \cref{Fig:Delta_sigma}. Finally, in \cref{Fig:D3}, it is obvious that the formation of N vacancies has a negligible contribution to the surface energy change of voids.

Given that the effect of nitrogen vacancies on surface energy reduction is negligible, and the contribution of $\text{O}_i^{\text{(s)}}$ is second order, the $\Delta \sigma$ formula (\cref{Eq:DeltaSigma1}) simplifies to a product of four terms: a geometric factor $\eta = ((1 - p)/p) \cdot R_v/3$, the oxygen number density $n_\text{O} = c_\text{O} / \Omega$, a kinetic correction factor $\alpha_1$, and an effective segregation energy $E_\text{seg}^\text{eff} = ( [\text{O}_\text{N}^{(\text{s})}] / [\text{O}_\text{N}^{(\text{b})}] ) \, E_\text{ad}(\text{O}_\text{N}^{(\text{s})})$. Specifically, we write:
\begin{equation}
\Delta \sigma \approx \left( \frac{(1 - p)}{p} \, \frac{R_v}{3} \right) \cdot \frac{c_\text{O}}{\Omega} \cdot \alpha_1 \cdot \left( \frac{ [ \text{O}_\text{N}^{(\text{s})} ] }{ [ \text{O}_\text{N}^{(\text{b})} ] } \, E_\text{ad}( \text{O}_\text{N}^{(\text{s})} ) \right) = \eta \, n_\text{O} \, \alpha_1 \, E_\text{seg}^\text{eff},
\label{Eq:DeltaSigmaSimple}
\end{equation}
where $c_\text{O}$ is the atomic fraction of oxygen, $\Omega = a^3/8$ is the atomic volume of the matrix (inverse of the atomic number density), $p$ is the porosity, and $R_v$ is the average void radius. The geometric factor $\eta$ captures the matrix-to-void volume ratio, $(1-p)/p$, and the void volume-to-surface ratio, $R_v/3$. That is, $\eta$ represents the average matrix volume per unit void surface area: $(V_\text{matrix}/V_\text{void}) \times (V_\text{void}/A_\text{void}) = V_\text{matrix} / A_\text{void}$, and gives a length scale that quantifies an effective thickness of matrix material supplying each unit of void surface. The kinetic correction factor $\alpha_1$ accounts for the finite diffusivity of oxygen within the void nucleation timescale (refer to \cref{Eq:alpha}). Finally, $E_\text{seg}^\text{eff}$ represents the net energy gain per oxygen atom that segregates to the void surface. This formulation makes explicit that the surface energy reduction is governed by the interplay of thermodynamics, kinetics, and microstructural geometry, and scales linearly with the number of oxygen atoms that successfully reach and adsorb on the available void surface area. Note that the temperature dependence of $|\Delta \sigma|$ is contained in $E_\text{seg}^{\text{eff}}$ and $\alpha_1$.

The temperature dependence of $|\Delta\sigma|$ is governed by the competition between oxygen adsorption kinetics and bulk solubility. At low temperatures, diffusion is too slow for significant oxygen transport to void surfaces. At high temperatures, the increased bulk solubility of oxygen disfavors surface segregation. As a result, $|\Delta\sigma|$ exhibits a maximum at an intermediate temperature $T_\text{peak}$.

A closed-form expression for this peak temperature is derived in \ref{App:Tpeak} by analyzing the piecewise monotonicity of $|\Delta\sigma|$. The maximum coincides with the kinetic saturation temperature $T^*$ at which oxygen diffusion first spans the effective capture length within the nucleation time, i.e., $\sqrt{D_1(T^*)\,t} = \lambda_v$. Solving this condition yields:
\begin{equation}
T_\text{peak} = \frac{ E_m^\mathrm{eff} }{ k\ln\!\left( D_{1,0}t/\lambda_v^2 \right) },
\label{Eq:TpeakMain}
\end{equation}
where $D_{1,0} = f z \lambda^2 \nu/6$ is the prefactor of the $\{\text{O}_\text{N} \! : \! v_\text{N}\}$-mediated diffusivity,
\begin{equation}
E_m^\mathrm{eff} = E_m + E_f(v_\text{N}^{(\text{b})}) + E_b(\{\text{O}_\text{N} \! : \! v_\text{N}\}^{(\text{b})})   
\end{equation}
is its effective activation energy, and $\lambda_v = 0.5\,n_v^{-1/3}-R_v$ is the effective capture length defined in \cref{Eq:lambda_eff}. Because $R_v$, $p$, and $t$ enter \cref{Eq:TpeakMain} only through the logarithm, $T_\text{peak}$ is controlled primarily by the effective activation energy $E_m^\mathrm{eff}$. The oxygen concentration $c_\text{O}$ does not enter \cref{Eq:TpeakMain}. Changing $c_\text{O}$ changes the magnitude of the surface-energy reduction but not the kinetic saturation temperature.

The porosity dependence of $|\Delta\sigma|$ is explained analytically in \ref{App:Tpeak:Porosity}. In the dominant $\text{O}_\text{N}^{(\text{s})}$ contribution, the geometric prefactor contains $(1-p)/p$, while the segregation ratio $[\text{O}_\text{N}^{(\text{s})}]/[\text{O}_\text{N}^{(\text{b})}]$ contains the surface-site factor $[\text{N}_\text{N}^{(\text{s})}]/[\text{N}_\text{N}^{(\text{b})}]$. Since $[\text{N}_\text{N}^{(\text{s})}] \sim p/(1-p)$, it cancels the leading $(1-p)/p$ dependence exactly. This cancellation should be kept separate from the remaining appearances of $p$ in the model. After cancellation, $|\Delta \sigma|_\text{peak}$ retains a residual direct dependence through $1/[\text{N}_\text{N}^{(\text{b})}]$, where $[\text{N}_\text{N}^{(\text{b})}]=1-\frac{3}{2}\frac{p}{1-p}\frac{a}{R_v}$, and an indirect dependence through the temperature-dependent factors evaluated at $T_\text{peak}(p)$. The latter arises because the capture length, $\lambda_v=0.5n_v^{-1/3}-R_v$, decreases with increasing porosity, lowering $T_\text{peak}$. Thus the cancellation is exact for the leading geometric factor, but only partial for the full peak magnitude. Numerical evaluation shows that the peak $|\Delta\sigma|$ increases by 23\% across $p=5$--20\% for $R_v=25$~nm and by 41\% for $R_v=1$~nm (\cref{Tab:App:por}); this is appreciable but still far smaller than the naive $4.75\times$ change implied by the isolated $(1-p)/p$ factor.

The dependence on void size is more significant. At fixed porosity,
\begin{equation}
\lambda_v = R_v \left[ 0.5 \left[ 4\pi/(3p) \right]^{1/3} - 1 \right] \sim R_v,
\end{equation}
so larger voids increase the denominator of \cref{Eq:TpeakMain}, shifting $T_\text{peak}$ to higher temperatures. This also reduces $|\Delta\sigma|_\text{peak}$ because the segregation drive is weaker at higher temperatures. For example, increasing $R_v$ from 1~nm to 10~nm shifts $T_\text{peak}$ upward by approximately 200~K (\cref{Fig:1500_0.05_1,Fig:1500_0.05_10}), consistent with \cref{Eq:TpeakMain}. Oxygen-induced surface stabilization is therefore most effective for small voids during the early stages of nucleation and growth.

\subsubsection{Parametric dependence and relevance to swelling models}

For the reference case of $w_\text{O}$ = 1500 ppm, $p$ = 5\%, and $R_v$ = 1 nm, the model predicts $T_\text{peak}=1524$~K and $\Delta \sigma_\text{min}=-0.438$~J/m$^2$ (see \cref{Fig:1500_0.05_1}). To assess the sensitivity of $|\Delta \sigma|$ to key parameters, we vary $w_\text{O}$ and $p$ while holding $R_v$ fixed at 25 nm, which is a typical size for both dislocations and intergranular bubbles \cite{Colin1983}. It has been shown that dislocation bubbles dominate bubble swelling in UN \cite{Ronchi1975,Ronchi1978,Rizk2025}. Although the assumptions of our methodology are more representative of bulk voids, we assume that the same surface energy reduction mechanism applies to intragranular (bulk and dislocation) and intergranular bubbles.

\begin{figure}[h!]
    \centering
    \includegraphics[width=0.5\textwidth]{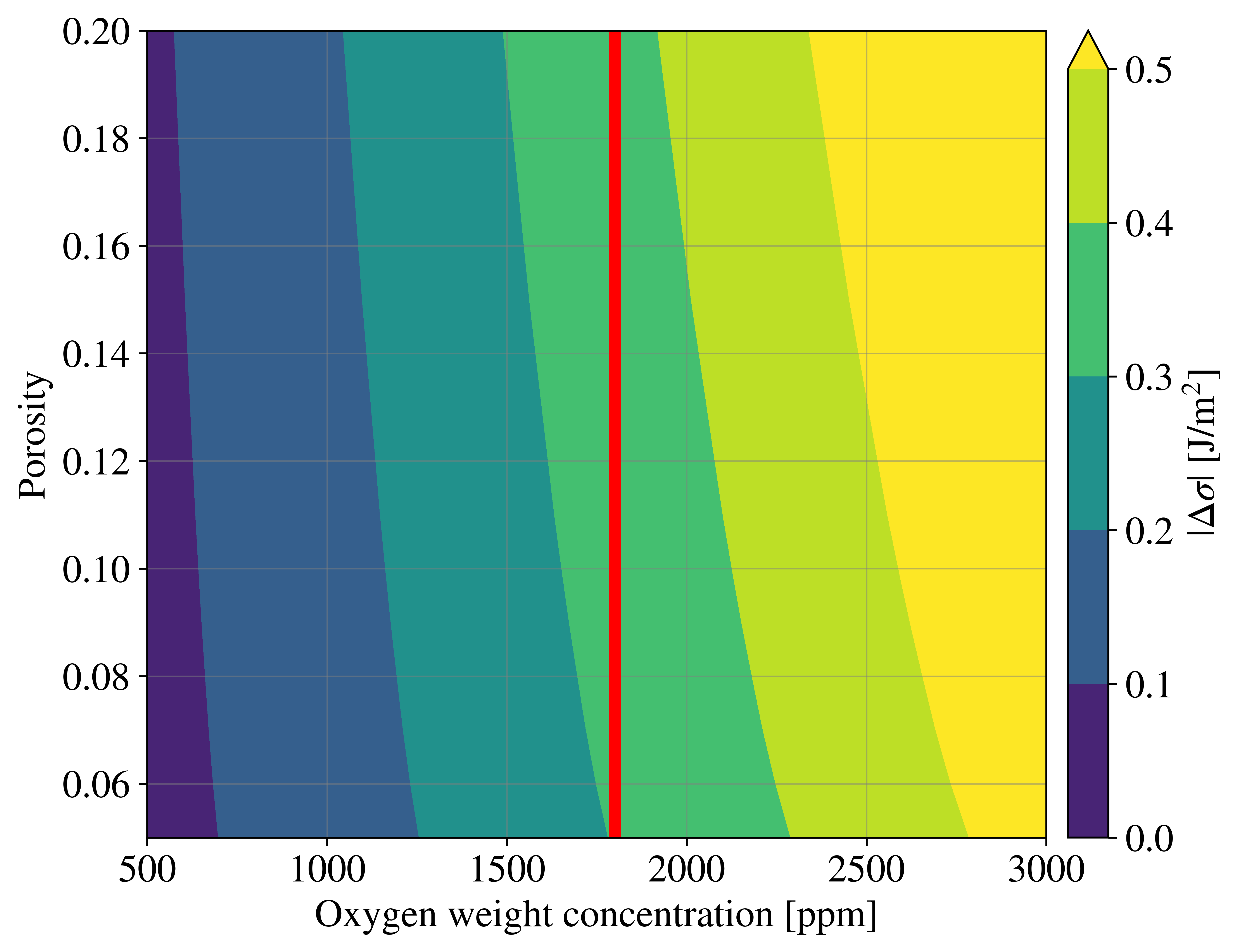}
    \caption{(Color online) Maximum surface energy reduction, $|\Delta \sigma|_\text{peak}$, as a function of porosity and oxygen concentration under intermediate stoichiometry for a fixed bubble radius $R_v$ = 25 nm, typical of dislocation and grain-boundary bubbles. The red line marks $w_\text{O}$ = 1800 ppm, the oxygen concentration in UN samples from Ronchi \textit{et al.} \cite{Ronchi1975,Ronchi1978}, used by Rizk \textit{et al.} \cite{Rizk2025} to validate their swelling model.}
    \label{Fig:p_w_plot}
\end{figure}

\Cref{Fig:p_w_plot} represents $|\Delta \sigma|_\text{peak}$ at intermediate stoichiometric conditions. For the fixed bubble radius $R_v$ = 25~nm, the peak temperature varies from approximately 1700 to 1820~K across $p=5$--20\% at intermediate stoichiometry. This variation is controlled by porosity through the capture length in \cref{Eq:TpeakMain}. By contrast, $T_\text{peak}$ is independent of $w_\text{O}$ in the saturated closed form. The color map shows a moderate porosity dependence of the peak magnitude: increasing $p$ from 5\% to 20\% raises $|\Delta\sigma|_\text{peak}$ by about 23\% for $R_v=25$~nm, still far less than the naive $(1-p)/p$ scaling would imply. The vertical red line marks the oxygen concentration in UN samples from Ronchi \textit{et al.} ($w_\text{O} = 1800$ ppm) \cite{Ronchi1975,Ronchi1978}, used by Rizk \textit{et al.} \cite{Rizk2025} to validate their swelling model. Fission gas bubbles with an average radius of 25~nm are predicted to experience a reduction in surface energy of about 0.30--0.37~J/m$^2$ at this oxygen content, approximately 20\% of our calculated surface energy in the oxygen-free case.

% This observation, however, requires some explanation. According to \cref{Eq:DeltaSigmaSimple}, $|\Delta \sigma| \sim c_\text{O} R_v$, i.e., increasing $R_v$ at fixed porosity reduces the total internal surface area available for oxygen segregation, which would naively suggest a lower required bulk oxygen concentration to maintain the same surface coverage, and hence the same $|\Delta \sigma|$. However, this geometric scaling is offset by two competing effects. First, the site availability for oxygen segregation decreases with increasing $R_v$, as the surface nitrogen vacancy concentration scales inversely with void radius, i.e., $[v_\text{N}^\text{(s)}] \sim [\text{N}_\text{N}^\text{(s)}] \sim 1/R_v$ (see \cref{Eq:VN-VNs,Eq:NNs1}). Second, the kinetic correction factor $\alpha_1 \sim 1/\lambda_v \sim 1/R_v$ (see \cref{Eq:alpha}). This is reasonable since fewer, and therefore farther apart, voids make it harder for an O atom to reach a surface within the nucleation time. These effects collectively result in a net scaling of $|\Delta \sigma| \sim c_{\text{O}} / R_v$, such that maintaining a constant surface energy reduction requires increasing $c_\text{O}$ (or equivalently $w_\text{O}$) as $R_v$ is increased.

In UN samples with an oxygen content of 1600 ppm and carbon content of 300 ppm, Mishchenko \textit{et al.} \cite{Mishchenko2021} found 1.5 wt.\% UO$_2$. Thus, having an oxygen content of, e.g., 1800 ppm will most likely lead to the separation of a UO$_2$ phase, unless there is some process by which this oxygen is forced to be in solution. There are many solubility limits of oxygen in UN reported in the literature, ranging from 650 ppm to 6640 ppm \cite{Javed1972,Jain1993,Konovalov2016,Jaques2015}. Lyubimov \textit{et al.} \cite{Lyubimov2014} conducted a thermodynamic modeling calculation and found that a carbon concentration of 720 ppm increases the oxygen solubility by a factor of 4--5. Based on this analysis, a carbon content of 720 ppm is large enough to keep an oxygen concentration of 3000 ppm in solution and suppress UO$_2$ phase separation, thus leading to $|\Delta \sigma| > 0.5$ J/m$^2$ (see \cref{Fig:p_w_plot}) and enhanced void swelling in UN.

\section{Discussion}
\label{Sec:Discuss}

This study presents a novel model that integrates first-principles calculations with defect chemistry to quantify the role of oxygen impurities in stabilizing voids and fission gas bubbles in UN. To our knowledge, this is the first model to establish a quantitative link between oxygen concentration and void nucleation thermodynamics in UN. By capturing how oxygen segregates to void surfaces and reduces the surface energy, this work advances the mechanistic understanding of swelling phenomena in oxygen-contaminated nuclear fuels. As shown in \cref{Sec:Oxygen}, the temperature of maximum $|\Delta \sigma|$ coincides with the experimentally reported breakaway swelling critical temperature for nitride fuels~\cite{Ronchi1978} and with the diffusion mechanism transition identified by Rizk \textit{et al.}~\cite{Rizk2025}, reinforcing the connection between oxygen-induced surface stabilization and the onset of accelerated swelling.

% More broadly, $T_c$ in MX-type fuels decreases with increasing burnup: for carbides, Colin \textit{et al.}~\cite{Colin1983} showed that $T_c$ decreases from $\sim$1150~K at low burnup to $\sim$1000~K at 11~at.\%, reflecting the fact that a higher fission gas inventory allows the critical gas concentration for accelerated bubble growth to be reached at progressively lower temperatures. For nitrides, $T_c$ is 200--300~K higher than for carbides at comparable burnup~\cite{Ronchi1978}. Thus, the coincidence between our predicted $|\Delta \sigma|$ peak and the experimentally reported $T_c$ range is expected to persist across burnup.

The model provides predictive capabilities by mapping the dependence of surface energy reduction on key physical parameters, including temperature, porosity, void radius, and oxygen concentration. Parametric analysis reveals that surface energy reduction is strongly influenced by oxygen concentration and void radius, while the porosity dependence is moderate and remains much smaller than the naive geometric prefactor would suggest. In particular, small bulk voids and bubbles experience a significant stabilization effect from oxygen at intermediate temperatures, highlighting the role of oxygen in the early stages of void or fission gas bubble formation. These insights may provide a physically grounded explanation for the association between oxygen content and accelerated swelling observed in experimental studies of mixed uranium carbide/nitride fuels~\cite{Rogozkin2003}. Furthermore, the generality of the approach allows for its extension to other materials systems and integration into mesoscale models and fuel performance codes, offering a path toward multiscale modeling of chemically assisted swelling in nuclear materials.

Rizk \textit{et al.} adopted a surface energy of 1.11 J/m$^2$, computed using the empirical potential developed by Kocevski \textit{et al.} \cite{Kocevski2022II}. In contrast, our DFT calculations yield a higher value of 1.59 J/m$^2$ for a planar surface. This 0.48 J/m$^2$ discrepancy, neglecting curvature corrections, likely stems from the known tendency of the Kocevski potential to underestimate defect energetics in UN~\cite{AbdulHameed2024}, a limitation that appears to manifest here as well. Substituting $\sigma$ = 1.59 J/m$^2$ into the model of Rizk \textit{et al.} would suppress both fission gas swelling and release, necessitating recalibration of other model parameters to preserve agreement with experiment. This discrepancy highlights the need for a mechanism that lowers surface energy in irradiated UN. We propose that oxygen adsorption at bubble surfaces fulfills this role, enabling the levels of swelling and gas release observed experimentally. A reduction in surface energy thus appears essential to reconcile the mechanistic model with experimental data. Notably, the oxygen concentration reported by Ronchi \textit{et al.} ($w_\text{O}$ = 1800 ppm) corresponds to a predicted surface energy reduction of 0.30--0.37 J/m$^2$ for $R_v=25$~nm over $p=5$--20\%, of the same order as the 0.48~J/m$^2$ gap between our DFT value and that used by Rizk \textit{et al.}

To quantify the impact of oxygen-induced surface energy reduction on void nucleation, we invoke classical nucleation theory. The free energy barrier for homogeneous void nucleation is \cite{Fultz2020}:
\begin{equation}
\Delta G^* = \frac{16 \pi \sigma^3}{3 (\Delta g_v)^2},
\end{equation}
where $\Delta g_v$ is the volumetric free energy driving force for void formation. Since $\Delta G^* \sim \sigma^3$, the ratio of nucleation barriers with and without oxygen is:
\begin{equation}
\frac{\Delta G^*_\text{O}}{\Delta G^*_\text{clean}} = \left( \frac{\sigma_\text{clean} + \Delta \sigma}{\sigma_\text{clean}} \right)^3 = \left( 1 + \frac{\Delta \sigma}{\sigma_\text{clean}} \right)^3.
\end{equation}
For $\sigma_\text{clean} = 1.59$~J/m$^2$ and $\Delta \sigma \approx -0.44$~J/m$^2$ (intermediate stoichiometry, $w_\text{O} = 1500$~ppm, $R_v = 1$~nm), this gives $\sigma_\text{O} \approx 1.15$~J/m$^2$ and:
\begin{equation}
\frac{\Delta G^*_\text{O}}{\Delta G^*_\text{clean}} = \left( \frac{1.15}{1.59} \right)^3 \approx 0.38,
\end{equation}
corresponding to a $\sim$62\% reduction in the nucleation barrier. Propagating the nucleation-time uncertainty ($t = 10$--100~h) shifts this to 57--65\%, because at the peak $\alpha_1 = 1$ and $|\Delta\sigma|_\text{peak}$ varies by only $-11\%$ to $+7\%$. A full propagation of the correlated DFT-energy ($\pm0.1$~eV) and nucleation-time uncertainties, detailed in \ref{App:UQ} and shown as a band in \cref{Fig:App:UQ:band}, yields a 95\% interval of 45--78\% for the barrier reduction. This substantial lowering of $\Delta G^*$ translates into a dramatically increased nucleation rate, $J \sim \exp(-\Delta G^*/kT)$ \cite{Fultz2020}, consistent with the experimentally observed enhancement of void swelling in oxygen-contaminated UN.

% An elevated surface energy would have significant consequences for the thermodynamic and kinetic evolution of fission gas bubbles in the mechanistic model. The equilibrium internal pressure of a bubble, given by the modified Young–Laplace relation $P_{\text{eq}} = 2\sigma / R_v - s_h$, scales linearly with $\sigma$. Thus, increasing the surface energy raises the equilibrium pressure required for mechanical stability at a given radius, resulting in more severe over-pressurization for the same gas content. This, in turn, enhances the driving force for vacancy absorption while simultaneously increasing the energetic penalty associated with surface formation. The net effect is a suppression of bubble nucleation and growth, particularly for dislocation-mediated intragranular bubbles that dominate swelling near the breakaway temperature. Moreover, grain boundary bubbles, which are already over-pressurized in the model, would experience even greater resistance to expansion, thereby delaying interconnection and fission gas release. These effects would collectively shift the Vitanza-type release threshold to higher temperatures and burnups, leading to unrealistically low gas release predictions and artificial retention of gas within the fuel matrix. 

Despite its contributions, the present model includes several simplifications that merit further investigation. First, the surface adsorption energy of oxygen is assumed to be independent of surface coverage, which may not hold at high impurity concentrations where impurity-impurity interactions become significant. Additionally, configurational and vibrational entropy contributions are neglected, which may affect the temperature dependence of defect formation and migration energies, particularly at elevated temperatures. Separately, because the main dataset uses PBE, we verified with a static DFT+$U$ test ($U_\text{eff}=1$~eV) that the bulk oxygen-defect ordering underlying the dominant $\text{O}_\text{N}^{(\text{s})}$ channel is preserved (\ref{App:UQ:U}).

Further, the kinetic parameter used to model oxygen segregation is introduced in an \textit{ad hoc} manner, limiting the model's capability to capture time-dependent or rate-limited processes. To assess sensitivity to this parameter, we note that $\alpha_i \sim \sqrt{t}$ (\cref{Eq:alpha}) on the kinetically limited low-temperature flank of $|\Delta \sigma|(T)$. The peak, however, occurs at the kinetic saturation temperature $T_\text{peak}$, where $\alpha_1 = 1$, so $|\Delta \sigma|_\text{peak}$ is only weakly (logarithmically) sensitive to $t$. Varying $t$ from 10 to 100~hours changes $|\Delta \sigma|_\text{peak}$ by only $-11$\% ($t = 10$~h) to $+7$\% ($t = 100$~h) relative to the baseline $t = 42$~hours (rising to $+28$\% at $t = 1000$~h), while $T_\text{peak}$ shifts downward by approximately 88~K per decade (through \cref{Eq:TpeakMain}) for the reference case of $R_v$ = 1~nm, $w_\text{O}$ = 1500 ppm, and $p$ = 0.05. The model conclusions are therefore robust to an order-of-magnitude uncertainty in $t$.

Also, the influence of sinks such as grain boundaries and second-phase precipitates, which may trap oxygen or alter defect transport, is also not incorporated.

Another limitation of the present model is that all surface properties are derived from planar (001) slab calculations, whereas real voids may expose higher-index facets such as $\{110\}$ and $\{111\}$, and exhibit curvature effects at small radii. Regarding curvature, if the total energy of a spherical void is written as \cite{Tolman1949,Chhapadia2011,Wang2021}:
\begin{equation}
E_v = 4 \pi R_v^2 \, \sigma \, \left(1 + \frac{\delta}{R_v} \right)    
\end{equation}
where $\delta$ is a Tolman-like length on the order of $\sim 1$~nm, the correction to the absolute surface energy (equal to $\delta/R_v$) is substantial for the smallest voids considered here: 100\% at $R_v$ = 1~nm, 50\% at $R_v$ = 2~nm, diminishing to 10\% at $R_v$ = 10~nm and 4\% at $R_v = 25$~nm. However, the central quantity predicted by our model is not $\sigma$ itself but the change in surface energy, $\Delta\sigma$, due to oxygen adsorption. As demonstrated in the derivation of \cref{Eq:DeltaSigma1}, the pristine surface energy cancels identically: $\Delta\sigma$ depends only on adsorption energies ($E_\mathrm{ad}$), defect formation energies ($E_f$), defect concentrations, and kinetic correction factors, none of which are functions of $\sigma$. Curvature corrections to the absolute surface energy therefore do not propagate into the predicted $\Delta\sigma$, and the conclusions regarding oxygen-induced surface energy reduction remain quantitatively unchanged irrespective of the curvature model adopted.

Regarding faceting, the (001) surface is the most stable low-index surface of UN \cite{Zhukovskii2009SS} and is expected to dominate the equilibrium shape of voids in rocksalt-structured compounds. While $\{110\}$ and $\{111\}$ facets may be present, particularly in non-equilibrium or irradiation-produced voids, their contribution to the total void surface area is expected to be secondary. In principle, the oxygen adsorption energies on these higher-index surfaces may differ from those on the (001) surface, which would introduce facet-dependent corrections to $\Delta\sigma$. To quantify this sensitivity, we repeated the surface calculations for the $\{110\}$ facet, as detailed in \ref{App:110}. Although oxygen binds somewhat less strongly on (110) while the surface nitrogen vacancy forms far more readily, the net effect is that the $\text{O}_\text{N}^{(\text{s})}$ segregation channel lowers the void surface energy at least as effectively as on (001). This confirms that the dominant contribution to $\Delta\sigma$, arising from the segregation of oxygen into surface nitrogen vacancies (O$_\mathrm{N}^{(\mathrm{s})}$), is governed primarily by the local coordination environment of the nitrogen sublattice rather than the global surface orientation, so the qualitative trends reported here are robust across facets. A systematic mapping of all low-index facets and their relative areas remains a natural extension of this work.

A further limitation is that the current model isolates oxygen as the sole impurity species. In reality, fission products and other impurities, such as carbon, krypton, and xenon, may co-segregate or interact with oxygen, altering the energetics and morphology of vacancy clusters. Notably, our calculations indicate that substitutional oxygen on nitrogen sites binds to uranium-site vacancies and noble gas atoms (i.e., Kr$_\text{U}$, Xe$_\text{U}$) with energies around $-0.4$~eV. Whether this binding promotes the stabilization of voids or instead alters gas mobility and clustering behavior is not yet fully understood. Future studies should assess how such interactions affect oxygen-assisted swelling in fission environments.

Finally, while the role of carbon in increasing oxygen solubility and suppressing UO$_2$ precipitation in UN is qualitatively known, the impact of carbon content on void stabilization was not explicitly addressed in this model. Given that carbon can modulate the chemical potential of oxygen and influence its incorporation and redistribution, a natural extension of this work is to explore the combined effects of carbon and oxygen on defect energetics and swelling behavior in UN. The framework developed here is directly applicable to this scenario.

\section{Conclusions}

This work develops a first-principles-based thermodynamic framework that quantifies how oxygen impurities interact with void surfaces in UN, leading to a reduction in the surface energy, $|\Delta \sigma|$. The model incorporates DFT-derived energetics, oxygen diffusivity, defect chemistry, and microstructural parameters to assess the extent of surface stabilization.

Key findings can be summarized as follows. Substitutional oxygen on nitrogen surface sites ($\text{O}_\text{N}^\text{(s)}$) is the dominant contributor to surface energy reduction. Oxygen in surface hollow sites ($\text{O}_i^\text{(s)}$) has a minor, and occasionally adverse, effect on $|\Delta \sigma|$. The dependence of $|\Delta \sigma|$ on key parameters is summarized below:
\begin{itemize}
    \item \textit{Oxygen concentration ($w_\text{O}$):} The surface energy reduction scales approximately linearly with oxygen content.
    \item \textit{Void radius ($R_v$):} $|\Delta \sigma|$ is most pronounced for small voids. Larger voids require higher temperatures to reach meaningful oxygen coverage; the peak magnitude falls by $\approx$44\% from $R_v$ = 1 to 25~nm, and at a fixed operating temperature the reduction falls even more steeply.
    \item \textit{Porosity ($p$):} The explicit geometric porosity factor is partly canceled by the surface-site fraction. The remaining porosity dependence is moderate because the capture length decreases with increasing $p$, but the peak magnitude changes far less than the isolated $(1-p)/p$ prefactor would suggest.
    \end{itemize}

\vspace{0.5cm}

For small voids and bubbles ($R_v$ = 1--10 nm), $|\Delta \sigma|$ exhibits three regimes in its dependence on temperature:
    \begin{itemize}
    \item \textit{Low temperatures:} $|\Delta \sigma| \approx 0$ due to limited oxygen diffusivity.
    \item \textit{Intermediate temperatures:} $|\Delta \sigma|$ reaches its maximum as the competing effects of oxygen solubility and diffusion kinetics are optimally balanced. This regime coincides with the onset of breakaway swelling in UN.
    \item \textit{High temperatures:} $|\Delta \sigma|$ decreases due to enhanced bulk solubility, which disfavors surface segregation.
    \end{itemize}

The reduction in surface energy can be approximately expressed as $|\Delta \sigma| \approx \eta \, n_\text{O} \, \alpha_1 \, |E_\text{seg}^\text{eff}|$ (\cref{Eq:DeltaSigmaSimple}), where $\eta$ is a geometric factor that depends on $p$ and $R_v$, $n_\text{O}$ is the oxygen number density, $\alpha_1$ is a kinetic correction factor, and $E_\text{seg}^\text{eff}$ is the effective segregation energy. The peak temperature is given by \cref{Eq:TpeakMain}. The reduction in $\sigma$ lowers the thermodynamic barrier for void and bubble formation. For the reference oxygen content $w_\text{O}$ = 1800 ppm in UN \cite{Ronchi1975}, our model predicts a surface energy reduction of about 0.30--0.37 J/m$^2$ for 25~nm dislocation and grain-boundary bubbles over $p=5$--20\%. This is of the same order as the discrepancy between our DFT-calculated surface energy (1.59 J/m$^2$) and the value adopted by Rizk \textit{et al.} \cite{Rizk2025} in their swelling model (1.11 J/m$^2$). The oxygen-induced reduction in $\sigma$ thus can offer a mechanistic explanation for the experimentally observed levels of swelling and fission gas release in UN. % Weak binding between $\text{O}_\text{N}^\text{(b)}$ and uranium vacancies ($v_\text{U}$) or noble gas atoms ($\text{Kr}_\text{U}$ and $\text{Xe}_\text{U}$) may contribute to the stabilization of void embryos.

Overall, this work provides a physically rigorous model for incorporating oxygen effects into mechanistic swelling models of UN. The framework is extendable to other chemical impurities such as carbon and can be integrated into mesoscale and fuel performance models to predict chemically-assisted bubble swelling in nuclear fuels.

\section{Acknowledgments}

This work was supported by the DOE-NE and Westinghouse Electric Company under contract DE-NE0008824. This research used resources provided by the Los Alamos National Laboratory Institutional Computing Program, which is supported by the U.S. Department of Energy National Nuclear Security Administration under Contract No. 89233218CNA000001. Los Alamos National Laboratory, an affirmative action/equal opportunity employer, is operated by Triad National Security, LLC, for the National Nuclear Security Administration of the U.S. Department of Energy under Contract No. 89233218CNA000001. This research also made use of the resources of the High-Performance Computing Center at Idaho National Laboratory, which is supported by the Office of Nuclear Energy of the U.S. Department of Energy and the Nuclear Science User Facilities under Contract No. DE-AC07-05ID14517. Mohamed AbdulHameed thanks William Neilson, Conor Galvin, and Kai Duemmler for the great help provided while conducting this research.

\FloatBarrier
\newpage

\appendix

\section{Derivation of the formulas for $[ \text{N}_\text{N}^\text{(s)} ]$ and $[ v_i^\text{(s)} ]$}
\label{App1}

In the following, $a$ denotes the UN lattice constant, $p$ the porosity (void volume fraction), $R_v$ the average void radius, and $V_0$ the total system volume. $[\text{N}_\text{N}^{(\text{s})}]$ is the site fraction of nitrogen atoms on void surfaces, and $[v_i^{(\text{s})}]$ is the site fraction of vacant interstitial sites on void surfaces, both expressed per formula unit.

The area associated with an N atom on the surface is $a^2/2$, where $a$ is the lattice constant. The total surface area of the internal voids is $N_v \times 4 \pi R_v^2$, where $N_v$ is the total number of voids, calculated as:
\begin{equation}
N_v = \frac{ p V_0 }{ \frac{4}{3} \pi R_v^3 },
\end{equation}
where $V_0$ is the total system volume. Then, the number of $\text{N}_\text{N}^\text{(s)}$ sites is:
\begin{equation}
\text{Number of } \text{N}_\text{N}^\text{(s)} = \frac{ N_v \times 4 \pi R_v^2 } { a^2 / 2 }. 
\label{Eq:1}
\end{equation}
To get the site fraction of $\text{N}_\text{N}^\text{(s)}$, we need to divide its total number by the number of formula units. The number of formula units is:
\begin{equation}
\text{Number of formula units} = \frac{V_0 (1 - p)}{ a^3 / 4 }, 
\label{Eq:2}
\end{equation}
where $V_0 (1 - p)$ is the occupied volume, and $a^3 / 4$ is the volume of a single formula unit. Dividing \cref{Eq:1} by \cref{Eq:2}, we arrive at:
\begin{equation}
[ \text{N}_\text{N}^\text{(s)} ] = \frac{p V_0}{ \frac{4}{3} \pi R_v^3 } \frac{4 \pi R_v^2}{a^2/2}  \frac{ a^3/4 }{ V_0 (1 - p) } = \frac{3}{2} \frac{p}{1-p} \frac{a}{R_v}.
\label{Eq:NNs}
\end{equation}

The same steps can be followed to get an expression for $[v_i^\text{(s)}]$, provided that we pay attention to the following differences: First, the area occupied by a vacant site on the surface is $a^2/4$. Then, the number of $v_i^\text{(s)}$ sites is:
\begin{equation}
\text{Number of } v_i^\text{(s)} = \frac{ N_v \times 4 \pi R_v^2 } { a^2 / 4 }.
\end{equation}
To get the site fraction of $v_i^\text{(s)}$, we need to divide its total number by the number of formula units in the occupied volume (i.e., \cref{Eq:2}), and then divide by 2 because there are 2 tetrahedral interstitial sites per formula unit, either in the bulk material or on the surface. In the case of $[ \text{N}_\text{N}^\text{(s)} ]$, there was one N atom (or site) per formula unit. Then, the site fraction of $v_i^\text{(s)}$ is: 
\begin{equation}
[ v_i^\text{(s)} ] = \frac{1}{2} \frac{p V_0}{ \frac{4}{3} \pi R_v^3 } \frac{4 \pi R_v^2}{ a^2/4}  \frac{ a^3/4 }{ V_0 (1 - p) } = \frac{3}{2} \frac{p}{1-p} \frac{a}{R_v},
\label{Eq:Vis}
\end{equation}
which is the same expression for $[ \text{N}_\text{N}^\text{(s)} ]$. Note that both \cref{Eq:NNs,Eq:Vis} are independent of $V_0$. As outlined in \cref{Sec:DefE}, the adsorption energy of O atoms in above-U sites is the same as that of O in a hollow site (\cref{Tab:Ead}). To account for this in our model, the available adsorption sites are doubled by multiplying $[ v_i^\text{(s)} ]$ by 2. In this way, the analysis is simplified by only considering $\text{O}_i^{\text{(s)}}$. The final expression reads:
\begin{equation}
[v_i^{\text{(s)}}] = \frac{3p}{1-p} \frac{a}{R_v}.
\end{equation}

\section{Derivation of the formula for surface energy change as a function of adsorbate atom concentration}
\label{App2}

Here, $\sigma_0$ is the pristine surface energy, $E_\text{ad}$ the adsorption energy, $E_f$ the defect formation energy, $N_\text{ad}$ the number of adsorbed atoms, $A$ the total surface area, and $\mu_\text{ad}$ the chemical potential of the adsorbate. $[\text{O}_\text{N}^{(\text{s})}]$ and $[\text{O}_i^{(\text{s})}]$ denote the site fractions of substitutional and interstitial oxygen on void surfaces, respectively, while $[\text{O}_\text{N}^{(\text{b})}]$ is the site fraction of substitutional oxygen in the bulk.

Suppose we have an impurity atom that got adsorbed on a pristine surface. The surface energy after absorption is: 
\begin{equation}
\sigma = \sigma_0 + \Delta \sigma,
\end{equation}
where $\sigma_0$ is the energy of the pristine surface before adsorption, and $\Delta \sigma$ is the surface energy change due to impurity adsorption. The change in the surface energy can be expressed as:
\begin{equation}
\Delta \sigma = \sigma_\text{surf+ad} - \sigma_\text{surf},
\end{equation}
where:
\begin{equation}
\sigma_\text{surf} = \frac{1}{A} \left[ E_\text{surf} - N_\text{U} \mu_\text{U} - N_\text{N} \mu_\text{N} \right],
\end{equation}
and:
\begin{equation}
\sigma_\text{surf+ad} = \frac{1}{A} \left[ E_\text{surf+ad} - N_\text{U} \mu_\text{U} - N_\text{N} \mu_\text{N} - N_\text{ad} \mu_\text{ad} \right].
\end{equation}
Then,
\begin{equation}
\Delta \sigma = \frac{1}{A} \left[ E_\text{surf+ad} - E_\text{surf} - N_\text{ad} \mu_\text{ad} \right].
\end{equation}

From the definition of the adsorption energy, i.e., \cref{Eq:Ead}, the change in the surface energy is:
\begin{equation}
\Delta \sigma = \frac{1}{A} \left[ N_\text{ad} ( E_\text{ad} + \mu_\text{ad} ) - N_\text{ad} \mu_\text{ad} \right] = \frac{1}{A} N_\text{ad} E_\text{ad}. 
\end{equation}

If we have a specific concentration of an impurity that can get adsorbed on different sites on the surface, with each site type having its own adsorption energy, then:
\begin{equation}
\Delta \sigma = \frac{1}{A} \left[ \sum_i N_\text{ad}^i E_\text{ad}^i \right].
\end{equation}

The same line of thought can be used to derive an expression for the surface energy change upon introducing vacancies to a previously pristine surface. For the general case where we introduce both adsorbate atoms and vacancies to a surface, the total surface energy change is:
\begin{equation}
\Delta \sigma = \frac{1}{A} \left[ \sum_i N_\text{ad}^i E_\text{ad}^i + \sum_j N_d^j E_f^j \right].
\label{Eq:DeltaSigma}
\end{equation}

This can be related to the relative concentrations calculated earlier by noticing that:
\begin{equation}
N_\text{ad}^i = N_\text{O} \frac{N_\text{ad}^i}{N_\text{O}} \approx N_\text{O} \frac { [ \text{O}_\text{X}^{\text{(s)}} ] } { [ \text{O}_\text{N}^{\text{(b)}} ] },   
\label{Eq:Nad}
\end{equation}
where $N_\text{O}$ is the total number of oxygen atoms in the sample and X $\in$ \{N, $i$\}. Finally, we can relate $N_d$, i.e., the total number of surface defects, to the concentrations calculated earlier by a similar argument. Assuming the surface defect is $v_\text{N}^{\text{(s)}}$, we have:
\begin{equation}
N_d = [ v_\text{N}^{\text{(s)}} ] \times \text{Number of N$_\text{N}^\text{(s)}$} = [ v_\text{N}^{\text{(s)}} ] \frac{A_v}{a^2 / 2}, 
\label{Eq:Nd}
\end{equation}
where we have made use of \cref{Eq:1} with $A_v$ being the total area of the void surface.

The total number of oxygen atoms in the system is not a useful parameter. Instead, we need the formulas to be expressed in terms of some measure of oxygen concentration. The number of voids is expressed in terms of both the void surface area and the void volume as:
\begin{equation}
N_v = \frac{A_v}{4 \pi R_v^2} = \frac{ p V_0 }{ \frac{4}{3} \pi R_v^3 }.
\label{Eq:Nv}
\end{equation}
From \cref{Eq:Nv}, the total void surface area is:
\begin{equation}
A_v = \frac{3 p V_0}{R_v} = \frac{3}{R_v} \frac{p}{1-p} V.
\label{Eq:Av}
\end{equation}
Then,
\begin{equation}
\frac {N_\text{O}} {A_v} = \frac {R_v} {3} \frac{1-p} {p} \frac {N_\text{O}} {V} = \frac {R_v} {3} \frac{1-p} {p} \frac {N_\text{O}} { 2 N_\text{UN} \cdot a^3/8 } \approx \frac{8}{3} \frac{1-p}{p} \frac{R_v}{a^3} c_\text{O}, 
\label{Eq:NOAv}
\end{equation}
where $N_\text{UN}$ is the number of formula units in the occupied volume, $a^3/8$ is the average volume per atom of UN, and $c_\text{O}$ is the atomic fraction of O atoms which is approximately $c_\text{O} \approx N_\text{O}/( 2N_\text{UN})$. The oxygen impurity concentration in UN is usually expressed as parts per million (ppm) or weight fractions. The relation between atomic fraction, $c_\text{O}$, and weight fraction, $w_\text{O}$, is:
\begin{equation}
c_\text{O} = \frac { M_\text{UN} w_\text{O} } { M_\text{UN} w_\text{O} + 2 M_\text{O} ( 1 - w_\text{O} ) },
\label{Eq:cO}
\end{equation}
where $M_\text{UN}$ and $M_\text{O}$ are the molar weights of UN and O, respectively. From \cref{Eq:cO}, $w_\text{O}$ can be expressed in terms of $c_\text{O}$ as:
\begin{equation}
w_\text{O} = \frac{ 2 M_\text{O} c_\text{O} }{ 2 M_\text{O} c_\text{O} + M_\text{UN} ( 1 - c_\text{O} ) }.
\end{equation}
Substituting \cref{Eq:Nad,Eq:NOAv,Eq:Nd} into \cref{Eq:DeltaSigma}, the change in surface energy becomes:
\begin{equation}
\Delta \sigma = \frac{8}{3} \frac{1-p}{p} \frac{R_v}{a^3} c_\text{O} \left[ \frac{ [ \text{O}_\text{N}^{\text{(s)}} ] }{ [ \text{O}_\text{N}^{\text{(b)}} ] }  E_\text{ad}( \text{O}_\text{N}^{\text{(s)}} ) + \frac{ [ \text{O}_i^{\text{(s)}} ] }{ [ \text{O}_\text{N}^{\text{(b)}} ] }  E_\text{ad}( \text{O}_i^{\text{(s)}} ) \right] + \frac{2}{a^2} [ v_\text{N}^{\text{(s)}} ] E_f ( v_\text{N}^{\text{(s)}} ).
\end{equation}

To take kinetics into account, the concentrations of the different oxygen surface defects must be corrected for the low oxygen diffusivity at low temperatures. Note that the increased oxygen solubility is inherently incorporated in the thermodynamic model. According to Zinkle and Lee \cite{Zinkle1990}, as a first approximation, the concentrations of oxygen surface defects can be multiplied by a parameter $\alpha$ that we term the kinetic correction, and that depends on the diffusivity of different oxygen defect types. Thus, the final expression for the change in surface energy becomes:
\begin{equation}
\Delta \sigma = \frac{8}{3} \frac{1-p}{p} \frac{R_v}{a^3} c_\text{O} \left[ \alpha_1 \frac{ [ \text{O}_\text{N}^{\text{(s)}} ] }{ [ \text{O}_\text{N}^{\text{(b)}} ] }  E_\text{ad}( \text{O}_\text{N}^{\text{(s)}} ) + \alpha_2 \frac{ [ \text{O}_i^{\text{(s)}} ] }{ [ \text{O}_\text{N}^{\text{(b)}} ] }  E_\text{ad}( \text{O}_i^{\text{(s)}} ) \right] + \frac{2}{a^2} [ v_\text{N}^{\text{(s)}} ] E_f ( v_\text{N}^{\text{(s)}} ),
\end{equation}
where $\alpha_1$ and $\alpha_2$ are the kinetic corrections for the defects $\text{O}_\text{N}^\text{(s)}$ and $\text{O}_i^{\text{(s)}}$, respectively. Conceptually, $\alpha_i$ stems from the fraction of diffusing oxygen atoms that encounter a void embryo during the void nucleation time, and, thus, can be defined as \cite{Zinkle1990}:
\begin{equation}
\alpha_i =
\begin{cases}
    \sqrt{D_i t}/\lambda_v & \text{if } \sqrt{D_i t}/\lambda_v < 1, \\
    1 & \text{if } \sqrt{D_i t}/\lambda_v \geq 1,
\end{cases}    
\end{equation}
where $D_i$ is the diffusivity of the relevant defect, $t$ is the void nucleation time, and $\lambda_v$ is the effective capture length, $\lambda_v=0.5\,n_v^{-1/3}-R_v$, with $n_v = p / ( \frac{4}{3} \pi R_v^3 ) $.

\FloatBarrier

\section{Closed-form expression for the temperature of maximum $|\Delta \sigma|$}
\label{App:Tpeak}
 
This appendix derives a closed-form expression for the temperature at which $|\Delta \sigma|$ is maximized. The derivation proceeds in four steps: we first state the assumptions, then factor the reduced $|\Delta \sigma|(T)$ into terms with known monotonicity, establish piecewise monotonicity on either side of a kinetic saturation temperature~$T^*$, and conclude that $T_\text{peak} = T^*$. The closed-form result for $T^*$ is given in \cref{Eq:App:TstarCompact}, and its implications for the porosity dependence of $|\Delta \sigma|$ are discussed in \ref{App:Tpeak:Porosity}.
 
\subsection{Assumptions}
\label{App:Tpeak:Assumptions}

The derivation rests on the following assumptions, each consistent with the modeling choices in \cref{Sec:Oxygen}:
\begin{enumerate}[label=(\roman*)]
    \item \textit{Dominant channel.} Only the $\text{O}_\text{N}^{(\text{s})}$ term in \cref{Eq:DeltaSigma1} is retained; the $\text{O}_i^{(\text{s})}$ and $v_\text{N}^{(\text{s})}$ contributions are neglected. This is justified by \cref{Fig:D}, where $\Delta \sigma(\text{O}_\text{N}^{(\text{s})})$ dominates by at least an order of magnitude.
    \item \textit{Fixed stoichiometry.} The vacancy formation energies $E_f(v_\text{N}^{(\text{b})})$ and $E_f(v_\text{N}^{(\text{s})})$ are held at their values for a given stoichiometric condition.
    \item \textit{Dilute defects.} Concentrations follow Arrhenius statistics with unit prefactor, e.g., $[v_\text{N}^{(\text{b})}] = \exp[-E_f(v_\text{N}^{(\text{b})})/(kT)]$ and $[\text{O}_\text{N}^{(\text{b})}] \approx c_\text{O}$.
    \item \textit{Single transport channel.} The kinetic correction $\alpha_1$ is computed from the $\{\text{O}_\text{N} : v_\text{N}\}$ diffusivity (\cref{Sec:Oxygen}).
    \item \textit{Existence of $T^*$.} The argument of the logarithm in \cref{Eq:App:TstarCompact} must exceed unity, which is verified for all parameter combinations considered in this work.
\end{enumerate}

\subsection{Factorized form}
\label{App:Tpeak:Factorization}

Retaining only the dominant $\text{O}_\text{N}^{(\text{s})}$ term, \cref{Eq:DeltaSigma1} takes the form:
\begin{equation}
|\Delta \sigma|(T) = \underbrace{\frac{8}{3}\frac{1-p}{p}\frac{R_v}{a^3}\,c_\text{O}}_{\displaystyle\mathcal{P}}\;\alpha_1(T)\;g(T)\;|E_\text{ad}(T)|,
\label{Eq:App:Factorization}
\end{equation}
where $\mathcal{P}$ is a $T$-independent prefactor. Three factors carry temperature dependence: the kinetic correction $\alpha_1$, the segregation ratio $g = [\text{O}_\text{N}^{(\text{s})}]/[\text{O}_\text{N}^{(\text{b})}]$, and the adsorption energy $E_\text{ad}$ evaluated with respect to the impurity chemical potential (\cref{Sec:uO}). We now derive the explicit $T$-dependence of each.

\paragraph{Kinetic factor} The kinetic correction $\alpha_1(T) = \min[\sqrt{D_1(T)\,t}/\lambda_v,\; 1]$ (\cref{Eq:alpha}) is continuous but non-differentiable at the \emph{saturation temperature} $T^*$ defined by $\sqrt{D_1(T^*)\,t} = \lambda_v$. The effective oxygen diffusivity via the $\{\text{O}_\text{N} : v_\text{N}\}$ mechanism is:
\begin{align}
D_1(T) & = f\,\frac{[\{\text{O}_\text{N}:v_\text{N}\}^{(\text{b})}]}{[\text{O}_\text{N}^{(\text{b})}]}\,D_{\{\text{O}_\text{N}:v_\text{N}\}} \\
& = \frac{f\,z\,\lambda^2\,\nu}{6}\,\exp\!\left[-\frac{E_m + E_f(v_\text{N}^{(\text{b})}) + E_b(\{\text{O}_\text{N}:v_\text{N}\}^{(\text{b})})}{kT}\right],
\label{Eq:App:D1}
\end{align}
where the second equality uses:
\begin{equation}
D_d = \frac{1}{6}z\lambda^2\nu\exp\left( - \frac{E_m}{kT} \right), 
\end{equation}
and
\begin{equation}
\frac {[\{\text{O}_\text{N}:v_\text{N}\}^{(\text{b})}]}  {[\text{O}_\text{N}^{(\text{b})}]} = [v_\text{N}^{(\text{b})}] \exp\left( - \frac{E_b(\{\text{O}_\text{N}:v_\text{N}\}^{(\text{b})})} {kT} \right).
\end{equation}
In the unsaturated regime ($T < T^*$):
\begin{align}
\alpha_1(T)
&=
C_\alpha
\exp\!\left[
-\frac{E_m^\mathrm{eff}}{2kT}
\right],
\label{Eq:App:alpha1Unsat}
\\
C_\alpha
&=
\frac{1}{\lambda_v}
\sqrt{
\frac{
f z \lambda^2 \nu t
}{6}
},
\label{Eq:App:CAlpha}
\\
E_m^\mathrm{eff}
&=
E_m
+
E_f(v_\text{N}^{(\text{b})})
+
E_b(\{\text{O}_\text{N}:v_\text{N}\}^{(\text{b})}).
\label{Eq:App:EmEff}
\end{align}

\paragraph{Segregation factor} Combining \cref{Eq:ONs_ONb} with the $K_5$ equilibrium (\cref{Eq:VN-VNs}):

\begin{align}
g(T) = & \frac{[\text{N}_\text{N}^{(\text{s})}]}{[\text{N}_\text{N}^{(\text{b})}]}\,\exp\!\left[+\frac{E_\text{seg}}{kT}\right], \\
E_\text{seg} = & -\bigl[E_f(v_\text{N}^{(\text{s})}) - E_f(v_\text{N}^{(\text{b})}) + E_\text{ad}(\text{O}_\text{N}^{(\text{s})}) - E_\text{inc}(\text{O}_\text{N}^{(\text{b})})\bigr],
\label{Eq:App:gT}
\end{align}

where both $E_\text{ad}$ and $E_\text{inc}$ in the definition of $E_\text{seg}$ contain $\mu_\text{O}$, which cancels in their difference. That is, $E_\text{seg}$ is independent of the choice of oxygen reference state and is constructed entirely from raw DFT total-energy differences. Note that the prefactor $[\text{N}_\text{N}^{(\text{s})}]/[\text{N}_\text{N}^{(\text{b})}]$ is $T$-independent.

% The segregation factor $g(T)$ therefore carries no chemical-potential dependence, and the explicit $T$-dependence associated with the impurity reference enters solely through $E_\text{ad}(T)$, as derived below. This separation avoids double-counting of chemical-potential contributions.

Substituting the DFT values from \cref{Tab:EfVN,Tab:Einc,Tab:Ead}:
\begin{equation}
E_\text{seg} = -\bigl[(-0.76) + (0.43)\bigr] = 0.33~\text{eV},
\label{Eq:App:EsegValue}
\end{equation}
where the segregation energy of $v_\text{N}$ is $E_f(v_\text{N}^{(\text{s})}) - E_f(v_\text{N}^{(\text{b})}) = -0.76$~eV (\cref{Tab:EfVN}) and $E_\text{ad}(\text{O}_\text{N}^{(\text{s})}) - E_\text{inc}(\text{O}_\text{N}^{(\text{b})}) = -6.80 - (-7.23) = 0.43$~eV (\cref{Tab:Ead,Tab:Einc}). Both differences are stoichiometry-independent, so $E_\text{seg} = 0.33$~eV for all conditions.

\paragraph{Adsorption-energy factor} The adsorption energy entering \cref{Eq:DeltaSigmaSimple} is referenced to the impurity oxygen chemical potential rather than to an external oxide or gas reservoir. For a fixed adsorption configuration,
\begin{equation}
E_\text{ad}(T) = E_\text{raw} - \mu_\text{O}(T),
\end{equation}
where:
\begin{equation}
E_\text{raw} = \frac{1}{N_\text{ad}} \left( E_\text{surf+ad} - E_\text{surf} \right)
\end{equation}
is independent of temperature. From \cref{Eq:OxygenTruePot},
\begin{equation}
\mu_\text{O}(T) = E(\text{O}_\text{N}) - E(v_\text{N}) + kT \ln\left( \frac{ [\text{O}_\text{N}^{(\text{b})}] } { [v_\text{N}^{(\text{b})}] } \right).
\end{equation}
Using $[\text{O}_\text{N}^{(\text{b})}]\approx c_\text{O}$ and $[v_\text{N}^{(\text{b})}] = \exp[-E_f(v_\text{N}^{(\text{b})})/(kT)]$ gives:
\begin{equation}
\mu_\text{O}(T) = E(\text{O}_\text{N}) - E(v_\text{N}) + E_f(v_\text{N}^{(\text{b})}) + kT \ln c_\text{O}.
\label{Eq:App:muOimp}
\end{equation}
Therefore,
\begin{equation}
E_\text{ad}(T) = E_\text{ad}^{(0)} - kT \ln c_\text{O},
\label{Eq:App:EadImpExplicit}
\end{equation}
with:
\begin{equation}
E_\text{ad}^{(0)} = E_\text{raw} - E(\text{O}_\text{N}) + E(v_\text{N}) - E_f(v_\text{N}^{(\text{b})}).
\end{equation}
Since $c_\text{O}\ll1$, $\ln c_\text{O}<0$, so increasing temperature makes $E_\text{ad}(T)$ less negative. In the temperature range where adsorption remains favorable, $E_\text{ad}(T)<0$, its magnitude is:
\begin{equation}
|E_\text{ad}(T)| = - E_\text{ad}^{(0)} + kT \ln c_\text{O},
\end{equation}
and therefore:
\begin{equation}
\frac{d|E_\text{ad}|}{dT} = k\ln c_\text{O} < 0.
\end{equation}
Thus the thermodynamic adsorption drive decreases with increasing temperature.

\subsection{Piecewise monotonicity and peak location}
\label{App:Tpeak:Monotonicity}

The function $|\Delta \sigma|(T)$ is continuous but not differentiable at $T = T^*$. We establish its monotonicity separately on each side.

\subsubsection{Unsaturated regime (\texorpdfstring{$T < T^*$}{T < T*})}

Substituting \cref{Eq:App:alpha1Unsat,Eq:App:gT,Eq:App:EadImpExplicit} into \cref{Eq:App:Factorization} and taking the logarithmic derivative gives:
\begin{equation}
\frac{d}{dT}\ln|\Delta \sigma| =
\underbrace{ \frac{1}{kT^2} \left[ \frac{E_m^\mathrm{eff}}{2} - E_\text{seg} \right] }_{\displaystyle > 0 \text{ if \cref{Eq:App:Inequality} holds}}
+ \underbrace{ \frac{k\ln c_\text{O}}{|E_\text{ad}(T)|} }_{\displaystyle < 0}.
\label{Eq:App:LogDerivUnsat}
\end{equation}
The first term is positive provided:
\begin{equation}
E_m^\mathrm{eff} = E_m + E_f(v_\text{N}^{(\text{b})}) + E_b(\{\text{O}_\text{N}:v_\text{N}\}^{(\text{b})}) > 2 E_\text{seg}.
\label{Eq:App:Inequality}
\end{equation}
The second term is negative because $\ln c_\text{O}<0$. Therefore, the sign of \cref{Eq:App:LogDerivUnsat} is determined by the competition between the positive kinetic term and the negative adsorption-energy term.

\begin{table}[h!]
\footnotesize
\centering
\caption{Verification of the monotonicity condition for the unsaturated $\mathrm{O_N}^{(\mathrm{s})}$ contribution to $|\Delta\sigma|$. The effective transport activation energy, $E_m+E_f(v_\text{N}^{(\text{b})})+E_b$, is much larger than $2E_\text{seg}$ for all stoichiometric limits, so the kinetic factor increases with temperature faster than the segregation factor decreases below the saturation temperature. The positive values in the last row show that the condition $\tfrac{1}{2}(E_m+E_f+E_b)>E_\text{seg}$ is comfortably satisfied.}
\begin{tabular}{lccc}
\hline
Energy [eV] & U-rich & Intermediate & N-rich \\
\hline
$E_m + E_f(v_\text{N}^{(\text{b})}) + E_b$ & 4.73 & 5.22 & 5.71 \\
$2E_\text{seg}$ & 0.66 & 0.66 & 0.66 \\
$\tfrac{1}{2}[E_m + E_f(v_\text{N}^{(\text{b})})+E_b] - E_\text{seg}$ & 2.04 & 2.28 & 2.53 \\
\hline
\end{tabular}
\label{Tab:App:monotonicity}
\end{table}

In \cref{Tab:App:monotonicity}, the inequality in \cref{Eq:App:Inequality} is evaluated for each stoichiometric condition using the DFT values in \cref{Tab:EfVN}, the migration energy $E_m$ = 3.06~eV, and $E_b(\{\text{O}_\text{N}:v_\text{N}\}^{(\text{b})})$ = 0.03~eV. The margin in the first term of \cref{Eq:App:LogDerivUnsat} is 2.04--2.52~eV, depending on stoichiometry. The negative contribution is bounded by $k|\ln c_\text{O}|/|E_\text{ad}|$. For $c_\text{O}\sim10^{-3}$--$10^{-2}$ and $|E_\text{ad}|\gtrsim1$~eV, this term is $\lesssim 6\times10^{-4}$~K$^{-1}$. The positive term remains larger over $T$ = 800--2500~K for the physically relevant range considered here. Therefore, \cref{Eq:App:LogDerivUnsat} remains positive throughout the unsaturated regime. Thus, within the reduced dominant-channel model, $|\Delta\sigma|$ increases monotonically with temperature for $T<T^*$.

\subsubsection{Saturated regime (\texorpdfstring{$T > T^*$}{T > T*})}

In this regime $\alpha_1 = 1$, so:
\begin{equation}
\frac{d}{dT}\ln|\Delta \sigma| = -\frac{E_\text{seg}}{kT^2} + \frac{k\ln c_\text{O}}{|E_\text{ad}(T)|}.
\label{Eq:App:LogDerivSat}
\end{equation}
Both terms are negative ($E_\text{seg} > 0$ and $\ln c_\text{O} < 0$), so $|\Delta \sigma|$ is \emph{strictly decreasing} for $T > T^*$, unconditionally. This conclusion is independent of the kinetic model.

\subsection{Closed-form expression for \texorpdfstring{$T^*$}{T*}}
\label{App:Tpeak:Formula}

Since $|\Delta \sigma|$ is strictly increasing for $T < T^*$ and strictly decreasing for $T > T^*$, continuity implies:
\begin{equation}
T_\text{peak} = T^*.
\label{Eq:App:TpeakIsTstar}
\end{equation}
Because $T^*$ is a kink (non-differentiable point) of $|\Delta \sigma|(T)$, this result is obtained via piecewise monotonicity rather than by setting a derivative to zero.

Solving the kinetic saturation condition,
\begin{equation}
\sqrt{D_1(T^*)\,t} = \lambda_v,
\end{equation}
with:
\begin{equation}
D_1(T) = D_{1,0} \exp\!\left( -\frac{E_m^\mathrm{eff}}{kT} \right),
\end{equation}
gives the compact form:
\begin{equation}
T_\text{peak} = T^* = \frac{ E_m^\mathrm{eff} }{ k\ln\!\left( \dfrac{D_{1,0}t}{\lambda_v^2} \right) }.
\label{Eq:App:TstarCompact}
\end{equation}
Here,
\begin{align}
E_m^\mathrm{eff} & = E_m + E_f(v_\text{N}^{(\text{b})}) + E_b(\{\text{O}_\text{N}:v_\text{N}\}^{(\text{b})}), \\
D_{1,0} & = \frac{f z \lambda^2 \nu}{6}, \\
\lambda_v & = 0.5\,n_v^{-1/3} - R_v.
\end{align}

The formula is meaningful only when the argument of the logarithm exceeds unity, $D_{1,0}t / \lambda_v^2 > 1$, or equivalently, $f z \lambda^2 \nu t > 6 \lambda_v^2$. Physically, this means that oxygen must be capable of diffusing across the effective capture length within the nucleation time. For the parameter ranges in this work, $R_v \in [1,25]$~nm, $p\in[0.05,0.20]$, and $w_\text{O}\in[500,3000]$~ppm, this condition is satisfied at all relevant temperatures.

\Cref{Eq:App:TstarCompact} was compared against direct numerical maximization of the full $ |\Delta\sigma(T)| $ from \cref{Eq:DeltaSigma1}, including the $\text{O}_i^{(\text{s})}$ and $v_\text{N}^{(\text{s})}$ contributions. The comparison was performed for representative parameter combinations at intermediate stoichiometry.

\begin{table}[h!]
\centering
\scriptsize
\caption{Comparison of $T_\text{peak}$ from the closed-form expression (\cref{Eq:App:TstarCompact}) and from direct numerical maximization of the full $\Delta\sigma(T)$ in \cref{Eq:DeltaSigma1}, at intermediate stoichiometry, together with the corresponding peak reduction $\Delta\sigma_\text{min}$.}
\label{Tab:App:Tpeak}
\begin{tabular}{lcccc}
\hline
($R_v, w_\text{O}, p$) & $T_\text{peak}$ (formula) [K] & $T_\text{peak}$ (numerical) [K] & Difference [K] & $\Delta\sigma_\text{min}$ [J/m$^2$] \\
\hline
(1 nm, 1500 ppm, 0.05) & 1523 & 1524 & $<1$ & $-0.438$ \\
(2 nm, 1500 ppm, 0.05) & 1578 & 1579 & $<1$ & $-0.384$ \\
(5 nm, 1500 ppm, 0.05) & 1657 & 1658 & $<1$ & $-0.327$ \\
(10 nm, 1500 ppm, 0.05) & 1723 & 1723 & $<1$ & $-0.290$ \\
(25 nm, 1500 ppm, 0.05) & 1817 & 1818 & $<1$ & $-0.246$ \\
(50 nm, 1500 ppm, 0.05) & 1896 & 1897 & $<1$ & $-0.216$ \\
(1 nm, 1500 ppm, 0.20) & 1440 & 1441 & $<1$ & $-0.615$ \\
(25 nm, 1800 ppm, 0.05) & 1817 & 1818 & $<1$ & $-0.304$ \\
(25 nm, 1800 ppm, 0.20) & 1701 & 1701 & $<1$ & $-0.372$ \\
\hline
\end{tabular}
\end{table}

Agreement is within 1~K for all cases in \cref{Tab:App:Tpeak}. Thus, although the closed-form result is derived from the reduced dominant-channel expression, it accurately predicts the peak temperature of the full numerical model over the parameter range considered here. The tabulated $\Delta\sigma_\text{min}$ additionally confirms that the peak reduction decreases monotonically with increasing void radius.

The closed-form expression gives the following parameter dependences:
\paragraph{Void radius} At fixed porosity, the capture length scales linearly with void radius,
\begin{equation}
\lambda_v = 0.5\,n_v^{-1/3}-R_v = R_v \left[ 0.5 \left( \frac{4\pi}{3p} \right)^{1/3} - 1 \right].
\end{equation}
Therefore, increasing $R_v$ increases $\lambda_v$, decreases $\ln(D_{1,0}t/\lambda_v^2)$, and shifts $T_\text{peak}$ upward.

\paragraph{Porosity} Increasing $p$ decreases $\lambda_v$, because voids are closer together at fixed $R_v$. This increases $\ln(D_{1,0}t/\lambda_v^2)$ and lowers $T_\text{peak}$. The response remains logarithmic, but it is non-negligible over $p=5$--20\% because $\lambda_v$ contains the porosity-dependent factor $0.5-(3p/4\pi)^{1/3}$.

\paragraph{Oxygen concentration} Oxygen concentration does not appear in \cref{Eq:App:TstarCompact}; changing $c_\text{O}$ changes the magnitude of $|\Delta\sigma|$, but not the kinetic saturation temperature.

\paragraph{Stoichiometry} Stoichiometry enters through $E_m^\mathrm{eff}$, specifically through $E_f(v_\text{N}^{(\text{b})})$. N-rich conditions increase $E_m^\mathrm{eff}$ and therefore shift $T_\text{peak}$ upward.

\subsection{Nucleation time dependence of \texorpdfstring{$T_\text{peak}$}{Tpeak}}
\label{App:Tpeak:tdep}

The compact form in \cref{Eq:App:TstarCompact},
\begin{equation}
T_\text{peak} = \frac{E_m^\mathrm{eff}} {k\ln\!\left(D_{1,0}t/\lambda_v^2\right)},
\end{equation}
shows that the nucleation time enters only through a logarithm. Differentiating with respect to $\ln t$ gives:
\begin{equation}
\frac{dT_\text{peak}}{d\ln t} = -\frac{ E_m^\mathrm{eff} } { k \left[ \ln\!\left(D_{1,0}t/\lambda_v^2\right) \right]^2 }
= -\frac{ kT_\text{peak}^2 } { E_m^\mathrm{eff} }.
\label{Eq:App:dTpeak_dlogt}
\end{equation}
Thus increasing $t$ lowers $T_\text{peak}$, but only logarithmically. For an order-of-magnitude change in $t$,
\begin{equation}
\Delta T_\text{peak} \approx -\frac{ kT_\text{peak}^2 } { E_m^\mathrm{eff} } \ln 10.
\label{Eq:App:deltaTpeak}
\end{equation}
For intermediate stoichiometry, $E_m^\mathrm{eff}=5.22$~eV. Using the reference value $T_\text{peak}=1524$~K for $R_v=1$~nm, $w_\text{O}=1500$~ppm, $p=0.05$, and $t=42$~h gives $\Delta T_\text{peak} \approx -88$~K per decade increase in $t$. Direct numerical evaluation validates this approximate expression (i.e., \cref{Eq:App:deltaTpeak}) and gives the same scale: $T_\text{peak}$ decreases from 1581~K at $t=10$~h to 1491~K at $t=100$~h, a shift of 90~K. Thus, even an order-of-magnitude uncertainty in $t$ changes $T_\text{peak}$ by only $\sim 90$~K because $t$ enters only through $\ln t$.

\subsection{Porosity dependence of \texorpdfstring{$|\Delta\sigma|$}{|Delta sigma|}}
\label{App:Tpeak:Porosity}

The explicit prefactor in \cref{Eq:DeltaSigma1} contains $(1-p)/p$, which decreases by a factor of 4.75 from $p=5\%$ to $p=20\%$. However, the full peak magnitude varies much less strongly because this factor is partly canceled as follows.

Using the dominant-channel expression and evaluating it at $T_\text{peak}$, where $\alpha_1=1$, gives:
\begin{equation}
|\Delta\sigma|_\text{peak} = \frac{8}{3} \frac{c_\text{O}}{a^3} \frac{1-p}{p} R_v \frac{ [\text{N}_\text{N}^{(\text{s})}] } { [\text{N}_\text{N}^{(\text{b})}] } \, \exp\!\left( \frac{E_\text{seg}}{kT_\text{peak}} \right) \left|E_\text{ad}(T_\text{peak})\right|.
\label{Eq:App:DsigPeak}
\end{equation}
The surface nitrogen site fraction is:
\begin{equation}
[\text{N}_\text{N}^{(\text{s})}] = \frac{3}{2} \frac{p}{1-p} \frac{a}{R_v},
\quad
[\text{N}_\text{N}^{(\text{b})}] = 1 - [\text{N}_\text{N}^{(\text{s})}].
\label{Eq:App:NNsNNb}
\end{equation}
Substituting \cref{Eq:App:NNsNNb} into \cref{Eq:App:DsigPeak} gives:
\begin{equation}
\frac{1-p}{p} R_v \frac{ [\text{N}_\text{N}^{(\text{s})}] } { [\text{N}_\text{N}^{(\text{b})}] }
= \frac{1-p}{p} R_v \frac{ \frac{3}{2} \frac{p}{1-p} \frac{a}{R_v} } { [\text{N}_\text{N}^{(\text{b})}]
}
= \frac{3a/2}{[\text{N}_\text{N}^{(\text{b})}]}.
\label{Eq:App:cancel}
\end{equation}
Thus, the leading $(1-p)/p$ dependence cancels exactly against the $p/(1-p)$ dependence of the surface-site fraction, and the explicit factor of $R_v$ cancels against $1/R_v$. Substitution gives:
\begin{equation}
|\Delta\sigma|_\text{peak} = \frac{4c_\text{O}}{a^2} \frac{1}{[\text{N}_\text{N}^{(\text{b})}]}
\exp\!\left( \frac{E_\text{seg}}{kT_\text{peak}} \right) \left|E_\text{ad}(T_\text{peak})\right|.
\label{Eq:App:DsigPeakReduced}
\end{equation}
This cancellation removes the leading geometric porosity dependence, but it does not make the peak magnitude independent of porosity. The residual porosity dependence enters through two channels. First,
\begin{equation}
\frac{1}{[\text{N}_\text{N}^{(\text{b})}]} = \frac{1}{1 - \frac{3}{2} \frac{p}{1-p} \frac{a}{R_v}},
\end{equation}
which is close to unity for large cavities but non-negligible for $R_v=1$~nm. Second, $T_\text{peak}$ depends on $p$ through $\lambda_v$:
\begin{equation}
T_\text{peak}(p) = \frac{ E_m^\mathrm{eff} } { k \ln \! \left[ \tfrac{ D_{1,0} t } { R_v^2
\left[ 0.5 \left[ 4\pi/(3p) \right]^{1/3} - 1 \right]^2 } \right] }.
\label{Eq:App:Tpeak_p}
\end{equation}
Increasing $p$ lowers $\lambda_v$, lowers $T_\text{peak}$, and therefore changes the thermodynamic factors $\exp[E_\text{seg}/(kT_\text{peak})]$ and $|E_\text{ad}(T_\text{peak})|$.

% \begin{table}[h!]
% \centering
% \footnotesize
% \caption{Porosity dependence of the peak $ |\Delta\sigma| $ at two void radii for $w_\text{O}=1500$~ppm and intermediate stoichiometry. Each row is evaluated at that porosity's own $T_\text{peak}$ from \cref{Eq:App:TstarCompact}.}
% \label{Tab:App:por}
% \begin{tabular}{lcccc}
% \hline
% & $p$ [\%] & $T_\text{peak}$ [K] & $1/[\text{N}_\text{N}^{(\text{b})}]$ & $ |\Delta\sigma| $ variation \\
% \hline
% \multirow{4}{*}{$R_v = 1$ nm}
% & 5  & 1524 & 1.040 & --- \\
% & 10 & 1488 & 1.088 & $+13\%$ \\
% & 15 & 1462 & 1.148 & $+26\%$ \\
% & 20 & 1441 & 1.223 & $+41\%$ \\
% \hline
% \multirow{4}{*}{$R_v = 25$ nm}
% & 5  & 1818 & 1.002 & --- \\
% & 10 & 1767 & 1.003 & $+9\%$ \\
% & 15 & 1731 & 1.005 & $+16\%$ \\
% & 20 & 1701 & 1.007 & $+23\%$ \\
% \hline
% \end{tabular}
% \end{table}

% \Cref{Tab:App:por} separates the two residual effects. For $R_v=25$~nm, the direct factor $1/[\text{N}_\text{N}^{(\text{b})}]$ changes by less than 1\%, so most of the residual porosity dependence comes through $T_\text{peak}(p)$. For $R_v=1$~nm, $1/[\text{N}_\text{N}^{(\text{b})}]$ changes by about 18\%, so both residual channels contribute. Overall, increasing $p$ from 5\% to 20\% raises the peak magnitude by 23\% for $R_v=25$~nm and by 41\% for $R_v=1$~nm. Thus, the leading geometric cancellation is exact, but the full peak magnitude is not porosity-independent; its residual porosity dependence is appreciable, especially for small cavities, while remaining far smaller than the naive $4.75\times$ scaling of the isolated $(1-p)/p$ prefactor.

\begin{table}[h!]
\centering
\scriptsize
\caption{Term-by-term porosity dependence of the peak $|\Delta\sigma|$ at two void radii for $w_\text{O}=1500$~ppm and intermediate stoichiometry. Each row is evaluated at that porosity's own $T_\text{peak}$. The columns $\Delta A$, $\Delta B$, and $\Delta C$ give the percent change relative to $p=5\%$ of $A=1/[\text{N}_\text{N}^{(\text{b})}]$, $B=\exp(E_\text{seg}/kT_\text{peak})$, and $C=|E_\text{ad}(T_\text{peak})|$, respectively. The reduced-product column is the multiplicative change in $ABC$ from \cref{Eq:App:DsigPeakReduced}. The full-model column is obtained from direct numerical evaluation of \cref{Eq:DeltaSigma1}.}
\label{Tab:App:por}
\begin{tabular}{lccccccc}
\hline
& $p$ [\%]
& $T_\text{peak}$ [K]
& $\Delta A$ [\%]
& $\Delta B$ [\%]
& $\Delta C$ [\%]
& $\Delta(ABC)$ [\%]
& Full $|\Delta\sigma|$ [\%] \\
\hline
\multirow{4}{*}{$R_v = 1$ nm}
& 5  & 1524 & ---   & ---   & ---  & ---   & --- \\
& 10 & 1488 & +4.6  & +6.2  & +1.2 & +12.5 & +12.8 \\
& 15 & 1462 & +10.4 & +11.2 & +2.1 & +25.3 & +25.9 \\
& 20 & 1441 & +17.6 & +15.4 & +2.9 & +39.6 & +40.5 \\
\hline
\multirow{4}{*}{$R_v = 25$ nm}
& 5  & 1818 & ---  & ---   & ---  & ---   & --- \\
& 10 & 1767 & +0.2 & +6.2  & +1.9 & +8.5  & +9.2 \\
& 15 & 1731 & +0.4 & +11.1 & +3.3 & +15.2 & +16.4 \\
& 20 & 1701 & +0.6 & +15.5 & +4.5 & +21.3 & +23.0 \\
\hline
\end{tabular}
\end{table}

\Cref{Tab:App:por} separates the residual porosity dependence into the three multiplicative factors in \cref{Eq:App:DsigPeakReduced}. The percentages in the individual columns should not be added. The reduced-product variation is obtained from the product of the three relative factors. For $R_v=25$~nm, the direct factor $A=1/[\text{N}_\text{N}^{(\text{b})}]$ changes by less than 1\% from $p=5$--20\%, so most of the residual porosity dependence comes from the lower $T_\text{peak}$, which increases the segregation factor $B=\exp(E_\text{seg}/kT_\text{peak})$ and the adsorption magnitude $C=|E_\text{ad}(T_\text{peak})|$. For $R_v=1$~nm, $A$ increases by 17.6\%, so both the direct site-balance residual and the $T_\text{peak}$-dependent thermodynamic factors contribute. The reduced dominant-channel expression predicts increases of 39.6\% and 21.3\% at $p=20\%$ for $R_v=1$ and 25~nm, respectively, while the full numerical model gives 40.5\% and 23.0\%. The small difference reflects the retained $\text{O}_i^{(\text{s})}$ and $v_\text{N}^{(\text{s})}$ contributions in the full model. Thus, the residual porosity dependence is non-negligible but moderate, and remains far smaller than the naive $4.75\times$ scaling of the isolated $(1-p)/p$ prefactor.

\section{Supplementary information}
\label{App3}

% \printnomenclature

\begin{figure}[h!]
\centering
\begin{subfigure}{0.45\textwidth}
    \includegraphics[width=\textwidth]{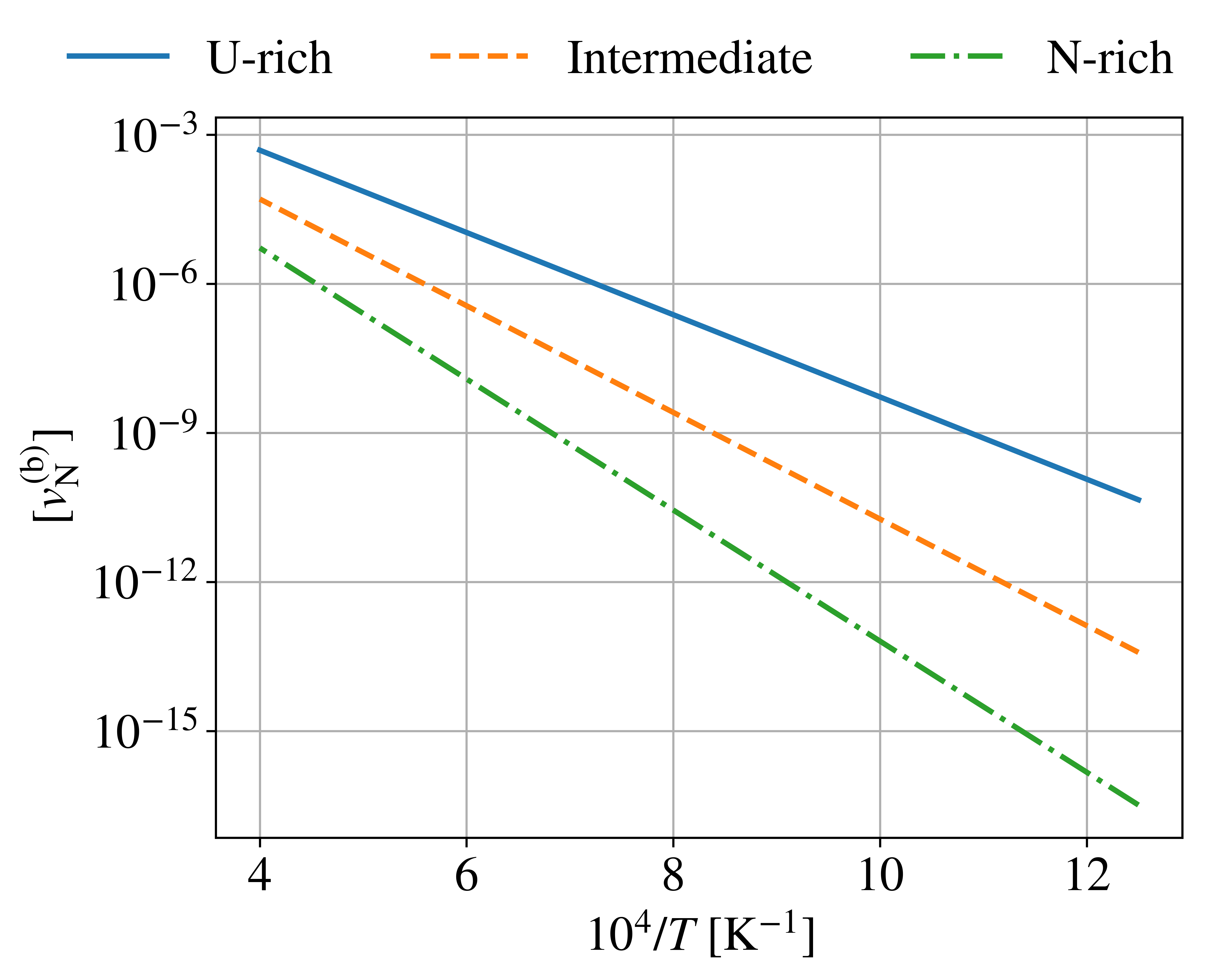}
    \caption{}
    \label{Fig:VN}
\end{subfigure}
\hfill
\begin{subfigure}{0.45\textwidth}
    \includegraphics[width=\textwidth]{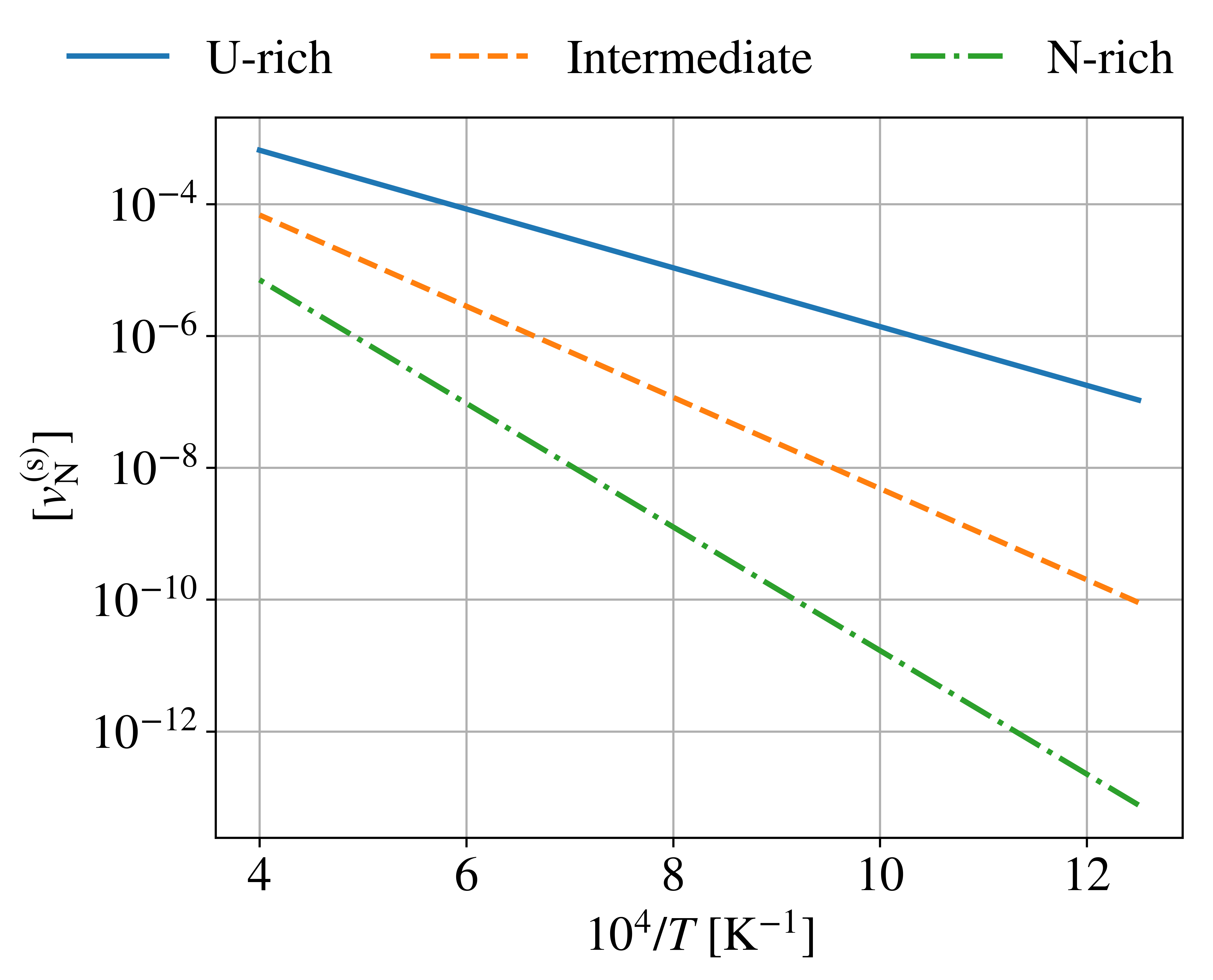}
    \caption{}
    \label{Fig:VNs}
\end{subfigure}
\caption{\textbf{(a)} Concentration of N vacancies in bulk UN calculated from the formation energies in \cref{Tab:EfVN}. \textbf{(b)} Concentration of N vacancies on the UN (001) surface calculated from the formation energies in \cref{Tab:EfVN}.}
\label{1}
\end{figure}

\begin{figure}[h!]
    \centering
    \includegraphics[width=0.45\textwidth]{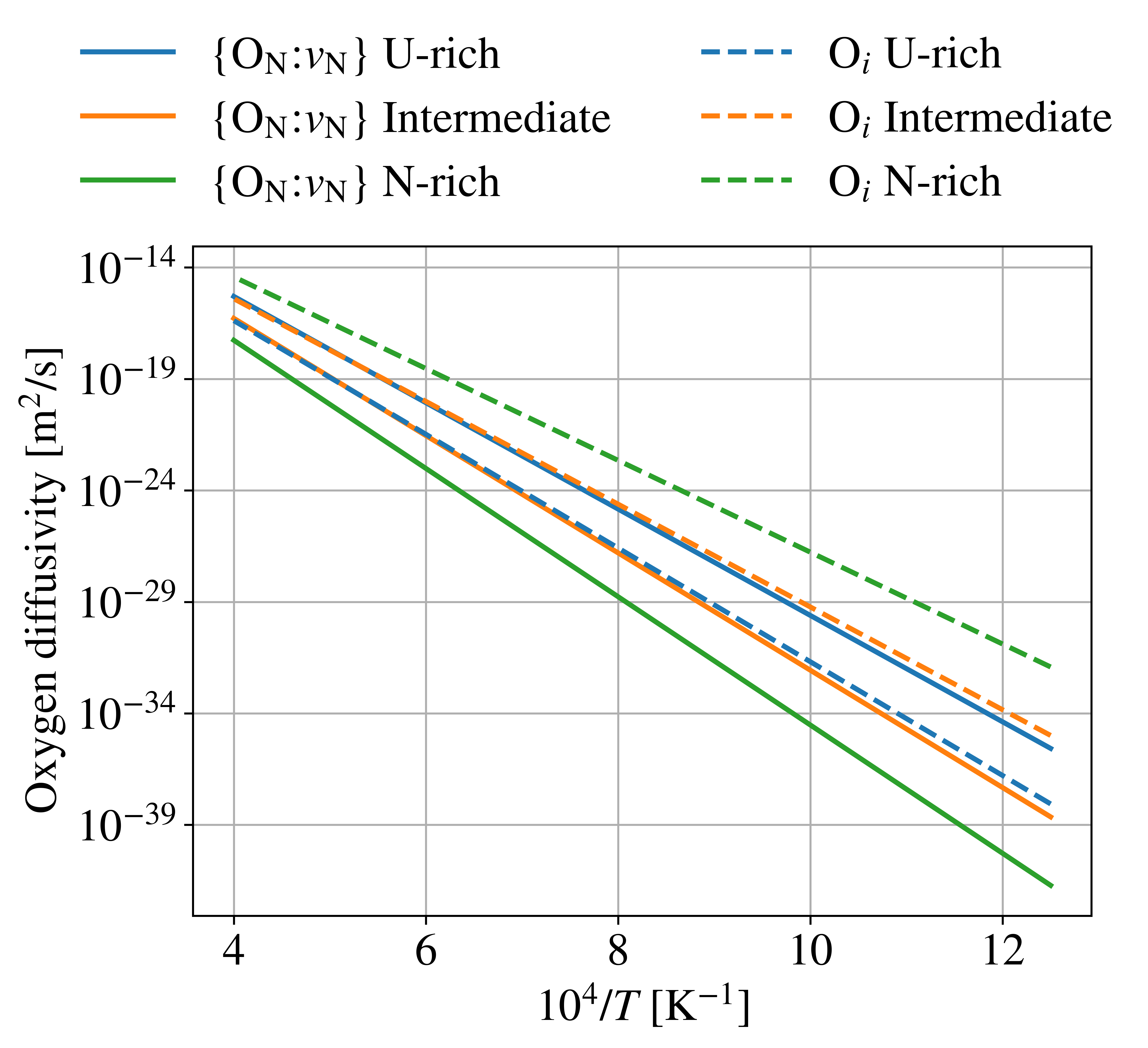}
    \caption{(Color online) Diffusivity of oxygen impurities in bulk UN by the defects $ \{\text{O}_\text{N} \! : \! v_\text{N}\} $ and $ \text{O}_i $ under U-rich (blue), Intermediate (orange), and N-rich (green) conditions.}
    \label{Fig:DO}
\end{figure}

\FloatBarrier

%%%%%%%%%%%%%%%%%%%%%%%%%%%%%%%%%%%%%%%%%%%%%%%%%%%%%%%%%%%%%

\section{Oxygen adsorption on the UN (110) surface}
\label{App:110}

The surface energetics in the main text are derived from planar (001) slabs, which is justified because (001) is the most stable low-index facet of rocksalt-structured UN and is expected to dominate the equilibrium void shape. To test the robustness of the dominant $\text{O}_\text{N}^{(\text{s})}$ channel against facet orientation, we repeat the surface calculations for the (110) facet. All bulk-derived quantities entering \cref{Eq:DeltaSigma1}---the chemical potentials $\mu_\text{U}$, $\mu_\text{N}$, $\mu_\text{O}$, the bulk incorporation energy $E_\text{inc}(\text{O}_\text{N}^{(\text{b})})$, the bulk vacancy formation energy $E_f(v_\text{N}^{(\text{b})})$, the binding energy $E_b$, the migration energy $E_m$, the diffusivity prefactor $D_{1,0}$, and the kinetic correction $\alpha_1$---are unchanged. Only the surface energetics ($\sigma$, $E_\text{ad}(\text{O}_\text{N}^{(\text{s})})$, $E_f(v_\text{N}^{(\text{s})})$) and the areal density of surface nitrogen sites are recomputed for the (110) orientation.

\subsection{Computational details}
\label{App:110:comp}

The (110) surface is modelled with a symmetric, stoichiometric 8-layer slab containing 96 U and 96 N atoms. The slab is a $3\times2$ supercell of the (110) surface unit cell, with in-plane lattice vectors of $3a_0=14.584$~\AA\ along $[001]$ and $2\sqrt{2}\,a_0=13.750$~\AA\ along $[1\bar{1}0]$, and a $34.375$~\AA\ cell dimension along the $[110]$ surface normal that provides a vacuum gap of approximately twice the slab thickness. Brillouin-zone sampling uses a $\Gamma$-centred $3\times3\times1$ Monkhorst-Pack mesh. All other settings (PBE functional, PAW potentials, 520~eV plane-wave cutoff, first-order Methfessel-Paxton smearing with a width of 0.1~eV) are identical to the (001) calculations. The electronic and ionic convergence criteria used for the (001) slabs (i.e., $10^{-4}$~eV and $10^{-2}$~eV/\AA\ respectively) led to convergence difficulties for the more corrugated, higher-energy (110) surface. We therefore adopted looser criteria of $5\times10^{-4}$~eV and $3\times10^{-2}$~eV/\AA\ for all (110) calculations.
% Upon relaxation, the surface layer rumples: the U sublattice relaxes outward and the N sublattice inward by $0.22$~\AA, without any change in periodicity or bonding topology (i.e.\ relaxation, not reconstruction).

\subsection{Surface energetics}
\label{App:110:energetics}

The surface energy $\sigma$ (\cref{Eq:Sigma}), the oxygen adsorption energy $E_\text{ad}(\text{O}_\text{N}^{(\text{s})})$ (\cref{Eq:Ead}, referenced to $\mu_\text{O}=-4.18$~eV), and the surface nitrogen vacancy formation energy $E_f(v_\text{N}^{(\text{s})})$ are evaluated exactly as for (001). The clean (110) surface energy, $\sigma=2.124$~J/m$^2$, is stoichiometry-independent and is $34\%$ higher than the (001) value of $1.586$~J/m$^2$, confirming that (110) is the less stable facet. Surface energies for the clean, vacancy-bearing, and oxygen-bearing slabs are collected in \cref{Tab:110:surfE}, using the same convention as \cref{Tab:SurfE}.

\begin{table}[h!]
\centering
\footnotesize
\caption{Surface energies of the UN (110) surface (J/m$^2$), computed as in \cref{Eq:Sigma,Tab:SurfE}. Values for the defected slabs are stoichiometry dependent because $N_\text{U}\neq N_\text{N}$.}
\begin{tabular}{lccc}
\hline
 & U-rich & Intermediate & N-rich \\
\hline
Clean (110)                                 & 2.124 & 2.124 & 2.124 \\
(110) with $v_\text{N}^{(\text{s})}$        & 2.127 & 2.146 & 2.166 \\
(110) with $\text{O}_\text{N}^{(\text{s})}$ & 1.875 & 1.894 & 1.914 \\
\hline
\end{tabular}
\label{Tab:110:surfE}
\end{table}

The derived defect energetics are compared with the (001) values in \cref{Tab:110:defE}. Oxygen binds $0.49$~eV \emph{less} strongly on (110) ($E_\text{ad}=-6.31$~eV versus $-6.80$~eV), but a surface nitrogen vacancy is far easier to form on the open (110) surface: the vacancy segregation energy $E_f(v_\text{N}^{(\text{s})})-E_f(v_\text{N}^{(\text{b})})=-1.57$~eV is twice the (001) value of $-0.76$~eV. Combining these through the segregation energy (\cref{Eq:App:gT}),
\begin{equation}
E_\text{seg}=-\bigl[E_f(v_\text{N}^{(\text{s})})-E_f(v_\text{N}^{(\text{b})}) + E_\text{ad}(\text{O}_\text{N}^{(\text{s})}) - E_\text{inc}(\text{O}_\text{N}^{(\text{b})})\bigr],
\label{Eq:110:Eseg}
\end{equation}
gives $E_\text{seg}(110)=0.65$~eV, again double the (001) value of $0.33$~eV. Both differences in \cref{Eq:110:Eseg} are stoichiometry independent, so $E_\text{seg}$ is a single value for all conditions.

\begin{table}[h!]
\centering
\footnotesize
\caption{Surface defect energetics on the (110) and (001) facets. The adsorption and incorporation energies use $\mu_\text{O}=-4.18$~eV. The bulk quantities $E_f(v_\text{N}^{(\text{b})})$ and $E_\text{inc}(\text{O}_\text{N}^{(\text{b})})$ are facet independent and repeated from \cref{Tab:EfVN,Tab:Einc}.}
\begin{tabular}{lcc}
\hline
Quantity [eV] & (001) & (110) \\
\hline
$E_\text{ad}(\text{O}_\text{N}^{(\text{s})})$               & $-6.80$ & $-6.31$ \\
$E_f(v_\text{N}^{(\text{s})})$ (U-rich/Int./N-rich)         & $0.88/1.37/1.86$ & $0.07/0.56/1.05$ \\
$E_f(v_\text{N}^{(\text{s})})-E_f(v_\text{N}^{(\text{b})})$ & $-0.76$ & $-1.57$ \\
$E_\text{seg}$                                              & $0.33$  & $0.65$ \\
\hline
\end{tabular}
\label{Tab:110:defE}
\end{table}

\subsection{Site-conserving oxygen adsorption model}
\label{App:110:model}

We retain only the dominant $\text{O}_\text{N}^{(\text{s})}$ channel of \cref{Eq:DeltaSigma1} and write it as
\begin{equation}
\Delta\sigma(\text{O}_\text{N}^{(\text{s})}) = \frac{8}{3} \frac{1-p}{p} \frac{R_v}{a^3} c_\text{O} \, \alpha_1 \,
\frac{[\text{O}_\text{N}^{(\text{s})}]}{[\text{O}_\text{N}^{(\text{b})}]} \,
E_\text{ad}(\text{O}_\text{N}^{(\text{s})}) = \alpha_1 \, n_{\text{O}_\text{N}} \, E_\text{ad}(\text{O}_\text{N}^{(\text{s})}),
\label{Eq:110:ON}
\end{equation}
where
\begin{equation}
n_{\text{O}_\text{N}} = \frac{N_\text{O}}{A_v} \, \frac{[\text{O}_\text{N}^{(\text{s})}]}{[\text{O}_\text{N}^{(\text{b})}]}
= \frac{8}{3} \frac{1-p}{p} \frac{R_v}{a^3}c_\text{O} \,
\frac{[\text{O}_\text{N}^{(\text{s})}]}{[\text{O}_\text{N}^{(\text{b})}]}
\label{Eq:110:nON}
\end{equation}
is the number of substitutional surface oxygen atoms per unit void area. Here, $N_\text{O}$ is the total number of oxygen atoms, $A_v = 3pV_0/R_v$ the total void surface area, $c_\text{O}$ is the atomic fraction of oxygen (\cref{Eq:cO}), and $[\text{O}_\text{N}^{(\text{s})}]$ and $[\text{O}_\text{N}^{(\text{b})}]$ are per-formula-unit site fractions.

Note that the prefactor $\tfrac{8}{3}\tfrac{1-p}{p}\tfrac{R_v}{a^3}c_\text{O}=N_\text{O}/A_v$ follows from \cref{Eq:NOAv} as the total oxygen content divided by the total spherical void area $A_v$. Both $N_\text{O}$ and $A_v$ are bulk and void-geometry quantities with no reference to the exposed crystallographic plane, so the prefactor is identical for the (001) and (110) facets. The facet enters the model only through the areal density of surface nitrogen sites,
\begin{equation}
n_\text{site} = \frac{1}{A_\text{site}},
\quad
A_\text{site} = \begin{cases}
                    a^2/2        & \{001\}, \\
                    a^2/\sqrt{2} & \{110\},
                \end{cases}
\label{Eq:110:nsite}
\end{equation}
where $A_\text{site}$ is the surface area per nitrogen site. On (001) each $a_0 \times a_0$ mesh exposes two nitrogen atoms per layer, giving $A_\text{site}=a^2/2$ (consistent with the site-counting of \cref{Eq:NNs1}); on (110) the $a_0 \times \sqrt{2} \, a_0$ mesh exposes two nitrogen atoms over an area $\sqrt{2}\,a_0^2$, giving $A_\text{site}=a^2/\sqrt{2}$, i.e., a surface nitrogen density lower by $\sqrt{2}$.

The physical requirement that oxygen cannot occupy more surface nitrogen sites than exist is expressed directly in terms of these two areal densities,
\begin{equation}
n_{\text{O}_\text{N}} \le n_\text{site},
\label{Eq:110:constraint}
\end{equation}
i.e., a surface coverage $\theta \equiv n_{\text{O}_\text{N}} / n_\text{site} \le 1$. The dilute mass-action expression \cref{Eq:ONs_ONb} contains no site-exclusion term and does not enforce \cref{Eq:110:constraint}. On (001) this is immaterial, because the coverage remains well below unity ($\theta = 0.24$--$0.38$ at $T_\text{peak}$; \cref{Tab:110:results}); on the strongly segregating (110) facet, however, the unconstrained $n_{\text{O}_\text{N}}$ exceeds $n_\text{site}$ by a factor of $2.2$--$5.6$. We therefore impose \cref{Eq:110:constraint} through a site-conserving occupancy,
\begin{equation}
n_{\text{O}_\text{N}} \to \min( n_{\text{O}_\text{N}}, n_\text{site}),
\end{equation}
so that the corrected $\text{O}_\text{N}^{(\text{s})}$ channel reads
\begin{equation}
\Delta \sigma ( \text{O}_\text{N}^{(\text{s})} ) = \alpha_1 \, \min \! \bigl( n_{\text{O}_\text{N}},\,n_\text{site} \bigr) \,
E_\text{ad}(\text{O}_\text{N}^{(\text{s})}).
\label{Eq:110:corrected}
\end{equation}

When $\theta < 1$ the minimum returns $n_{\text{O}_\text{N}}$ and \cref{Eq:110:corrected} is identical to \cref{Eq:110:ON}, hence to the $\text{O}_\text{N}^{(\text{s})}$ term of \cref{Eq:DeltaSigma1}. The (001) facet is in this regime at every temperature, so \emph{the present correction leaves the (001) surface model and every (001) result reported in the main text unchanged.} When $\theta=1$ the minimum returns $n_\text{site}$ and
\begin{equation}
\Delta \sigma(\text{O}_\text{N}^{(\text{s})}) = \frac{\alpha_1}{A_\text{site}} \, E_\text{ad}(\text{O}_\text{N}^{(\text{s})}),
\label{Eq:110:sat}
\end{equation}
a complete oxygen monolayer on the surface nitrogen sublattice, modulated by the kinetic factor $\alpha_1$. The larger segregation energy of (110) ($E_\text{seg}=0.65$~eV versus $0.33$~eV) raises $[\text{O}_\text{N}^{(\text{s})}]/[\text{O}_\text{N}^{(\text{b})}]$, and hence $n_{\text{O}_\text{N}}$, above $n_\text{site}$ over the temperature range of interest, placing the (110) facet in the saturated regime of \cref{Eq:110:sat}.

\subsection{Results and comparison}
\label{App:110:results}

\Cref{Tab:110:results} reports the corrected $\text{O}_\text{N}^{(\text{s})}$ channel for the reference case $R_v=1$~nm, $p=5\%$, $w_\text{O}=1500$~ppm. Since $T_\text{peak}$ (\cref{Eq:TpeakMain}) depends only on bulk transport and void geometry, it is identical for the two facets ($1381$, $1524$, $1666$~K for the U-rich, intermediate, and N-rich conditions, respectively). On the (110) surface, the nitrogen sublattice is fully occupied by oxygen ($\theta=1$), yielding $|\Delta\sigma|=0.18$, $0.60$, and $1.01$~J/m$^2$ under U-rich, intermediate, and N-rich conditions, respectively. Compared with the corresponding sub-monolayer values for (001), $0.35$, $0.45$, and
$0.50$~J/m$^2$, the reduction is larger for the intermediate and N-rich cases but smaller under U-rich conditions. In the latter case, the weaker oxygen adsorption on (110) more than offsets the increase in surface coverage. In every case, the oxygen-covered (110) surface retains a positive effective surface energy, $\sigma_\text{eff}=\sigma_\text{clean}+\Delta\sigma>0$, so the reduction stays within the domain of the surface-energy model.

\begin{table}[h!]
\centering
\footnotesize
\caption{Corrected $\text{O}_\text{N}^{(\text{s})}$-channel surface-energy
reduction (\cref{Eq:110:corrected}) for the reference case $R_v=1$~nm, $p=5\%$,
$w_\text{O}=1500$~ppm. $T_\text{peak}$ is identical for both facets. $\theta$ is
the coverage $n_{\text{O}_\text{N}}/n_\text{site}$ at $T_\text{peak}$: the (001)
facet remains dilute ($\theta<1$) while (110) is saturated ($\theta=1$).
$\sigma_\text{eff}$ uses the respective clean surface energy ($1.586$ and
$2.124$~J/m$^2$).}
\begin{tabular}{llcccc}
\hline
Facet & Stoich. & $T_\text{peak}$ [K] & $\theta$ &
$\Delta\sigma_\text{min}$ [J/m$^2$] & $\sigma_\text{eff}$ [J/m$^2$] \\
\hline
\multirow{3}{*}{(001)}
 & U-rich       & 1381 & 0.38 & $-0.35$ & 1.23 \\
 & Intermediate & 1524 & 0.29 & $-0.45$ & 1.14 \\
 & N-rich       & 1666 & 0.24 & $-0.50$ & 1.09 \\
\hline
\multirow{3}{*}{(110)}
 & U-rich       & 1381 & 1.00 & $-0.18$ & 1.94 \\
 & Intermediate & 1524 & 1.00 & $-0.60$ & 1.53 \\
 & N-rich       & 1666 & 1.00 & $-1.01$ & 1.11 \\
\hline
\end{tabular}
\label{Tab:110:results}
\end{table}

The main-text conclusion is therefore reinforced rather than weakened by facet variation: the $\text{O}_\text{N}^{(\text{s})}$ segregation channel lowers the void surface energy on (110) at least as effectively as on (001), and the reduction remains largest under the N-rich conditions relevant to irradiation. The one qualitative difference is that the strong nitrogen-vacancy segregation on the open (110) facet saturates the surface nitrogen sublattice, so the dilute expression \cref{Eq:ONs_ONb} is replaced by the site-conserving form \cref{Eq:110:corrected}. The dominant (001) facet remains safely in the dilute regime.

\section{Uncertainty quantification and sensitivity analysis}
\label{App:UQ}

The model in \cref{Sec:Oxygen} uses DFT-derived energetic quantities to predict the oxygen-induced reduction in void surface energy, $\Delta\sigma(T)$. Since these quantities enter both thermodynamic and kinetic factors, their uncertainty can propagate nonlinearly into the predicted surface-energy reduction. This appendix provides an uncertainty propagation for the reference case used in the main text: $R_v=1$~nm, $p=5\%$, $w_\mathrm{O}=1500$~ppm, intermediate stoichiometry, and a nominal annealing time of $42$~h. For this reference case, the deterministic baseline model gives $\Delta\sigma_\mathrm{min}=-0.438$~J/m$^{2}$ at $T=1524$~K. With $\sigma_\mathrm{clean}$ = 1.590~J/m$^{2}$, this corresponds to an effective surface energy $\sigma_\mathrm{eff}=\sigma_\mathrm{clean}+\Delta\sigma$ = 1.152~J/m$^{2}$, and a void-nucleation barrier reduction of $61.9\%$.

\subsection{Monte Carlo propagation}

The uncertainty propagation is restricted to the dominant $\mathrm{O_N}^{(\mathrm{s})}$ surface channel. The interstitial oxygen channel is retained in the full $\Delta\sigma(T)$ expression, but its parameters are not varied in the uncertainty propagation. This keeps the analysis focused on the channel responsible for the main surface-energy reduction.

Six energetic inputs are varied: $E_\mathrm{inc}(\mathrm{O_N}^{(\mathrm{b})})$, $E_\mathrm{ad}(\mathrm{O_N}^{(\mathrm{s})})$, $E_f(v_\mathrm{N}^{(\mathrm{b})})$, $E_f(v_\mathrm{N}^{(\mathrm{s})})$, $E_b(\mathrm{O_N}:v_\mathrm{N})$, and $E_m$. Note that $E_m$, the migration energy for the vacancy-assisted $\mathrm{O_N}:v_\mathrm{N}$ mechanism, enters twice: once through the attempt-frequency prefactor and once through the Arrhenius exponential. The annealing time is also treated as uncertain, but it is not a DFT uncertainty. It is sampled separately as a kinetic/model parameter over the range $10$--$100$~h.

Each DFT-derived energetic input is assigned a bounded uncertainty of $\pm0.1$~eV. The errors are not sampled independently for all energies because the quantities entering the model are not independent measurements. They are differences of related DFT total energies and therefore share systematic errors. For example, $E_\mathrm{inc}(\mathrm{O_N}^{(\mathrm{b})})$ and $E_\mathrm{ad}(\mathrm{O_N}^{(\mathrm{s})})$ both involve oxygen-containing UN configurations. A systematic error in the DFT description of oxygen bonding or in the oxygen chemical reference should therefore move both quantities in the same direction. Similarly, $E_f(v_\mathrm{N}^{(\mathrm{b})})$ and $E_f(v_\mathrm{N}^{(\mathrm{s})})$ both involve nitrogen-vacancy formation in UN and are expected to share a vacancy-related systematic component.

The oxygen-related pair and the vacancy-related pair are therefore perturbed using a shared systematic term plus a smaller residual term:
\begin{equation}
E_i = E_i^\mathrm{DFT} + s_g + r_i,
\label{Eq:App:UQ:errormodel}
\end{equation}
where $s_g$ is the shared shift for group $g$, and $r_i$ is the residual term specific to input $i$. The shared terms are sampled from:
\begin{equation}
s_g \sim U(-0.08,0.08)~\mathrm{eV},
\end{equation}
and the residual terms are sampled from:
\begin{equation}
r_i \sim U(-0.02,0.02)~\mathrm{eV}.
\end{equation}
The resulting total perturbation is clipped to remain within $\pm0.1$~eV. The binding energy $E_b$ and the migration energy $E_m$ are sampled independently from $U(-0.1,0.1)$~eV. The annealing time is sampled from a log-uniform distribution between $10$ and $100$~h.

This correlation structure preserves the common-mode cancellation expected in differences such as $E_\mathrm{ad}-E_\mathrm{inc}$ and $E_f(v_\mathrm{N}^{(\mathrm{s})})-E_f(v_\mathrm{N}^{(\mathrm{b})})$. The uncertainty model, therefore, allows the absolute DFT energies to vary while avoiding artificial decorrelation of chemically related quantities.

For each Monte Carlo sample, the six energetic inputs and the annealing time are perturbed according to the model above. The full $\Delta\sigma(T)$ curve is then recomputed over the temperature range $800$--$2500$~K. From the ensemble of sampled curves, the median and 2.5--97.5 percentile band are reported at each temperature.

The correlated Monte Carlo calculation used $2\times10^4$ samples for the scalar statistics and $5\times10^3$ samples for the plotted band. The sampled perturbations produced the intended correlation structure. The oxygen pair had a correlation coefficient of $0.94$, and the vacancy pair also had a correlation coefficient of $0.94$. Cross-correlations between unrelated groups were negligible.

The propagated uncertainty is summarized in \cref{Tab:App:UQ:mc}. Note that $E_m$ broadens the peak-temperature interval because it enters the uncancelled kinetic activation term in $D_1$.

\begin{table}[h!]
\centering
\footnotesize
\caption{Correlated Monte Carlo uncertainty propagation for the reference case. The 95\% percentile intervals correspond to the 2.5--97.5 percentile range of the sampled distribution.}
\label{Tab:App:UQ:mc}
\begin{tabular}{lcc}
\hline
Quantity & Median & 95\% percentile interval \\
\hline
$T_\mathrm{peak}$ & $1534~\mathrm{K}$ & $1464$ to $1609~\mathrm{K}$ \\
$\Delta\sigma_\mathrm{min}$ & $-0.428~\mathrm{J~m^{-2}}$ & $-0.632$ to $-0.291~\mathrm{J~m^{-2}}$ \\
$\sigma_\mathrm{eff}$ & $1.162~\mathrm{J~m^{-2}}$ & $0.958$ to $1.299~\mathrm{J~m^{-2}}$ \\
Barrier reduction & $60.9\%$ & $45.4$ to $78.1\%$ \\
\hline
\end{tabular}
\end{table}
The main conclusion is unchanged by the correlated Monte Carlo uncertainty propagation. Oxygen still produces a substantial reduction in the void surface energy over the full uncertainty band. No Monte Carlo sample violates the physical condition
\begin{equation}
\sigma_\mathrm{eff}=\sigma_\mathrm{clean}+\Delta\sigma>0,
\end{equation}
so the calculation remains inside the domain of the classical surface-energy model.

\begin{figure}[h!]
    \centering
    \includegraphics[width=0.75\textwidth]{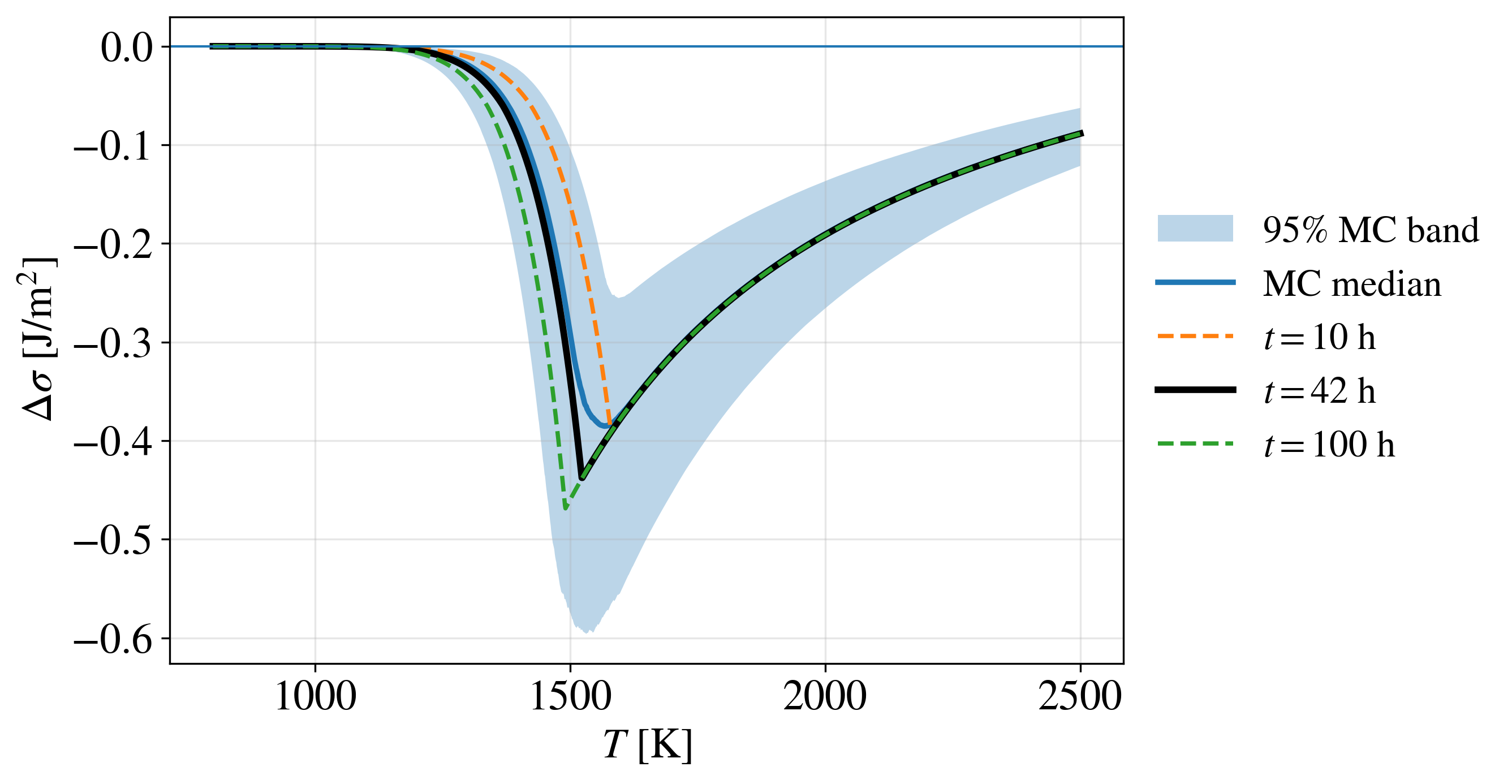}
    \caption{Correlated Monte Carlo uncertainty band for $\Delta\sigma(T)$ at the reference case. The shaded region is the 2.5--97.5 percentile range, the solid curve is the Monte Carlo median, and the baseline curves show the deterministic model at $t=10$, $42$, and $100$~h. The energetic uncertainty is bounded by $\pm0.1$~eV, with shared systematic shifts applied to the oxygen-related and vacancy-related energy pairs. The migration energy $E_m$ is also varied and is updated consistently in both the Arrhenius exponent and the attempt-frequency prefactor.}
    \label{Fig:App:UQ:band}
\end{figure}

The correlated uncertainty model is a physically motivated epistemic uncertainty propagation. It is not a claim that the DFT errors are known statistical random variables. The central assumption is that related DFT energy differences share systematic errors. This is appropriate here because $\Delta\sigma(T)$ depends strongly on adsorption-incorporation and surface-bulk differences, not on unrelated absolute energies.

The result should therefore be interpreted as follows. The model prediction is robust to bounded, correlated shifts of the dominant DFT energetic inputs. The shared part of the error largely cancels in the energy differences controlling the magnitude of $\Delta\sigma$. The uncancelled kinetic terms, especially $E_m$, mainly affect the peak temperature. That is, the peak \textit{magnitude} is controlled primarily by segregation-like differences between surface and bulk oxygen states, whereas the peak \textit{position} is controlled more by the kinetic activation term in $D_1$, which contains $E_m$, $E_f(v_\mathrm{N}^{(\mathrm{b})})$, and $E_b$.

\subsection{Hubbard $U$ sensitivity}
\label{App:UQ:U}

Although the main dataset in this work is based on PBE, we performed a limited DFT+$U$ test to check whether the bulk oxygen-defect ordering changes when a Hubbard correction is applied. This test is motivated by the fact that DFT+$U$ calculations in UN are numerically delicate: the reported imaginary phonons in DFT+$U$ UN \cite{Kocevski2022I} may reflect, at least in part, metastable occupation-matrix states and the difficulty of locating the correct electronic minimum, rather than only an intrinsic physical instability. We therefore repeated a subset of the bulk calculations using DFT+$U$ with $U_\text{eff}$~=~1~eV and use these results only as a robustness check for converged bulk defect energetics. The incorporation energies used the same oxide-referenced oxygen chemical potential, $\mu_\mathrm{O}=-4.18$ eV. The resulting values are summarized in \cref{Tab:BulkUSens}.

\begin{table}[h!]
\centering
\caption{Converged bulk oxygen-defect energetics from PBE and DFT+$U$ calculations. Incorporation energies are reported using $\mu_\mathrm{O}=-4.18$ eV. Negative binding energies indicate favorable binding.}
\footnotesize
\begin{tabular}{lcc}
\hline
Quantity & PBE/$U_\text{eff}=0$ eV & $U_\text{eff}=1$ eV \\
\hline
$E_\text{inc}(\mathrm{O_N^{(b)}})$ & $-7.213$ & $-7.028$ \\
$E_\text{inc}(\text{O}_i^{(\text{b})})$ & $-2.942$ & $-2.760$ \\
$E_\text{inc}(\text{O}_i^{(\text{b})})-E_\text{inc}(\mathrm{O_N^{(b)}})$ & $+4.271$ & $+4.269$ \\
$E_b({\mathrm{O_N}:v_\mathrm{N}}^{(b)})$ & $+0.027$ & $-0.010$ \\
\hline
\end{tabular}
\label{Tab:BulkUSens}
\end{table}

The PBE and $U_\text{eff}=1$ eV results give essentially the same qualitative picture. In both cases, substitutional oxygen on the nitrogen sublattice is more stable than interstitial oxygen by approximately $4.27$ eV. The ${\mathrm{O_N}:v_\mathrm{N}}^{(b)}$ binding energy also remains very close to zero, changing from $+0.027$ eV in PBE to $-0.010$ eV at $U_\text{eff}=1$ eV. Thus, the conclusion that oxygen predominantly occupies nitrogen sublattice sites, while the neighboring nitrogen vacancy is only weakly bound to $\mathrm{O_N}$, is not altered by applying a modest Hubbard correction.

The commonly adopted $U_\text{eff}$ value for UN is approximately $1.9$ eV \cite{Claisse2016b,Kocevski2022I}. However, we do not report quantitative defect energetics at $U_\text{eff} > 1$~eV because the electronic minimization becomes substantially more sensitive to metastable occupation states. A complete treatment of larger $U_\text{eff}$ values would require a dedicated minimization strategy, such as occupation-matrix control or gradual $U$-ramping from a converged lower-$U_\text{eff}$ solution. Therefore, the DFT+$U$ comparison is used here only as a limited robustness check of the bulk oxygen-defect ordering.

% This appendix is an initial uncertainty assessment. The $\pm0.1$~eV bounds and the $0.08/0.02$~eV shared/residual split are physically motivated but not yet calibrated from direct Hubbard-$U$ sensitivity calculations. They are used here to test whether the prediction is stable under a reasonable bounded perturbation of the dominant energetic inputs. The planned $U_\mathrm{eff}=1$ and $2$~eV calculations will replace this assumed uncertainty model with measured energy shifts. In particular, those calculations will determine how much of the DFT+$U$ response is common-mode within the oxygen-related pair and within the vacancy-related pair, and how much is residual. The present correlated Monte Carlo framework is designed so that these calibrated shifts can be inserted directly once the new calculations are complete.

\FloatBarrier
\newpage

\bibliographystyle{elsarticle-num}
\bibliography{ref}

\begin{thebibliography}{10}
\expandafter\ifx\csname url\endcsname\relax
  \def\url#1{\texttt{#1}}\fi
\expandafter\ifx\csname urlprefix\endcsname\relax\def\urlprefix{URL }\fi
\expandafter\ifx\csname href\endcsname\relax
  \def\href#1#2{#2} \def\path#1{#1}\fi

\bibitem{Wallenius2020}
J.~Wallenius, Nitride fuels, Comprehensive Nuclear Materials 5 (2020) 88--101.
\newblock \href {https://doi.org/10.1016/B978-0-12-803581-8.11694-7} {\path{doi:10.1016/B978-0-12-803581-8.11694-7}}.

\bibitem{Uno2020}
M.~Uno, T.~Nishi, M.~Takano, Thermodynamic and thermophysical properties of the actinide nitrides, Comprehensive Nuclear Materials: Second Edition (2020) 202--231\href {https://doi.org/10.1016/B978-0-12-803581-8.11749-7} {\path{doi:10.1016/B978-0-12-803581-8.11749-7}}.

\bibitem{AbdulHameed2024b}
M.~AbdulHameed, B.~Beeler, A.~Claisse, Atomistic investigation of plastic deformation and dislocation motion in uranium mononitride, Applied Sciences 15 (2024) 2666.
\newblock \href {https://doi.org/https://doi.org/10.3390/app15052666} {\path{doi:https://doi.org/10.3390/app15052666}}.

\bibitem{AbdulHameed2025}
M.~AbdulHameed, B.~Beeler, C.~O. Galvin, M.~W. Cooper, N.~Elamrawy, A.~Claisse, \href{https://www.sciencedirect.com/science/article/pii/S0022311525005471}{Molecular-dynamics study of diffusional creep in uranium mononitride}, Journal of Nuclear Materials 617 (2025) 156153.
\newblock \href {https://doi.org/https://doi.org/10.1016/j.jnucmat.2025.156153} {\path{doi:https://doi.org/10.1016/j.jnucmat.2025.156153}}.
\newline\urlprefix\url{https://www.sciencedirect.com/science/article/pii/S0022311525005471}

\bibitem{AbdulHameed2026}
M.~AbdulHameed, B.~Beeler, The contribution of nitrogen {Frenkel-pair} formation to the high-temperature heat capacity of uranium mononitride, Computational Materials Science 272 (2026) 114827.
\newblock \href {https://doi.org/https://doi.org/10.1016/j.commatsci.2026.114827} {\path{doi:https://doi.org/10.1016/j.commatsci.2026.114827}}.

\bibitem{Rizk2025}
J.~T. Rizk, M.~W. Cooper, P.-C.~A. Simon, A.~J. Schneider, D.~A. Andersson, S.~R. Novascone, C.~Matthews, Mechanistic nuclear fuel performance modeling of uranium nitride, Journal of Nuclear Materials 606 (2025) 155604.
\newblock \href {https://doi.org/https://doi.org/10.1016/j.jnucmat.2024.155604} {\path{doi:https://doi.org/10.1016/j.jnucmat.2024.155604}}.

\bibitem{Ronchi1975}
C.~Ronchi, C.~Sari, Swelling analysis of highly rated {MX-type} {LMFBR} fuels. {I}. {Restructuring} and porosity behaviour, Journal of Nuclear Materials 58 (1975).
\newblock \href {https://doi.org/10.1016/0022-3115(75)90100-2} {\path{doi:10.1016/0022-3115(75)90100-2}}.

\bibitem{Ronchi1978}
C.~Ronchi, I.~L. Ray, H.~Thiele, J.~van~de Laar, Swelling analysis of highly-rated {MX-type} {LMFBR} fuels: {II}. {Microscopic} swelling behaviour, Journal of Nuclear Materials 74 (1978).
\newblock \href {https://doi.org/10.1016/0022-3115(78)90359-8} {\path{doi:10.1016/0022-3115(78)90359-8}}.

\bibitem{Rogozkin2003}
B.~D. Rogozkin, N.~M. Stepennova, A.~A. Proshkin, Mononitride fuel for fast reactors, Atomic Energy 95 (2003).
\newblock \href {https://doi.org/10.1023/B:ATEN.0000007886.86817.32} {\path{doi:10.1023/B:ATEN.0000007886.86817.32}}.

\bibitem{Schuler2017}
T.~Schuler, D.~A. Lopes, A.~Claisse, P.~Olsson, Transport properties of {C} and {O} in {UN} fuels, Physical Review B 95 (2017).
\newblock \href {https://doi.org/10.1103/PhysRevB.95.094117} {\path{doi:10.1103/PhysRevB.95.094117}}.

\bibitem{Zinkle1987a}
S.~J. Zinkle, L.~E. Seitzman, W.~G. Wolfer, Stability of vacancy clusters in metals: {I}. energy calculations for pure metals, Philosophical Magazine A 55 (1987).

\bibitem{Zinkle1987b}
S.~J. Zinkle, W.~G. Wolfer, G.~L. Kulcinski, L.~E. Seitzman, Stability of vacancy clusters in metals: {II}. effect of oxygen and helium on void formation in metals, Philosophical Magazine A 55 (1987).

\bibitem{Zinkle1990}
S.~J. Zinkle, E.~H. Lee, Effect of oxygen on vacancy cluster morphology in metals, Metallurgical Transactions A 21 (1990).
\newblock \href {https://doi.org/10.1007/BF02656525} {\path{doi:10.1007/BF02656525}}.

\bibitem{Igata1998}
N.~Igata, A.~Ryazanov, D.~N. Korolev, Effect of light impurities on the early stage of swelling in austenitic stainless steel, Journal of Nuclear Materials 258-263 (1998).
\newblock \href {https://doi.org/10.1016/S0022-3115(98)00404-8} {\path{doi:10.1016/S0022-3115(98)00404-8}}.

\bibitem{McLean1957}
D.~McLean, Grain Boundaries in Metals, Monographs on the Physics and Chemistry of Materials, Clarendon Press, Oxford, 1957.

\bibitem{Kotomin2008}
E.~A. Kotomin, Y.~A. Mastrikov, First principles modeling of oxygen impurities in {UN} nuclear fuels, Journal of Nuclear Materials 377 (2008) 492--495.
\newblock \href {https://doi.org/10.1016/j.jnucmat.2008.04.015} {\path{doi:10.1016/j.jnucmat.2008.04.015}}.

\bibitem{Kotomin2009}
E.~A. Kotomin, Y.~A. Mastrikov, S.~N. Rashkeev, P.~V. Uffelen, Implementing first principles calculations of defect migration in a fuel performance code for {UN} simulations, Journal of Nuclear Materials 393 (2009) 292--299.
\newblock \href {https://doi.org/10.1016/j.jnucmat.2009.06.016} {\path{doi:10.1016/j.jnucmat.2009.06.016}}.

\bibitem{Zhukovskii2009JNM}
Y.~F. Zhukovskii, D.~Bocharov, E.~A. Kotomin, Chemisorption of molecular oxygen on the {UN} (001) surface: Ab initio calculations, Journal of Nuclear Materials 393 (2009).
\newblock \href {https://doi.org/10.1016/j.jnucmat.2009.07.010} {\path{doi:10.1016/j.jnucmat.2009.07.010}}.

\bibitem{Zhukovskii2009SS}
Y.~F. Zhukovskii, D.~Bocharov, E.~A. Kotomin, R.~A. Evarestov, A.~V. Bandura, First principles calculations of oxygen adsorption on the {UN} (001) surface, Surface Science 603 (2009).
\newblock \href {https://doi.org/10.1016/j.susc.2008.10.019} {\path{doi:10.1016/j.susc.2008.10.019}}.

\bibitem{Bocharov2011JNM}
D.~Bocharov, D.~Gryaznov, Y.~F. Zhukovskii, E.~A. Kotomin, Ab initio modeling of oxygen impurity atom incorporation into uranium mononitride surface and sub-surface vacancies, Journal of Nuclear Materials 416 (2011).
\newblock \href {https://doi.org/10.1016/j.jnucmat.2010.11.090} {\path{doi:10.1016/j.jnucmat.2010.11.090}}.

\bibitem{Bocharov2011SS}
D.~Bocharov, D.~Gryaznov, Y.~F. Zhukovskii, E.~A. Kotomin, {DFT} calculations of point defects on {UN} (001) surface, Surface Science 605 (2011) 396--400.
\newblock \href {https://doi.org/10.1016/j.susc.2010.11.007} {\path{doi:10.1016/j.susc.2010.11.007}}.

\bibitem{Bocharov2013}
D.~Bocharov, D.~Gryaznov, Y.~F. Zhukovskii, E.~A. Kotomin, Ab initio simulations of oxygen interaction with surfaces and interfaces in uranium mononitride, Journal of Nuclear Materials 435 (2013).
\newblock \href {https://doi.org/10.1016/j.jnucmat.2012.12.031} {\path{doi:10.1016/j.jnucmat.2012.12.031}}.

\bibitem{Lopes2016}
D.~A. Lopes, A.~Claisse, P.~Olsson, Ab-initio study of {C} and {O} impurities in uranium nitride, Journal of Nuclear Materials 478 (2016).
\newblock \href {https://doi.org/10.1016/j.jnucmat.2016.06.008} {\path{doi:10.1016/j.jnucmat.2016.06.008}}.

\bibitem{Zergoug2018}
T.~Zergoug, S.~E. Abaidia, A.~Nedjar, Oxygen diffusion and migration in clean and defective uranium nitride ({UN}) (001) surfaces, Computational Materials Science 144 (2018).
\newblock \href {https://doi.org/10.1016/j.commatsci.2017.12.003} {\path{doi:10.1016/j.commatsci.2017.12.003}}.

\bibitem{Sikorski2021}
E.~L. Sikorski, B.~J. Jaques, L.~Li, First-principles magnetic treatment of the uranium nitride (100) surface and effect on corrosion initiation, Journal of Applied Physics 130 (2021).
\newblock \href {https://doi.org/10.1063/5.0056904} {\path{doi:10.1063/5.0056904}}.

\bibitem{Kocevski2022I}
V.~Kocevski, D.~A. Rehn, M.~W.~D. Cooper, D.~A. Andersson, First-principles investigation of uranium mononitride ({UN}): Effect of magnetic ordering, spin-orbit interactions and exchange correlation functional, Journal of Nuclear Materials 559 (2022).
\newblock \href {https://doi.org/10.1016/j.jnucmat.2021.153401} {\path{doi:10.1016/j.jnucmat.2021.153401}}.

\bibitem{Lyubimov2014}
D.~Y. Lyubimov, A.~V. Androsov, G.~S. Bulatov, K.~N. Gedgovd, Thermodynamic modeling of oxygen dissolution in uranium mononitride at 900-1400 {K}, Radiochemistry 56 (2014).
\newblock \href {https://doi.org/10.1134/S1066362214050087} {\path{doi:10.1134/S1066362214050087}}.

\bibitem{Mishchenko2021}
Y.~Mishchenko, K.~D. Johnson, D.~Jädernäs, J.~Wallenius, D.~A. Lopes, Uranium nitride advanced fuel: an evaluation of the oxidation resistance of coated and doped grains, Journal of Nuclear Materials 556 (2021).
\newblock \href {https://doi.org/10.1016/j.jnucmat.2021.153249} {\path{doi:10.1016/j.jnucmat.2021.153249}}.

\bibitem{Kresse1993}
G.~Kresse, J.~Hafner, Ab initio molecular dynamics for liquid metals, Physical Review B 47 (1993).
\newblock \href {https://doi.org/10.1103/PhysRevB.47.558} {\path{doi:10.1103/PhysRevB.47.558}}.

\bibitem{Kresse1996a}
G.~Kresse, J.~Furthmüller, Efficiency of ab-initio total energy calculations for metals and semiconductors using a plane-wave basis set, Computational Materials Science 6 (1996).
\newblock \href {https://doi.org/10.1016/0927-0256(96)00008-0} {\path{doi:10.1016/0927-0256(96)00008-0}}.

\bibitem{Kresse1996b}
G.~Kresse, J.~Furthmüller, Efficient iterative schemes for ab initio total-energy calculations using a plane-wave basis set, Physical Review B - Condensed Matter and Materials Physics 54 (1996).
\newblock \href {https://doi.org/10.1103/PhysRevB.54.11169} {\path{doi:10.1103/PhysRevB.54.11169}}.

\bibitem{Perdew1996}
J.~P. Perdew, K.~Burke, M.~Ernzerhof, Generalized gradient approximation made simple, Physical Review Letters 77 (1996).
\newblock \href {https://doi.org/10.1103/PhysRevLett.77.3865} {\path{doi:10.1103/PhysRevLett.77.3865}}.

\bibitem{Kresse1999}
G.~Kresse, D.~Joubert, From ultrasoft pseudopotentials to the projector augmented-wave method, Physical Review B - Condensed Matter and Materials Physics 59 (1999).
\newblock \href {https://doi.org/10.1103/PhysRevB.59.1758} {\path{doi:10.1103/PhysRevB.59.1758}}.

\bibitem{Methfessel1989}
M.~Methfessel, A.~T. Paxton, High-precision sampling for {Brillouin-zone} integration in metals, Physical Review B 40 (1989).
\newblock \href {https://doi.org/10.1103/PhysRevB.40.3616} {\path{doi:10.1103/PhysRevB.40.3616}}.

\bibitem{Hayes1990I}
S.~L. Hayes, J.~K. Thomas, K.~L. Peddicord, Material property correlations for uranium mononitride {I}. physical properties, Journal of Nuclear Materials 171 (1990) 262--270.

\bibitem{Monkhorst1976}
H.~J. Monkhorst, J.~D. Pack, Special points for {Brillouin-zone} integrations, Physical Review B 13 (1976).
\newblock \href {https://doi.org/10.1103/PhysRevB.13.5188} {\path{doi:10.1103/PhysRevB.13.5188}}.

\bibitem{Henkelman2000a}
G.~Henkelman, H.~Jónsson, Improved tangent estimate in the nudged elastic band method for finding minimum energy paths and saddle points, Journal of Chemical Physics 113 (2000).
\newblock \href {https://doi.org/10.1063/1.1323224} {\path{doi:10.1063/1.1323224}}.

\bibitem{Henkelman2000b}
G.~Henkelman, B.~P. Uberuaga, H.~Jónsson, Climbing image nudged elastic band method for finding saddle points and minimum energy paths, Journal of Chemical Physics 113 (2000).
\newblock \href {https://doi.org/10.1063/1.1329672} {\path{doi:10.1063/1.1329672}}.

\bibitem{Stukowski2010}
A.~Stukowski, Visualization and analysis of atomistic simulation data with {OVITO}--the open visualization tool, Modelling and Simulation in Materials Science and Engineering 18 (2010).
\newblock \href {https://doi.org/10.1088/0965-0393/18/1/015012} {\path{doi:10.1088/0965-0393/18/1/015012}}.

\bibitem{Momma2008}
K.~Momma, F.~Izumi, {VESTA}: A three-dimensional visualization system for electronic and structural analysis, Journal of Applied Crystallography 41 (2008).
\newblock \href {https://doi.org/10.1107/S0021889808012016} {\path{doi:10.1107/S0021889808012016}}.

\bibitem{AbdulHameed2024}
M.~AbdulHameed, B.~Beeler, C.~O. Galvin, M.~W. Cooper, Assessment of uranium nitride interatomic potentials, Journal of Nuclear Materials (2024) 155247\href {https://doi.org/https://doi.org/10.1016/j.jnucmat.2024.155247} {\path{doi:https://doi.org/10.1016/j.jnucmat.2024.155247}}.

\bibitem{Huang2020}
G.~Y. Huang, G.~Pastore, B.~D. Wirth, First-principles study of intrinsic point defects and {Xe} impurities in uranium monocarbide, Journal of Applied Physics 128 (2020).
\newblock \href {https://doi.org/10.1063/5.0021951} {\path{doi:10.1063/5.0021951}}.

\bibitem{Yang2021}
L.~Yang, N.~Kaltsoyannis, Incorporation of {Kr} and {Xe} in uranium mononitride: A density functional theory study, Journal of Physical Chemistry C 125 (2021) 26999--27008.
\newblock \href {https://doi.org/10.1021/acs.jpcc.1c08523} {\path{doi:10.1021/acs.jpcc.1c08523}}.

\bibitem{Woodward1998}
C.~Woodward, S.~Kajihara, L.~H. Yang, Site preferences and formation energies of substitutional {Si}, {Nb}, {Mo}, {Ta}, and {W} solid solutions in {L1$_0$} {Ti-Al}, Physical Review B 57 (1998).

\bibitem{Finnis2005}
M.~W. Finnis, A.~Y. Lozovoi, A.~Alavi, The oxidation of {NiAl}: What can we learn from ab initio calculations? (2005).
\newblock \href {https://doi.org/10.1146/annurev.matsci.35.101503.091652} {\path{doi:10.1146/annurev.matsci.35.101503.091652}}.

\bibitem{Mansur1983}
L.~K. Mansur, W.~A. Coghlan, Mechanisms of helium interaction with radiation effects in metals and alloys: A review, Journal of Nuclear Materials 119 (1983).
\newblock \href {https://doi.org/10.1016/0022-3115(83)90047-8} {\path{doi:10.1016/0022-3115(83)90047-8}}.

\bibitem{Saha2018}
U.~Saha, K.~Devan, A.~Bachchan, G.~Pandikumar, S.~Ganesan, Neutron radiation damage studies in the structural materials of a 500 {MWe} fast breeder reactor using {DPA} cross-sections from {ENDF/B-VII.1}, Pramana - Journal of Physics 90 (2018).
\newblock \href {https://doi.org/10.1007/s12043-018-1536-y} {\path{doi:10.1007/s12043-018-1536-y}}.

\bibitem{Surh2004}
M.~Surh, J.~Sturgeon, W.~Wolfer, Incubation period for void swelling and its dependence on temperature, dose rate, and dislocation structure evolution, in: Effects of Radiation on Materials: 21st International Symposium, ASTM International, 2004.
\newblock \href {https://doi.org/10.1520/STP11245S} {\path{doi:10.1520/STP11245S}}.

\bibitem{Sargeant2021}
E.~Sargeant, F.~Illas, P.~Rodríguez, F.~Calle-Vallejo, Importance of the gas-phase error correction for {O2} when using {DFT} to model the oxygen reduction and evolution reactions, Journal of Electroanalytical Chemistry 896 (2021).
\newblock \href {https://doi.org/10.1016/j.jelechem.2021.115178} {\path{doi:10.1016/j.jelechem.2021.115178}}.

\bibitem{Linstrom2024}
P.~J. Linstrom, W.~G. Mallard, \href{https://doi.org/10.18434/T4D303}{NIST Chemistry WebBook, NIST Standard Reference Database Number 69}, National Institute of Standards and Technology, 2024.
\newblock \href {https://doi.org/10.18434/T4D303} {\path{doi:10.18434/T4D303}}.
\newline\urlprefix\url{https://doi.org/10.18434/T4D303}

\bibitem{Kroger1956}
F.~A. Kröger, H.~J. Vink, Relations between the concentrations of imperfections in crystalline solids, Solid State Physics - Advances in Research and Applications 3 (1956).
\newblock \href {https://doi.org/10.1016/S0081-1947(08)60135-6} {\path{doi:10.1016/S0081-1947(08)60135-6}}.

\bibitem{Cooper2023}
M.~W. Cooper, J.~Rizk, C.~Matthews, V.~Kocevski, G.~T. Craven, T.~Gibson, D.~A. Andersson, Simulations of self- and {Xe} diffusivity in uranium mononitride including chemistry and irradiation effects, Journal of Nuclear Materials 587 (2023).
\newblock \href {https://doi.org/10.1016/j.jnucmat.2023.154685} {\path{doi:10.1016/j.jnucmat.2023.154685}}.

\bibitem{Kocevski2022II}
V.~Kocevski, M.~W. Cooper, A.~J. Claisse, D.~A. Andersson, Development and application of a uranium mononitride ({UN}) potential: Thermomechanical properties and {Xe} diffusion, Journal of Nuclear Materials 562 (2022).
\newblock \href {https://doi.org/10.1016/j.jnucmat.2022.153553} {\path{doi:10.1016/j.jnucmat.2022.153553}}.

\bibitem{Claisse2016}
A.~Claisse, T.~Schuler, D.~A. Lopes, P.~Olsson, Transport properties in dilute {UN(X)} solid solutions ({X = Xe, Kr}), Physical Review B 94 (2016).
\newblock \href {https://doi.org/10.1103/PhysRevB.94.174302} {\path{doi:10.1103/PhysRevB.94.174302}}.

\bibitem{Olander2017}
D.~R. Olander, A.~T. Motta, Light Water Reactor Materials Volume I: Fundamentals, American Nuclear Society, 2017.

\bibitem{Laughlin2014}
D.~E. Laughlin, K.~Hono (Eds.), Physical Metallurgy, 5th Edition, Vol.~1, Elsevier, 2014.

\bibitem{Turos1990}
A.~Turos, S.~Fritz, H.~Matzke, Defects in ion-implanted uranium nitride, Physical Review B 41 (1990).
\newblock \href {https://doi.org/10.1103/PhysRevB.41.3968} {\path{doi:10.1103/PhysRevB.41.3968}}.

\bibitem{Johnson2018}
K.~D. Johnson, D.~A. Lopes, Grain growth in uranium nitride prepared by spark plasma sintering, Journal of Nuclear Materials 503 (2018).
\newblock \href {https://doi.org/10.1016/j.jnucmat.2018.02.041} {\path{doi:10.1016/j.jnucmat.2018.02.041}}.

\bibitem{Nichenko2014}
S.~Nichenko, D.~Staicu, Thermal conductivity of porous {UO$_2$}: Molecular dynamics study, Journal of Nuclear Materials 454 (2014).
\newblock \href {https://doi.org/10.1016/j.jnucmat.2014.08.009} {\path{doi:10.1016/j.jnucmat.2014.08.009}}.

\bibitem{Matzke1986}
H.~Matzke, Science of advanced {LMFBR} fuels, Elsevier Science Publishers B.V., 1986.

\bibitem{Colin1983}
M.~Colin, M.~Coquerelle, I.~L. Ray, C.~Ronchi, C.~T. Walker, H.~Blank, The sodium-bonding pin concept for advanced fuels. {Part} {I}: Swelling of carbide fuel up to 12% burnup, Nuclear Technology 63 (1983).
\newblock \href {https://doi.org/10.13182/NT83-A33271} {\path{doi:10.13182/NT83-A33271}}.

\bibitem{Javed1972}
N.~A. Javed, Oxygen solubility in uranium mononitride phase, Journal of the Less Common Metals 29 (1972).
\newblock \href {https://doi.org/10.1016/0022-5088(72)90186-5} {\path{doi:10.1016/0022-5088(72)90186-5}}.

\bibitem{Jain1993}
G.~C. Jain, C.~Ganguly, Experimental evaluation of oxygen solubility in {UN}, {PuN} and {(U,~Pu)N}, Journal of Nuclear Materials 202 (1993).
\newblock \href {https://doi.org/10.1016/0022-3115(93)90394-E} {\path{doi:10.1016/0022-3115(93)90394-E}}.

\bibitem{Konovalov2016}
I.~I. Konovalov, B.~A. Tarasov, E.~M. Glagovsky, Structural-phase state and creep of mixed nitride fuel, in: IOP Conference Series: Materials Science and Engineering, Vol. 130, 2016.
\newblock \href {https://doi.org/10.1088/1757-899X/130/1/012030} {\path{doi:10.1088/1757-899X/130/1/012030}}.

\bibitem{Jaques2015}
B.~J. Jaques, J.~Watkins, J.~R. Croteau, G.~A. Alanko, B.~Tyburska-Püschel, M.~Meyer, P.~Xu, E.~J. Lahoda, D.~P. Butt, Synthesis and sintering of {UN-UO$_2$} fuel composites, Journal of Nuclear Materials 466 (2015).
\newblock \href {https://doi.org/10.1016/j.jnucmat.2015.06.029} {\path{doi:10.1016/j.jnucmat.2015.06.029}}.

\bibitem{Fultz2020}
B.~Fultz, Phase Transitions in Materials, 2nd Edition, Cambridge University Press, 2020.

\bibitem{Tolman1949}
R.~C. Tolman, The effect of droplet size on surface tension, The Journal of Chemical Physics 17 (1949).
\newblock \href {https://doi.org/10.1063/1.1747247} {\path{doi:10.1063/1.1747247}}.

\bibitem{Chhapadia2011}
P.~Chhapadia, P.~Mohammadi, P.~Sharma, Curvature-dependent surface energy and implications for nanostructures, Journal of the Mechanics and Physics of Solids 59 (2011).
\newblock \href {https://doi.org/10.1016/j.jmps.2011.06.007} {\path{doi:10.1016/j.jmps.2011.06.007}}.

\bibitem{Wang2021}
D.~Wang, Z.~Hu, G.~Peng, Y.~Yin, Surface energy of curved surface based on {Lennard‐Jones} potential, Nanomaterials 11 (2021).
\newblock \href {https://doi.org/10.3390/nano11030686} {\path{doi:10.3390/nano11030686}}.

\bibitem{Claisse2016b}
A.~Claisse, M.~Klipfel, N.~Lindbom, M.~Freyss, P.~Olsson, {GGA$+U$} study of uranium mononitride: A comparison of the {U}-ramping and occupation matrix schemes and incorporation energies of fission products, Journal of Nuclear Materials 478 (2016) 119--124.
\newblock \href {https://doi.org/10.1016/j.jnucmat.2016.06.007} {\path{doi:10.1016/j.jnucmat.2016.06.007}}.

\end{thebibliography}

\end{document}